\documentclass{aa}

\usepackage{graphicx}
\usepackage{amsmath}
\usepackage{amssymb}
\usepackage{txfonts}
\usepackage{hyperref}
\usepackage[dvipsnames]{xcolor}
\usepackage{soul}

\newcommand{\ros}{{\it ROSAT}}
\newcommand{\chan}{{\it Chandra}}
\newcommand{\xmm}{{\it XMM-Newton}}
\newcommand{\gaia}{{\it Gaia}}

\newcommand{\eROS}{eROSITA}
\newcommand{\newat}{{\it NewAthena}}
\newcommand{\swift}{{\it Swift}}
\newcommand{\fermi}{{\it Fermi-LAT}}

\newcommand{\fast}{FAST}
\newcommand{\nh}{N_{\rm H}}
\newcommand{\fluxcgs}{erg\,s$^{-1}$\,cm$^{-2}$}
\newcommand{\lumcgs}{erg\,s$^{-1}$}

\def \msev{M7}

\def \carINS{\object{2XMM~J104608.7$-$594306}}

\def \joeoe{\object{4XMM J181844.3$-$120751}}
\def \jottt{\object{4XMM J123337.8$+$374127}}
\def \jofzt{\object{4XMM J140340.4$-$603007}}
\def \jzozt{\object{4XMM J010331.0$+$250540}}
\def \jonfs{\object{4XMM J194744.5$+$274220}}
\def \jztto{\object{4XMM J022141.5$-$735632}}
\def \jztoo{\object{4XMM J031146.5$+$412110}}
\def \jttfo{\object{4XMM J225139.6$-$162748}}
\def \joofz{\object{4XMM J114051.9$-$641848}}
\def \josff{\object{4XMM J175437.8$-$294148}}
\def \si{J1818-12}
\def \liu{J1233+37}
\def \shi{J1403-60}
\def \shiyi{J0103+25}
\def \ershi{J1947+27}
\def \ershiyi{J0221-73}
\def \ershisan{J0311+41}
\def \sanshier{J2251-16}

\def \sanshiwu{J1140-64}

\def \sanshijiu{J1754-29}

\begin{document} 

\title{Isolated neutron star candidates from the fourth generation XMM-Newton catalogues}

\author{A.~M.~Pires\inst{1,2}
   \and C.~Motch\inst{3}
   \and J.~Kurpas\inst{2}
   \and A.~D.~Schwope\inst{2}
   \and Baoda~Li\inst{4}
   \and Dejiang~Yin\inst{4}
   \and Li~Ji\inst{5}
   \and Jianzhong~Liu\inst{1}
   \and Lei~Qian\inst{6,7,8,9}
   \and I.~Traulsen\inst{2}
   \and Liyun~Zhang\inst{4}
   \and Zhongli~Zhang\inst{10}}

\offprints{A.~M.~Pires}

\institute{
	Center for Lunar and Planetary Sciences, Institute of Geochemistry, Chinese Academy of Sciences, 99 West Lincheng Rd., 550081 Guiyang, China
	\email{adriana@mail.gyig.ac.cn}
	\and
	Leibniz-Institut f\"ur Astrophysik Potsdam (AIP), An der Sternwarte 16, 14482 Potsdam, Germany 
	\and
	Universit\'e de Strasbourg, CNRS, Observatoire Astronomique de Strasbourg, UMR 7550, F-67000 Strasbourg, France
	\and
	College of Physics, Guizhou University, Guiyang, 550025, China
	\and
	Purple Mountain Observatory, Chinese Academy of Sciences, 10 Yuanhua Road, Nanjing, 210023, China
	\and
	Guizhou Radio Astronomical Observatory, Guizhou University, Guiyang, China
	\and
	National Astronomical Observatories, Chinese Academy of Sciences, 20A Datun Rd., Chaoyang District, Beijing 100101, China
	\and
	College of Astronomy and Space Sciences, University of Chinese Academy of Sciences, Chinese Academy of Sciences, Beijing, China
	\and
	CAS Key Laboratory of FAST, National Astronomical Observatories, Chinese Academy of Sciences, Beijing 100101, China
	\and
	Shanghai Astronomical Observatory, Chinese Academy of Sciences, Shanghai 200030, China}

\date{Received ...; accepted ...}


\titlerunning{Isolated neutron star candidates from XMM-Newton}

\authorrunning{A.~M.~Pires et al.}

\abstract{
	X-ray thermally emitting isolated neutron stars (XINSs) are a rare population, offering insights into neutron star cooling, magnetic field evolution, and Galactic demographics.
	}{
	Using over two decades of observations from the European Space Agency's \xmm\ Observatory, we aim to identify more distant XINSs, extending the population beyond the solar neighbourhood.
	}{
	We searched the 4XMM-DR9 and 4XMM-DR12 releases of the \xmm\ serendipitous source catalogue for absorbed XINS candidates down to $10^{-14}$\,\fluxcgs\ in the 0.5--1\,keV band, selecting sources with soft X-ray spectra and no catalogued optical, ultraviolet, or infrared counterparts. The selected candidates were followed up with \xmm\ and \fast, complemented by data from the \textit{SRG}/\eROS\ All-Sky Survey, \chan, and optical surveys. The observational results were compared with predictions from a population synthesis model.
	}{
	Of the ten sources analysed, the sample comprises five compelling XINS candidates, one previously confirmed XINS (\jztto), two extragalactic contaminants, and two ambiguous sources with limited photon statistics. The five compelling candidates exhibit soft ($kT \sim 80$--$100$\,eV), moderately absorbed, and stable X-ray emission, consistent with distant XINSs. They lie primarily in the Galactic plane, with possible associations with open clusters at $\sim$1.8\,kpc and with an supernova remant--pulsar window nebula system at $\sim$6\,kpc. The population synthesis model predicts a median of $20 \pm 5$ XINSs within the 4XMM-DR12 footprint, of which $6_{-3}^{+2}$ exceed our flux threshold. This is comparable to the number of XINSs identified in the \xmm\ data, provided that additional candidates are validated. The simulations further indicate that a substantial fraction of the population ($13_{-4}^{+6}$ XINSs, $\sim$70\%) remains below the selection threshold. These cooler, more absorbed, and more distant neutron stars are likely present in the 4XMM-DR12 catalogue but remain difficult to identify because of source confusion and limitations in catalogue cross-matching.
	}{
	Firm classification of XINS candidates identified in this work requires deep optical and additional X-ray follow-up. While pulsations may be difficult to detect in the faintest sources with current facilities, future missions such as \newat\ will enable more detailed studies of these distant populations. Improvements to population synthesis models, incorporating more accurate prescriptions for cooling, absorption, and birthrate, together with results from \eROS\ and \xmm, will enable a better understanding of the Galactic evolution and population properties of XINSs.}

\maketitle

\nolinenumbers

\section{Introduction\label{sec_intro}}

After 25 years in space, the European Space Agency's \xmm\ Observatory \citep{2001A&A...365L...1J} has conducted nearly 14,000 targeted observations. Thanks to its large field of view and exceptional sensitivity, each observation typically uncovers between 35 and 100 additional X-ray sources beyond the primary target. The \xmm\ Survey Science Centre (SSC; \citealt{2001A&A...365L..51W}) is responsible for reprocessing this extensive archive and generating comprehensive source catalogues \citep{2009A&A...493..339W, 2016A&A...590A...1R, 2020A&A...641A.136W}. The most recent release, 4XMM-DR14, includes over a million detections, almost 700,000 of which are unique sources. These detections, which often involve hundreds of photon events, benefit from the observatory's deep exposures, with a typical exposure time of 25\,ks (68th percentile). In comparison, the \textit{SGR/}\eROS\ All-Sky Survey (eRASS; \citealt{2021A&A...647A...1P, 2024A&A...682A..34M}) identified approximately 930,000 X-ray sources across half the sky during its first six months of operation, although with a median of just eight photon counts per source in the 0.2--2.3\,keV energy range. While \xmm\ surveys a smaller fraction of the sky than all-sky surveys (e.g.~\eROS), its competitive ability to discover and characterise rare, extreme, and previously unknown objects makes it an invaluable tool in the field.

In the low-luminosity regime ($L_{\rm X} \lesssim 10^{33}$\,\lumcgs), source populations in the Milky Way are significantly affected by interstellar extinction, limiting our understanding primarily to nearby objects. 
This is particularly true for the seven radio-quiet, thermally emitting isolated neutron stars (XINSs) discovered through \ros\ observations \citep{1999A&A...349..389V}. These middle-aged, cooling neutron stars, collectively known as the Magnificent Seven \citep[\msev;][]{2007Ap&SS.308..181H,2009ASSL..357..141T}, form a local group with similar spectral and timing characteristics \citep[see][for recent reviews]{2023Univ....9..273P,2024mbhe.confE..55R}. Unlike other X-ray detected isolated neutron stars, their emission is predominantly thermal, believed to originate from the stellar surface. This makes them invaluable for testing models of neutron star emission and cooling, particularly in the context of crustal heating driven by magnetic field decay \citep{2015SSRv..191..171P,2020MNRAS.496.5052P,2021ApJ...914..118D,2024MNRAS.533..201A}. 

Furthermore, the \msev\ XINSs are as numerous in the local volume as young radio and gamma-ray pulsars, suggesting that they may constitute an overlooked component of the Galactic neutron star population \citep{2003A&A...406..111P,2008AIPC..983..331K,2008MNRAS.391.2009K}. Identifying similar sources beyond the local solar neighbourhood is essential to understanding their intrinsic properties and their relationship to other classes of Galactic neutron stars.

Searches for XINSs typically focus on identifying sources with no catalogued counterparts and unusually high X-ray-to-optical flux ratios, which distinguish them from more common X-ray emitters. The process begins with the selection of blank field soft X-ray sources, supported by deep, large-scale photometric surveys and efficient catalogue cross-matching. Confirming the XINS nature of these candidates and ruling out contaminants requires advanced observational facilities and dedicated follow-up campaigns. As a result, only a few XINSs have been identified since the \ros\ era, and the confirmed sources exhibit significant diversity in their properties \citep{2008ApJ...672.1137R, 2009A&A...498..233P, 2022A&A...666A.148P, 2023A&A...674A.155K, 2024A&A...683A.164K, 2025A&A...694A.160K, 2026A&A...705A.148K}. The nature of several other proposed candidates \citep[e.g.][]{2009A&A...504..185P, 2022MNRAS.509.1217R, 2024ApJ...961...36D, 2024A&A...687A.251K} remains uncertain, necessitating deeper optical limits and further multiwavelength investigation.

In this work we present the properties of a sample of ten X-ray sources that we identified in the 4XMM-DR9 and 4XMM-DR12 catalogues as faint XINS candidates. Through dedicated follow-up \xmm\ observations, we assessed their long-term flux stability, refined their spectral properties, and improved their localisation. To support their characterisation, we incorporated ancillary X-ray data from \chan\ and \eROS\ (when available), optical imaging from public surveys, and radio data from a follow-up programme with the Five-hundred-meter Aperture Spherical Radio Telescope \citep[\fast;][]{2019SCPMA..6259502J}. 

From this analysis, five sources emerged as compelling XINS candidates, together with one previously confirmed XINS \citep[\jztto;][]{2022A&A...666A.148P}. These results were then compared with predictions from a population synthesis model of cooling isolated neutron stars.

The paper is organised as follows. Section~\ref{sec_selection} describes the selection criteria for viable candidates. Section~\ref{sec_obsdatared} summarises the observational data and reduction procedures. Section~\ref{sec_analysis} presents the analysis and characterisation of the sample, including astrometry, spectroscopy, and timing. In Section~\ref{sec_disc}, we discuss the properties of the compelling and confirmed subsets of sources and their implications for the Galactic XINS population. The paper concludes in Section~\ref{sec_summary}. Additional details are provided in the Appendices. 
Preliminary results, including the science validation of tools and catalogues developed as part of the XMM2ATHENA project \citep{2023AN....34420102W}, were reported in \citet{2025AN....34640116P}.

\section{Selection of candidates\label{sec_selection}}

For each publication of the \xmm\ catalogue, the SSC provides multiwavelength X-ray source identifications and spectral energy distributions from a set of external catalogues selected for their broad sky coverage \citep{2020A&A...641A.136W}.  
The ARCHES\footnote{\url{https://www.arches-fp7.eu}} statistical multi-catalogue cross-matching tool \citep{2017A&A...597A..89P}, developed as part of the Astronomical Resource Cross-matching for High Energy Studies project \citep{2017ASPC..512..165M}, was used to evaluate the identification probability of an X-ray source with a single or a combination of entries in the ultraviolet, optical, infrared, and radio domains. At the time of the 4XMM-DR9 processing, the SSC used GALEX GR6+7 in the ultraviolet, APASSDR9, SDSS DR12, \gaia\ DR2, SkyMapper DR1.1, and Pan-STARRS PS1 in the optical, and 2MASS, ALLWISE, and AKARI/FIS in the infrared. The \xmm\ DR12 release added the XMM-SUSS5.0 UV catalogue \citep[see][for a more detailed description of the cross-matching process and references]{2020A&A...641A.136W}.  
Very high densities of optical or infrared background sources lead to multiple low-probability matches, which are uninformative. Consequently, the LMC, SMC, and M31 regions were excluded from the cross-matching process, as were extended X-ray sources.

Our main selection criteria are designed to identify X-ray sources similar to the \ros-discovered \msev\ XINSs, i.e., those with very faint optical counterparts (V $\geq$ 25, and therefore not present in any of the external catalogues listed above), and with intrinsically soft, blackbody-like X-ray spectra ($kT\sim100$\,eV).  
The first constraint was implemented by selecting only X-ray sources with a probability of no identification greater than 50\% ($P_{\rm no\text{-}id} > 50\%$) with respect to the external catalogues covering the ultraviolet to infrared domains. The X-ray spectral constraint involved selecting a region in the HR$_2$--HR$_3$ hardness ratio\footnote{Hardness ratios (HR) are ratios of the difference to total counts in the five \xmm\ energy bands; see caption of Table~\ref{tab_4xmm}, for details.} space populated by the known \msev\ sources. Since HR$_1$ is primarily influenced by interstellar absorption, no constraint was applied to its value \citep[see][for a detailed description of the source selection and screening]{2022A&A...666A.148P}. 

Only 268 sources with EP\_8\_DET\_ML $>$ 100 and X-ray fluxes greater than $10^{-14}$\,\fluxcgs\ in the 0.5--1\,keV range satisfied the initial selection criteria. Lower maximum-likelihood values and X-ray fluxes generally result in HR$_3$ uncertainties significantly larger than 0.3, increasing the probability of falsely identifying a hard X-ray component.  
We then discarded sources with clear SIMBAD\footnote{\url{https://simbad.u-strasbg.fr}} identifications, those likely due to high proper motion stars, sources relatively close to optical objects but suspected of uncertain astrometry due to their position at the edge of the EPIC field of view, and sources for which faint counterparts were found in deeper optical surveys.  
Among an additional group of 15 sources with 0.5--1\,keV fluxes below our threshold of $10^{-14}$\,\fluxcgs\ but consistent with it within the uncertainties, we retained \josff, an XINS candidate also singled out by \citet{2022MNRAS.509.1217R}. This source was observed repeatedly, lies closest to the flux limit, and has a high detection likelihood, resulting in the most accurate X-ray spectral parameters in this subsample.  
After this screening, nine good candidates remained in the 4XMM-DR9 catalogue, and applying the same procedure to 4XMM-DR12 yielded one additional candidate.

Table~\ref{tab_4xmm} summarises the properties of the ten XINS candidates, including their catalogued fluxes and hardness ratios. We also provide the total hydrogen column density along the line-of-sight \citep{2024arXiv240303127D}, $N_{\rm H}^{\rm Gal}$, and the number of detections by \xmm, both serendipitous and dedicated. Any additional coverage from \chan\ or \eROS\ is indicated in the table for each source.

\section{Observations and data reduction\label{sec_obsdatared}}

\subsection{XMM-Newton}

Among the XINS candidates identified in the 4XMM-DR9 and 4XMM-DR12 catalogues, two (\si\footnote{For convenience, we refer to the sources in this programme by shortened forms of their IAU designations in Table~\ref{tab_4xmm} by truncating the right ascension and declination to the format JHHMM$\pm$DD.} and \sanshijiu) were serendipitously observed more than once by \xmm. Dedicated second-epoch observations of five other candidates (\liu, \shi, \shiyi, \ershiyi, and \sanshier), each initially observed only once, were conducted under follow-up fulfil programmes 088419, 090126, and 092282 between July 2021 and January 2024. The remaining three (\ershi, \ershisan, and \sanshiwu) have not yet been re-observed by \xmm.

For the follow-up programme, we adopted the full-frame data mode with the thin filter for both the EPIC pn \citep{2001A&A...365L..18S} and MOS \citep{2001A&A...365L..27T} cameras. The medium filter and/or window mode was applied when necessary to minimise optical and X-ray loading at the aimpoint for targets at low Galactic latitudes or near bright sources. 

All datasets were reprocessed using the latest calibration files with SAS~21\footnote{\texttt{xmmsas\_20230412\_1735-21.0.0}} using the meta-tasks \texttt{epchain} and \texttt{emchain}. Standard procedures outlined in the EPIC cameras' analysis guidelines\footnote{\url{https://xmmweb.esac.esa.int/docs/documents/CAL-TN-0018.pdf}} were followed. Event lists were filtered to exclude periods of high background activity, as well as bad CCD pixels and columns. Source centroids and optimal extraction regions for each EPIC imaging exposure were defined with the SAS task \texttt{eregionanalyse} in the 0.2--2\,keV energy band. Background regions were selected away from the source, preferably on the same CCD as the target. 

For the spectral analysis, we used datasets filtered for good-time intervals (GTIs), selecting single- and double-pattern events for the pn camera and all valid CCD patterns for MOS. The energy channels were regrouped to ensure a minimum of five or twenty counts per spectral bin, depending on the source brightness and the statistical quality of the dataset. The oversampling of the instrumental resolution was limited to a factor of three.

Table~\ref{tab_obsdet} summarises the observations included in the analysis. It lists the observation date, instrumental set-up, target off-axis angle, net exposure time, percentage of GTIs, and total EPIC target counts in the 0.2--2\,keV energy band.

\subsection{SRG/eROSITA}

Three XINS candidates (\sanshiwu, \shi, and \ershiyi) lie in the western Galactic hemisphere of the eRASS footprint and appear in its first six-month data release \citep{2024A&A...682A..34M}. The \eROS\ data for \ershiyi\ were previously analysed by \citet{2022A&A...666A.148P} using an earlier pipeline (c020); we include them here to verify consistency with the updated calibration.

For these three candidates, we retrieved event lists from the sky tiles covering their positions across four to five all-sky scans (eRASS1--4/5), processed with pipeline version c030 and the latest eSASS calibration files (\texttt{eSASSusers\_240410\_0\_4}; \citealt{2022A&A...661A...1B}). Table~\ref{tab_erass} summarises the corresponding observations and derived detection properties for these three sources. Data were analysed in the 0.2--5\,keV band, including all valid photon patterns and active telescope modules. Cumulative exposures, corrected for vignetting and high-background intervals, range from 1.8 to 2.4\,ks, yielding 35--350 source counts.

Source characterisation was performed using maximum-likelihood point-spread function (PSF) fitting \citep{2022A&A...661A...1B} simultaneously in three energy bands (0.2--0.6, 0.6--2.3, and 2.3--5\,keV), adopting photon mode. The eSASS tasks \texttt{evtool}, \texttt{expmap}, \texttt{ermask}, \texttt{erbox}, \texttt{erbackmap}, \texttt{ermldet}, and \texttt{catprep} were used to create images, exposure and background maps, detector masks, and source catalogues for each individual scan and the cumulative event files.

Using the auto option of \texttt{srctool}, we defined optimised source and background extraction regions from both individual and cumulative eRASS:4 and eRASS:5 event lists\footnote{The notation eRASS:\textit{n} denotes the cumulative dataset obtained from the first $n$ \eROS\ all-sky surveys \citep{2024A&A...682A..34M}.} (Fig.~\ref{fig_eros}). Background regions excluded all X-ray sources within the corresponding annuli. These regions were used to extract spectra along with the associated response matrices and ancillary files, which were regrouped to a minimum of five counts per spectral bin.

\begin{table*}[t]
\small
\caption{Target parameters from \xmm\ spectral PSF fitting and astrometry
\label{tab_psfloc}}
\centering
\begin{tabular}{@{}lccrrccrrccrc@{}}
\hline\hline
Target & \multicolumn{1}{c}{Counts} & \multicolumn{1}{c}{Flux} & \multicolumn{1}{c}{RA\,\tablefootmark{a}} & \multicolumn{1}{c}{Dec\,\tablefootmark{a}} & Error\,\tablefootmark{b} & Shift\,\tablefootmark{c} & \multicolumn{1}{c}{$l$\,\tablefootmark{a}} & \multicolumn{1}{c}{$b$\,\tablefootmark{a}} & $\mathcal{E}$\,\tablefootmark{d} & $\mathcal{P_{\rm var}}$\,\tablefootmark{e} & \multicolumn{1}{c}{$N_{\rm H}$\,\tablefootmark{f}} & $\Gamma$\,\tablefootmark{f}\\
& \multicolumn{1}{c}{($\times10^3$)} & \multicolumn{1}{c}{($\times10^{-14}$\,cgs)} & \multicolumn{1}{c}{(\degr)} & \multicolumn{1}{c}{(\degr)} & (\arcsec) & (\arcsec) & \multicolumn{1}{c}{(\degr)} & \multicolumn{1}{c}{(\degr)} & (\arcsec) & (\%) & & \\
\hline
J1140 & 0.78(3) & $28.3\pm1.7$ & 175.21650 & $-64.31353$ & 0.72 & 0.1 & 295.376 & $-2.482$ & 0 & $-$ & $3.96(17)$ & $5^\star$\\
J1818\tablefootmark{$\ast$} & 2.03(5) & $8.13(21)$ & 274.68463 & $-12.13096$ & 0.27 & 0.2 & 18.424 & $+1.599$ & 0 & 75 & $5.62(15)$ & $5^\star$  \\ 
J1233 & 2.63(7) & $8.8(4)$ & 188.40736 & $+37.68931$ & 0.50 & 5.8 & 141.359 & $+78.794$ & 6.1(7) & 92 & $5.09(15)$ & $5^\star$\\
J1403 & 1.54(5) & $2.59(9)$ & 210.91862 & $-60.50203$ & 0.42 & 0.3 & 311.714 & $+1.132$ & 0 & 49 & $4.30(15)$ & $5^\star$\\
J0103 & 0.28(2) & $2.0(3)$ & 15.87942 & $+25.09448$ & 0.70 & 0.4 & 126.389 & $-37.696$ & 0 & 0.4 & $2.44(28)$ & 4.2(2)\\
J1947 & 1.08(4) & $2.55(14)$ & 296.93574 & $+27.70581$ & 0.47 & 0.1 & 63.734 & $+1.167$ & 0 & $-$ & $5.02(20)$ & $5^\star$\\
J0221 & 4.19(7) & $20.8(4)$ & 35.42317 & $-73.94265$ & 0.34 & 1.5 & 294.757 & $-41.722$ & 0 & 11 & $0.12(9)$ & $5^\star$\\
J0311 & 0.13(2) & $3.2\pm1.4$ & 47.94373 & $+41.35296$ & 1.05 & 0.4 & 149.360 & $-14.207$ & 0 & $-$ & $2.8(4)$ & 4.5(3)\\
J2251 & 0.40(3) & $1.95(17)$ & 342.91515 & $-16.46373$ & 0.60 & 1.4 & 47.755 & $-60.319$ & 0 & 7 & $1.14(14)$ & $5^\star$\\
J1754 & 0.57(4) & $1.56(23)$ & 268.65756 & $-29.69689$ & 0.38 & 0.4 & 0.354 & $-2.076$ & 0 & 47 & $3.52(24)$ & $5^\star$\\
\hline
\end{tabular}
\tablefoot{
The sources are sorted by decreasing flux in the 0.5--1\,keV energy band (\texttt{SC\_EP\_2\_FLUX} in the 4XMM-DR9 and 4XMM-DR12 catalogues). Errors (significant figures in brackets) indicate $1\sigma$ confidence levels. Total EPIC counts and fluxes are given for the 0.2--12\,keV energy band.
\tablefoottext{a}{Astrometrically corrected equatorial (RA, Dec) and Galactic ($l,b$) coordinates of the target.}
\tablefoottext{b}{Positional error (90\% confidence level) in arcseconds, including astrometric uncertainties.}
\tablefoottext{c}{Offset between astrometrically corrected coordinates and the catalogue position, in arcseconds.}
\tablefoottext{d}{Extent parameter $\mathcal{E}$ in arcseconds from PSF fitting.}
\tablefoottext{e}{Probability $\mathcal{P}_{\rm var}$ that flux measurements across epochs are consistent with a constant value (not applicable to single-epoch targets).}
\tablefoottext{f}{Spectral parameters (hydrogen column density $\nh$ in units of $10^{21}$\,cm$^{-2}$ and photon index $\Gamma$) obtained during the enhanced source-detection stage assuming an absorbed power-law model.}
\tablefoottext{$\star$}{Parameter pegged at the boundary of the allowed range.}
\tablefoottext{$\ast$}{Due to the maximum event limit imposed by \texttt{emldetect}, only pn and MOS2 events from the four observations with the highest count statistics were included in the stacked analysis.}
}
\end{table*}

\subsection{\chan}

Two of the XINS candidates, \ershi\ and \sanshijiu, are listed in the \chan\ Source Catalogue \citep{2024ApJS..274...22E} and have archival ACIS observations available (Table~\ref{tab_chanobs}). The data for these two sources were reprocessed with CIAO~v4.17 and CALDB~4.12 using the \texttt{chandra\_repro} script, applying the latest calibrations and excluding high-background intervals. For each observation, we generated images for the active chips, together with aspect histograms, instrument maps, and exposure maps using the standard CIAO tools \texttt{asphist}, \texttt{mkinstmap}, and \texttt{mkexpmap} (Fig.~\ref{fig_chanpsf}).

Source detection was performed on each chip at full spatial resolution with the \texttt{wavdetect} algorithm, using exposure and PSF maps computed at 2.3\,keV. For these two sources, detection was carried out in multiple energy bands and with varying significance thresholds to optimise sensitivity. The source lists from individual chips were then merged to produce a final catalogue for each observation. 

Source photons were extracted from circular regions encompassing $\sim$90\% of the PSF (radii of 5\arcsec--8\arcsec, depending on source position and off-axis angle), with background taken from concentric annuli (15\arcsec--50\arcsec) centred on the \texttt{wavdetect} source position. Spectra and response files (ARF and RMF) were generated following the standard CIAO threads using \texttt{dmextract}, \texttt{mkacisrmf}, \texttt{mkarf}, and \texttt{arfcorr}, and were grouped to a minimum of five counts per bin.

\subsection{FAST}

As part of a programme to follow up XINS candidates from \xmm\ and \textit{SRG}/\eROS\ in the radio with \fast\ (PID: PT2024\_0061), the sources \liu\ and \ershi\ were observed on October 4 and 6, 2024, for 1200\,s and 1800\,s, respectively. Observations were performed in \texttt{Tracking} mode using the central beam of \fast's $L$-band 19-beam receiver \citep{2020RAA....20...64J}, covering the 1.0--1.5\,GHz frequency range with 1024 channels (channel width of 0.488\,MHz). The data were sampled with 8-bit precision in four polarisations at 49.152\,$\mu$s resolution, and recorded in \texttt{PSRFITS} search mode format \citep{2004PASA...21..302H}. The system temperature was approximately 24\,K, and the beam size at 1.4\,GHz was 2.9\arcmin. Both sources lie well within the central beam.

\section{Sample characterisation\label{sec_analysis}}

\subsection{Stacked detection and astrometry\label{sec_sourcedetection}}

\begin{figure*}[t]
\begin{center}
\includegraphics[width=\textwidth]{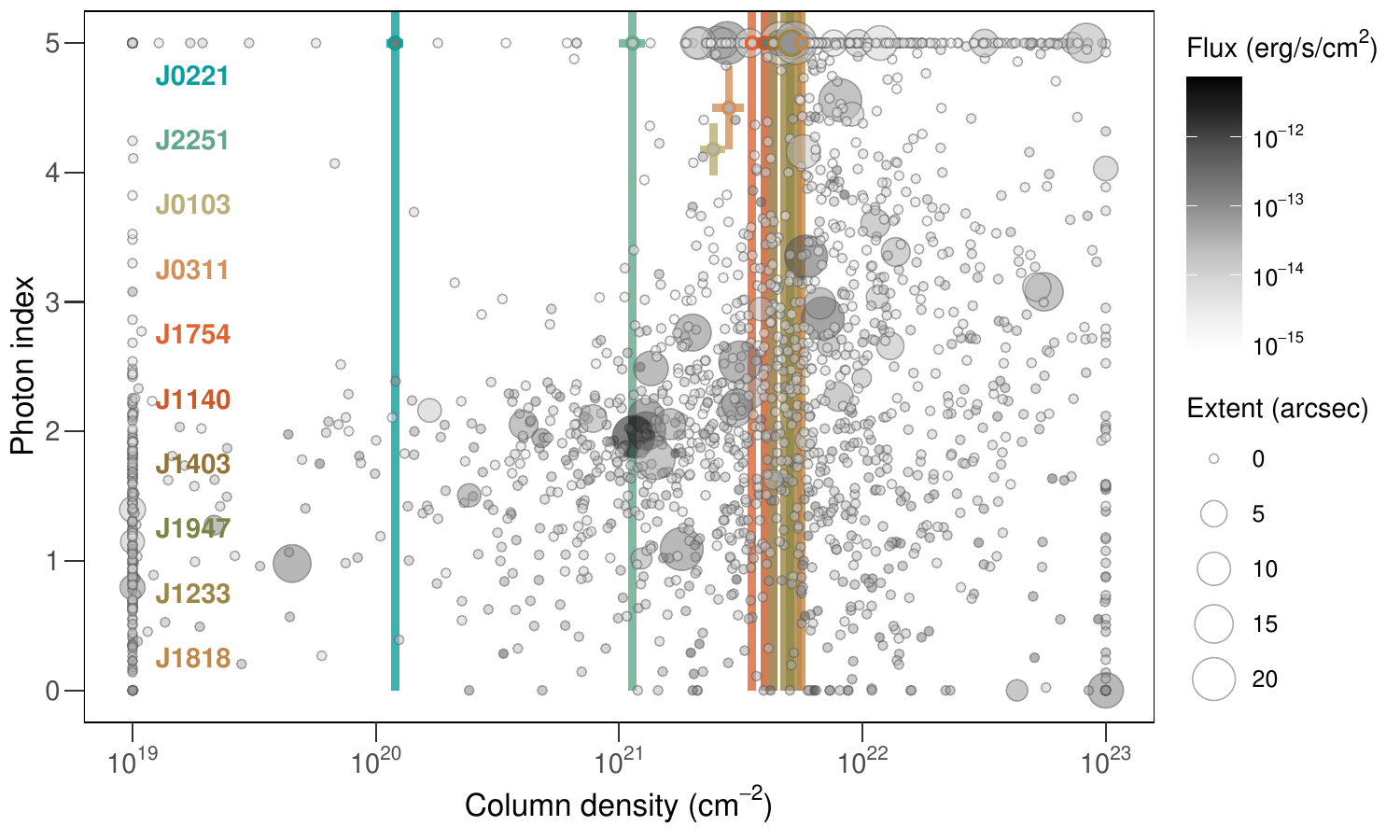}
\end{center}
\caption{Spectral parameters ($\Gamma$ vs.\ $\nh$) for the 1902 X-ray sources detected across the ten stacked sky fields in enhanced detection mode \citep{2023AN....34420102W,2025AN....34640116P}. The size and colour of each marker represent the source extent and flux, respectively. The ten XINS candidates are highlighted with error bars and colour-coded labels, sorted by increasing column density. For all candidates except \shiyi\ and \ershisan, $\Gamma$ is unconstrained and reaches the upper bound of the allowed fitting range (see also Table~\ref{tab_psfloc}).\label{fig_nhgamma}}
\end{figure*}

Source detection was initially performed on each individual \xmm\ observation using the maximum likelihood PSF-fitting method implemented in \texttt{edetect\_stack} \citep{2020A&A...641A.137T}. For each observation, we generated images, background maps, and exposure maps in the five standard \xmm\ energy bands. Detector masks were also generated for the three EPIC cameras. 

To refine the astrometry, we used the SAS task \texttt{eposcorr} to cross-correlate the EPIC source lists with optical objects from the Guide Star Catalogue~2.4.2 \citep[GSC;][]{2008AJ....136..735L} within a 15\arcmin\ radius of the average pointing position. Table~\ref{tab_obsdet} presents the small positional offsets in right ascension ($\Delta\alpha$) and declination ($\Delta\delta$), measured in arcseconds, derived from boresight corrections based on $N_{\rm ref}$ potential X-ray--optical matches for each observation. These offsets were applied to the attitude files, after which source detection and astrometric calibration were repeated using reference coordinates centred on the target positions. For fields observed multiple times, all available observations were then combined (stacked) to refine the source parameters. For the field of \si, observed six times and located near bright X-ray sources linked to massive star clusters in \object{Ser OB2}, stacked detection was limited to the pn camera (30 images) owing to the maximum event count limit imposed by \texttt{emldetect}.

As part of an enhanced source detection method for overlapping observations developed within the XMM2ATHENA project \citep{2023AN....34420102W,2025AN....34640116P}, we also performed spectral PSF fitting for each XINS candidate. This approach assumes a simple spectral model and constant X-ray emission over time, improving sensitivity to faint sources and allowing direct spectral extraction during source detection.

The results of this approach are summarised in Table~\ref{tab_psfloc}. For each XINS candidate, we list the total EPIC counts and flux in the 0.2--12\,keV band, the astrometrically corrected equatorial and Galactic coordinates, the total positional error, the positional shifts relative to the 4XMM-DR9 and DR12 catalogue positions, and the extent parameter. For sources detected across multiple epochs, we also report the variability probability ($\mathcal{P}_{\rm var}$), which quantifies the likelihood that their flux remains consistent over time.

Table~\ref{tab_psfloc} additionally provides the hydrogen column density ($\nh$) and photon index ($\Gamma$) derived from spectral PSF fitting using an absorbed power-law model. In the current implementation of the enhanced source-detection method, the spectral parameters are allowed to vary within $\log(\nh)=19$--$23$ and $\Gamma=1$--$5$.

Figure~\ref{fig_nhgamma} shows the spectral values of the 1902 sources detected across the ten stacked fields, with the XINS candidates highlighted by error bars and colour-coded labels. Owing to their soft spectra, all candidates except \shiyi\ and \ershisan\ exhibit very steep photon indices that reach the upper bound of the allowed range.

We further verified the localisation of the targets using complementary data. Following a similar procedure as for the EPIC datasets, we applied astrometric corrections to our \chan\ and concatenated \eROS\ source catalogues, aligning the X-ray source positions with the GSC. Overall, source characterisation and astrometry are consistent across datasets (Fig.~\ref{fig_chanpsf}; see the absolute angular separations between EPIC and ancillary positions in Tables~\ref{tab_erass} and \ref{tab_chanobs}).

\subsection{Optical depth and counterparts\label{sec_optlim}}

\begin{figure*}[t]
\begin{center}
\includegraphics[width=0.33\textwidth]{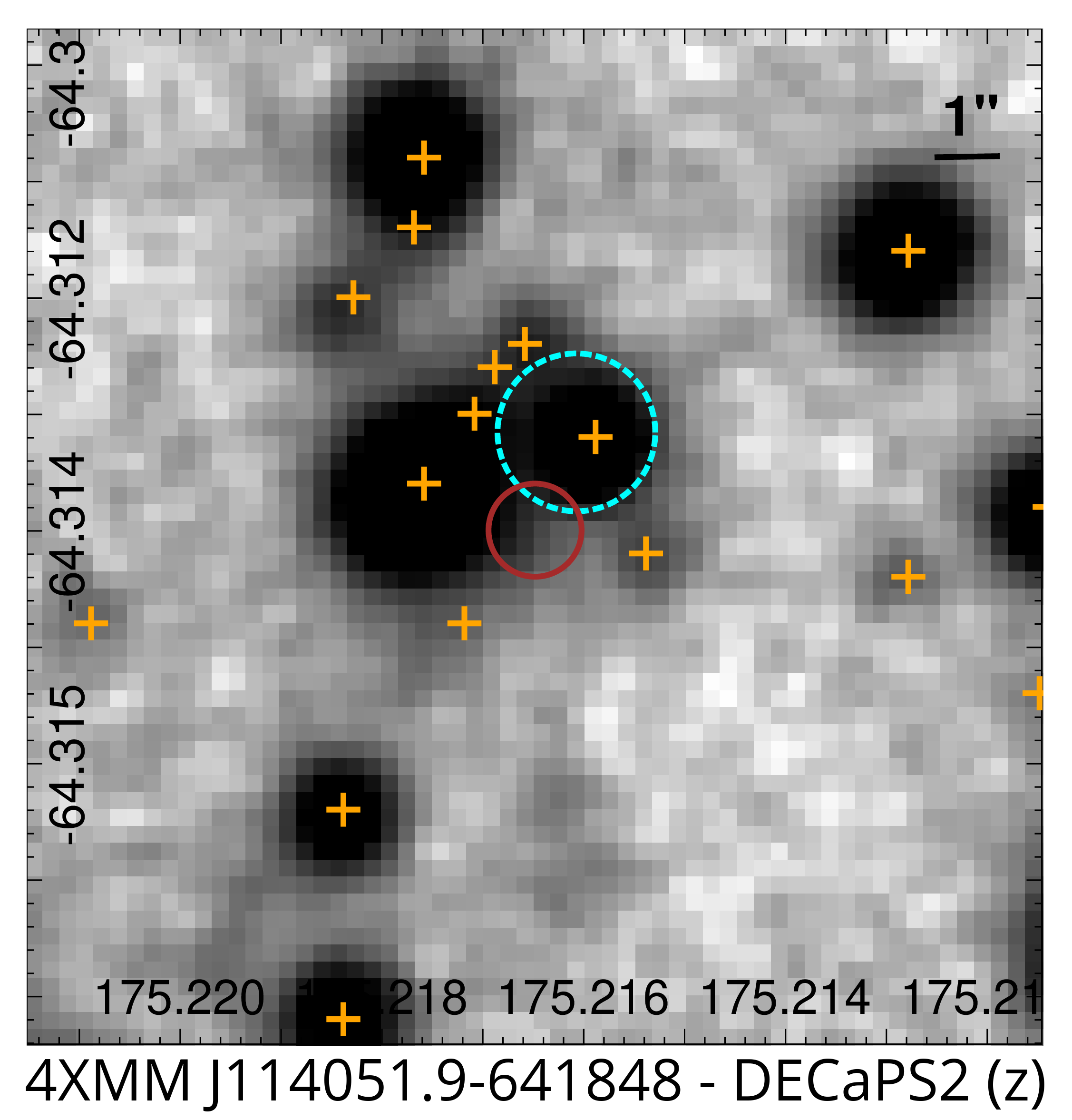}\hfill
\includegraphics[width=0.33\textwidth]{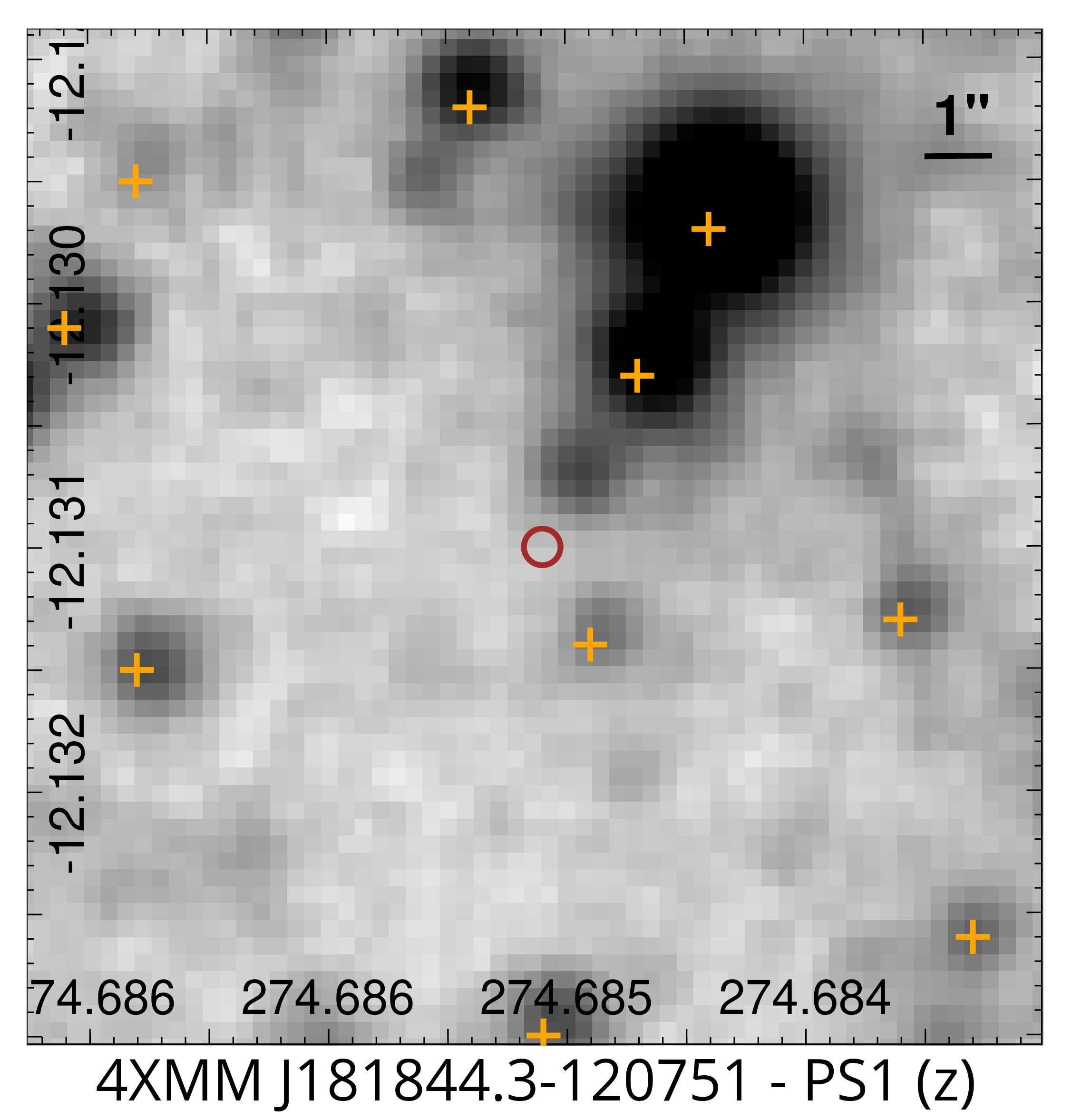}\hfill
\includegraphics[width=0.33\textwidth]{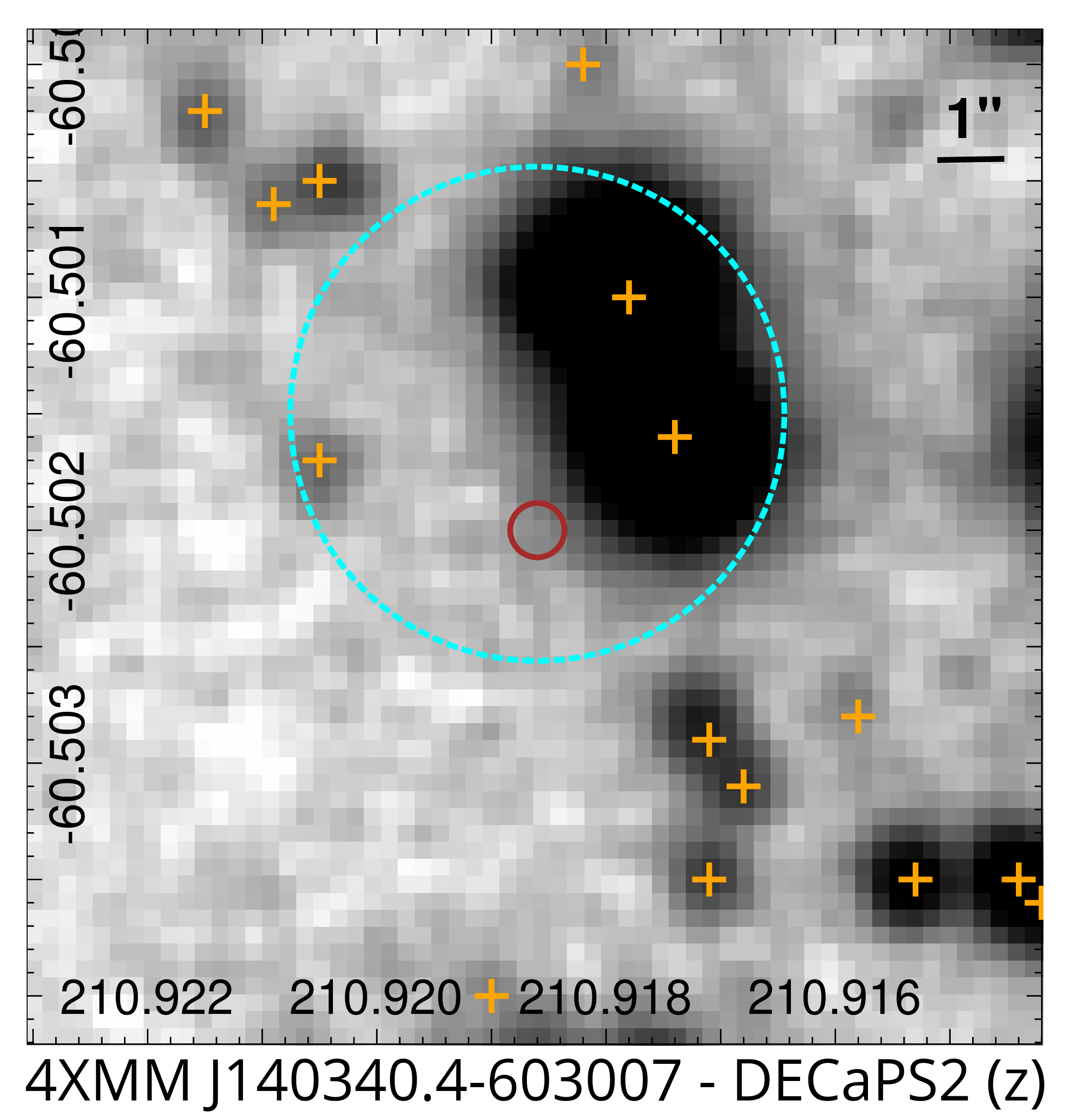}\\
\includegraphics[width=0.33\textwidth]{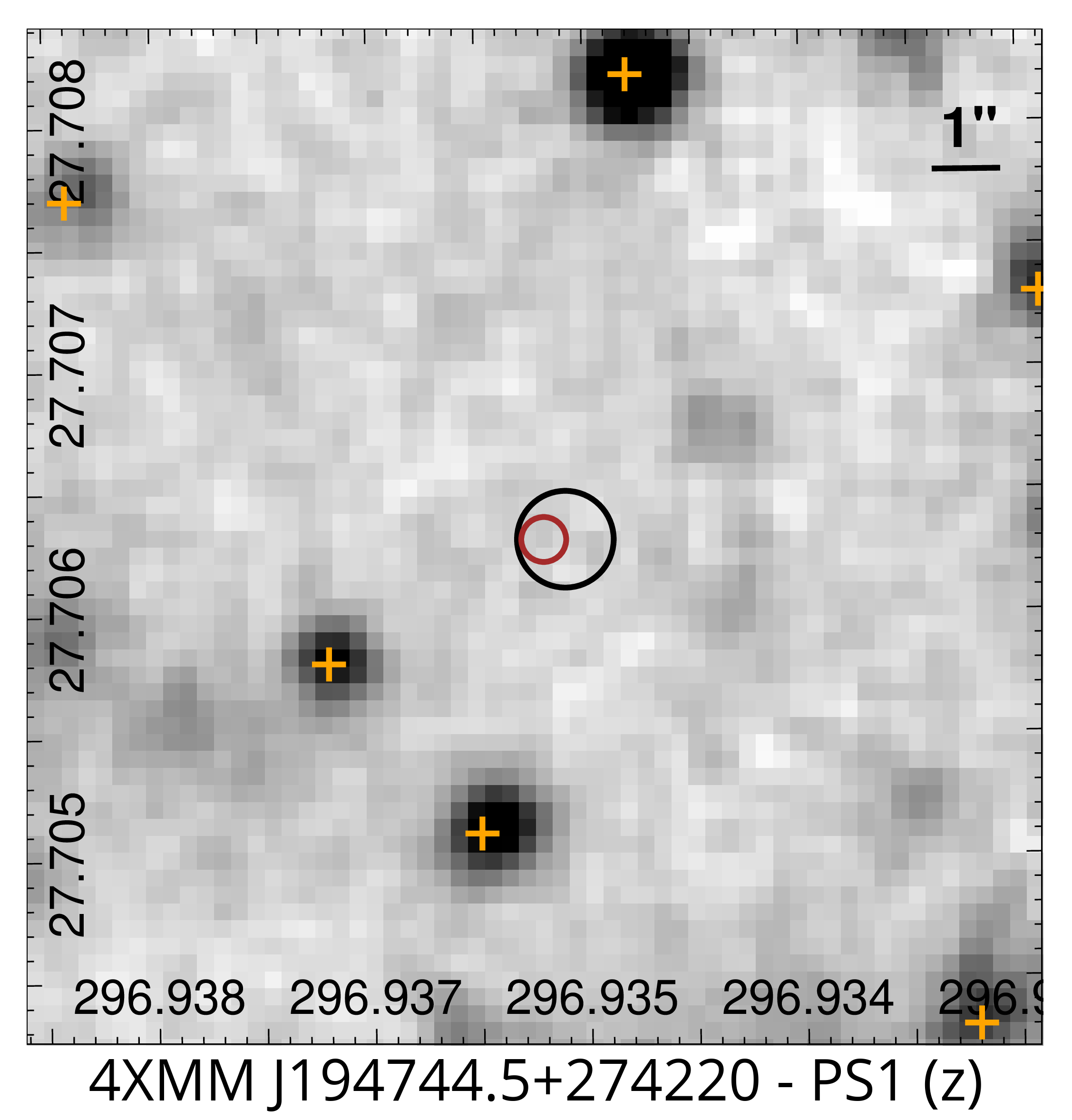}\hfill
\includegraphics[width=0.33\textwidth]{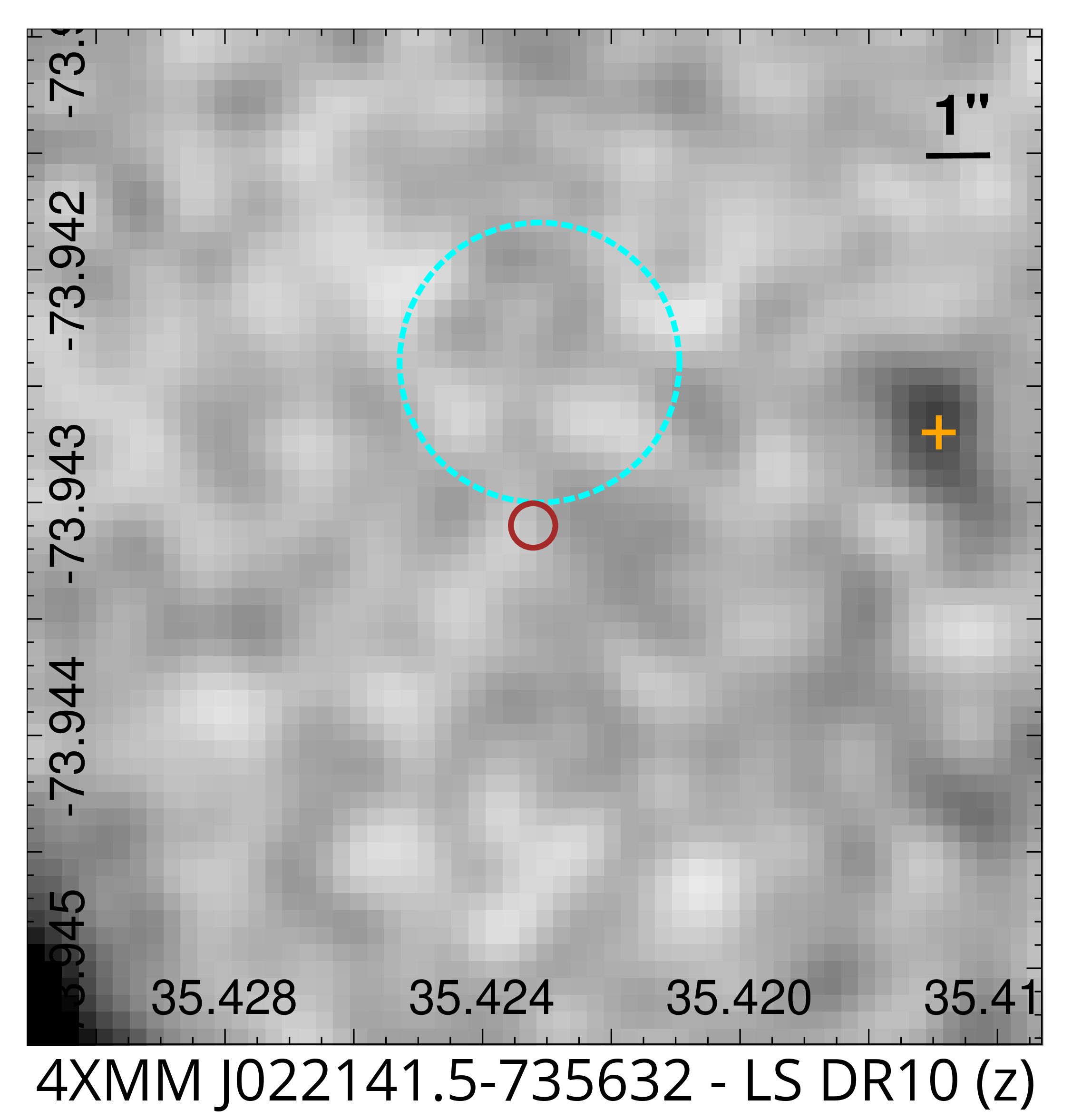}\hfill
\includegraphics[width=0.33\textwidth]{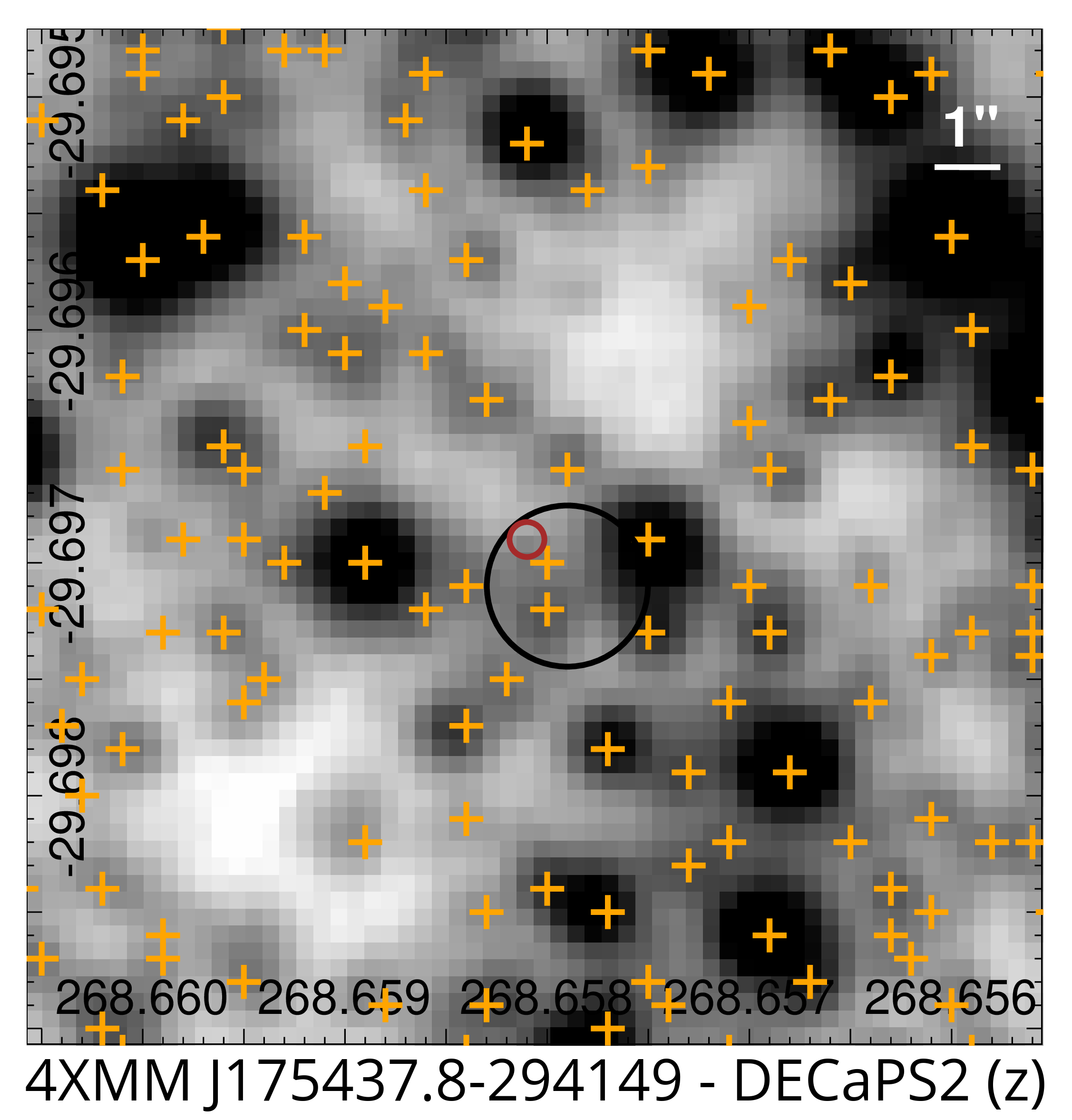}
\end{center}
\caption{Optical finding charts in the $z$ filter for the compelling 4XMM XINS candidates (see also Fig.~\ref{fig_fcextra} for the subset of contaminants and poorly constrained candidates). The solid brown circles indicate the 99.994\% confidence level uncertainties in the (stacked) \xmm\ source positions (Table~\ref{tab_psfloc}). The dashed cyan and solid black circles show the corresponding uncertainties, when available, for detections within the \eROS\ footprint or with \chan, respectively. The crosses mark the positions of catalogued optical objects in the $z$ band from the deepest available photometric survey.\label{fig_fcxins}}
\end{figure*}

\begin{table}[t]
\caption{Field optical depth and counterpart properties
\label{tab_optlim}}
\small
\centering
\begin{tabular}{@{}llcccccc@{}}
\hline\hline\noalign{\vskip 0.4ex}
Field & & \multicolumn{2}{c}{Survey\,\tablefootmark{a}} & $A_{\rm V}$\,\tablefootmark{b} & \multicolumn{2}{c}{Counterpart\,\tablefootmark{c}} & $X/O$\,\tablefootmark{d} \\
\cline{2-4}\cline{6-7}\noalign{\vskip 0.4ex}
& \multicolumn{2}{r}{(arcmin$^{-2}$)} & $g$ (mag) & (mag) & (\arcsec) & $g$ (mag) & \\
\hline\noalign{\vskip 0.4ex}
J1140 & DE & 40 & 24.3 & 2.21 & $-$ & $-$ & $>3.51$ \\
J1818 & PS & 48 & 22.8 & 2.77 & $-$ & $-$ & $>2.01$ \\ 
J1233 & LS &  8 & 24.5 & 0.08 & 0.5 & 18.5 & $-0.47$ \\
J1403 & DE & 25 & 23.8 & 1.78 & $-$ & $-$ & $>2.07$ \\
J0103 & LS & 15 & 24.7 & 0.20 & $-$ & $-$ & $>2.03$ \\
J1947 & PS & 47 & 22.9 & 2.27 & $-$ & $-$ & $>1.72$ \\
J0221 & LS &  6 & 25.4 & 0.14 & $-$ & $-$ & $>3.20$ \\
J0311 & PS &  5 & 22.7 & 0.36 & $-$ & $-$ & $>1.10$ \\
J2251 & LS & 15 & 24.8 & 0.12 & 0.6 & 23.7 & $\sim1.92$ \\
J1754 & DE & 42 & 24.0 & 1.02 & $-$ & $-$ & $>1.64$ \\
\hline
\end{tabular}
\tablefoot{
\tablefoottext{a}{Wide-area public surveys used in this work are: LS = Legacy Survey DR10, DE = DECaPS2, and PS = Pan-STARRS PS1. Depending on the field, we report the local number density of catalogued objects (per square arcminute) and the corresponding survey depth (magnitudes).}
\tablefoottext{b}{Visual-band extinction (480--600\,nm) towards each source \citep{2024arXiv240303127D,2024A&A...685A..82E,2017ApJ...835...29Y,2016A&A...596A.109P}.}
\tablefoottext{c}{Angular separation (arcseconds) and magnitude of any optical source located within the 99.994\% confidence X-ray error circle.}
\tablefoottext{d}{Logarithmic ratio of the X-ray flux (0.2--12\,keV) to the optical flux in the visual band, corrected for photoelectric absorption and interstellar extinction using each source's best-fit spectral parameters (Tables~\ref{tab_spec} and \ref{tab_spec_extra}). Values correspond either to the detection limit or to an identified counterpart.}
}
\end{table}

To identify possible counterparts and assess the photometric depth in each XINS candidate field, we retrieved magnitudes of optical sources catalogued in \gaia\ Data Release 3 \citep{2023A&A...674A...1G,2016A&A...595A...1G}, Legacy Survey Data Release 10 \citep[LS;][]{2019AJ....157..168D}, the Panoramic Survey Telescope and Rapid Response System \citep[Pan-STARRS PS1;][]{2016arXiv161205560C}, and the Dark Energy Camera Plane Survey \citep[DECaPS2;][]{2023ApJS..264...28S}, within a $30\arcmin \times 30\arcmin$ region centred on each target. We filtered the photometric data to retain only sources with finite magnitudes and well-defined errors in the most populated band. For LS and DECaPS2, we further restricted the selection to sources with a signal-to-noise ratio greater than five. In the absence of counterparts, the limiting magnitude was estimated as the 99.994th percentile ($4\sigma$) of the magnitude distribution, excluding both the faintest and brightest sources to avoid skewing.

Optical counterparts are confidently identified for \liu\ and \sanshier\ (individual cases are discussed in Section~\ref{sec_nature} and Appendix~\ref{sec_contaminants}). The X-ray-to-optical flux ratios of these counterparts are consistent with those of extragalactic sources. For the remaining candidates, no optical counterpart is found within the current limits.

The resulting optical depth for each field (quoted in the $g$ band), together with the number density of catalogued sources, the angular separation and magnitude of any optical object within the 99.994\% confidence-level X-ray error circle, and the corresponding logarithmic X-ray-to-optical flux ratio ($X/O$, corrected for photoelectric absorption and interstellar extinction $A_{\rm V}$; \citealt{2024arXiv240303127D}), are summarised in Table~\ref{tab_optlim}. Corresponding $z$-band finding charts are shown in Figs.~\ref{fig_fcxins} and \ref{fig_fcextra}.

\subsection{Spectral analysis and inter-epoch variability\label{sec_spec}}

\begin{table*}[!t]
\small
\caption{Results of spectral analysis
\label{tab_spec}}
\centering
\begin{tabular}{@{}lc@{}rrr@{\hspace{10pt}}rr@{\hspace{10pt}}r@{\hspace{10pt}}rc@{\hspace{10pt}}c@{\hspace{3pt}}cr@{}}
\hline\hline\noalign{\vskip 0.4ex}
Model & NHP\,\tablefootmark{a} & $C$ || $\chi^2$\,\tablefootmark{b} & \multicolumn{1}{c}{$\nh$\,\tablefootmark{c}} & \multicolumn{1}{c}{$d_{N_{\rm H}}$\,\tablefootmark{d}} & \multicolumn{4}{c}{Continuum model parameters} & $z$ | $Z_{\odot}^{\rm O}$ & \multicolumn{2}{c}{\texttt{gabs} (eV)} & $\log(F_{\rm X})$\,\tablefootmark{e}\\
\cline{6-9}\cline{11-12}
\noalign{\vskip 0.4ex}
& (\%) & & & \multicolumn{1}{c}{(kpc)} & $kT_{\rm eV}$ || $T^{\rm eff}_{\rm 10^5\,K}$ & $R_{\rm km}^{\rm em}$ || $D_{\rm kpc}$ & $kT_{\rm eV}$ || $\Gamma$ & $R_{\rm km}^{\rm em}$ || $Z_{\rm Z_\odot}$ & & $\epsilon_1$ & $\epsilon_2$ & \\
\hline\noalign{\vskip 0.4ex}
\multicolumn{4}{l}{\joofz} & \multicolumn{9}{r}{(six epochs)\quad[0.2--1.1\,keV]\quad$\mathcal{C}=970\pm30$\quad$\mathcal{B}=5\%$} \\
\hline\noalign{\vskip 0.4ex}
\texttt{apecG} & 74 & 96\,(106) & $1.8_{-0.4}^{+0.5}$ & $0.2$ & $-$ & $-$ & $295_{-8}^{+8}$ & 1\,\tablefootmark{$\dagger$} & 0 & $-$ & $-$ & $-12.23_{-0.02}^{+0.02}$\\
\texttt{bb} & 98 & 78\,(106) & $7.7_{-1.2}^{+1.2}$ & $3.6_{-0.8}^{+2.1}$ & $84.5_{-1.8}^{+1.8}$ & $47_{-4}^{+5}$ & $-$ & $-$ & 0 & $-$ & $-$ & $-10.59_{-0.04}^{+0.04}$\\
\texttt{nsa} & 97 & 81\,(106) & $10.2_{-1.1}^{+1.2}$ & $7.4_{-1.4}^{+0.6}$ & $3.40_{-0.09}^{+0.10}$ & $<0.16$ & $-$ & $-$ & 0 & $-$ & $-$ & $-9.35_{-0.06}^{+0.06}$\\
\hline\noalign{\vskip 0.4ex}
\multicolumn{4}{l}{\joeoe} & \multicolumn{9}{r}{(six epochs)\quad[0.2--2\,keV]\quad$\mathcal{C}=3110\pm60$\quad$\mathcal{B}=8$\,\%} \\
\hline\noalign{\vskip 0.4ex}
\texttt{apecG} & 45 & 139\,(138) & $7.7_{-0.5}^{+0.5}$ & $1.8_{-0.2}^{+0.1}$ & $-$ & $-$ & $285_{-7}^{+8}$ & 1\,\tablefootmark{$\dagger$} & 0 & $-$ & $-$ & $-12.00_{-0.02}^{+0.03}$\\
\texttt{apecX} & 83 & 121\,(137) & $11.9_{-0.6}^{+0.6}$ & $5.9_{-0.9}^{+4}$ & $-$ & $-$ & $312_{-4}^{+5}$ & $<0.02$ & 1.4\,\tablefootmark{$\dagger$} & $-$ & $-$ & $-10.13_{-0.01}^{+0.04}$\\
\texttt{bb} & 79 & 124\,(138) & $9.6_{-0.7}^{+0.7}$ & $2.0_{-0.1}^{+1.3}$ & $100.0_{-1.2}^{+1.2}$ & $13.0_{-0.7}^{+0.8}$ & $-$ & $-$ & 0 & $-$ & $-$ & $-11.18_{-0.15}^{+0.16}$\\
\texttt{bb$\ast$g} & 91 & 115\,(136) & $3.7_{-1.6}^{+2.1}$ & $1.3_{-1.1}^{+0.2}$ & $131_{-3}^{+3}$ & $1.77_{-0.15}^{+0.17}$ & $-$ & $-$ & 0 & $510_{-17}^{+15}$ & $-$ & $-12.53_{-0.04}^{+0.04}$\\
\texttt{bb+bb} & 94 & 111\,(136) & $12.5_{-2.0}^{+2.9}$ & $10_{-7}^{+13}$ & $74.0_{-3}^{+2.6}$ & $210_{-30}^{+50}$ & $150_{-26}^{+40}$ & $3.0_{-2.0}^{+5}$ & 0 & $-$ & $-$ & $-10.14_{-0.06}^{+0.06}$\\
\texttt{bb+pl} & 95 & 111\,(136) & $15.8_{-3}^{+2.1}$ & $>12$ & $64.6_{-2.2}^{+3}$ & $790_{-250}^{+230}$ & $9.3_{-0.8}^{+0.5}$ & $-$ & 0 & $-$ & $-$ & $-8.00_{-0.05}^{+8}$\\
\texttt{nsa} & 87 & 119\,(138) & $12.6_{-0.6}^{+0.4}$ & $10_{-4}^{+2}$ & $3.97_{-0.04}^{+0.06}$ & $0.015_{-0.001}^{+0.001}$ & $-$ & $-$ & 0 & $-$ & $-$ & $-9.93_{-0.04}^{+0.03}$\\
\texttt{nsa$\ast$g} & 93 & 113\,(136) & $5.8_{-1.8}^{+3}$ & $1.5_{-0.1}^{+0.4}$ & $5.95_{-0.20}^{+0.19}$ & $0.21_{-0.03}^{+0.03}$ & $-$ & $-$ & 0 & $520_{-18}^{+15}$ & $-$ & $-11.48_{-0.08}^{+0.08}$\\
\hline\noalign{\vskip 0.4ex}
\multicolumn{4}{l}{\jofzt} & \multicolumn{9}{r}{(three epochs)\quad[0.4--2\,keV]\quad$\mathcal{C}=1380\pm40$\quad$\mathcal{B}=10$\,\%} \\
\hline\noalign{\vskip 0.4ex}
\texttt{apecX} & 85 & 43\,(54) & $8.1_{-0.6}^{+0.7}$ & $4.4_{-0.3}^{+0.1}$ & $-$ & $-$ & $310_{-8}^{+8}$ & 0.2\,\tablefootmark{$\dagger$} & 1.4\,\tablefootmark{$\dagger$} & $-$ & $-$ & $-10.82_{-0.05}^{+0.04}$\\
\texttt{bb} & 93 & 39\,(54) & $6.2_{-0.8}^{+0.9}$ & $3.6_{-0.7}^{+0.4}$ & $98.1_{-1.8}^{+1.8}$ & $7.6_{-0.6}^{+0.6}$ & $-$ & $-$ & 0 & $-$ & $-$ & $-12.02_{-0.04}^{+0.04}$\\
\texttt{nsa} & 95 & 38\,(54) & $9.0_{-0.9}^{+0.6}$ & $4.6_{-0.2}^{+0.1}$ & $3.95_{-0.07}^{+0.09}$ & $0.037_{-0.004}^{+0.005}$ & $-$ & $-$ & 0 & $-$ & $-$ & $-10.78_{-0.05}^{+0.04}$\\
\hline\noalign{\vskip 0.4ex}
\multicolumn{4}{l}{\jonfs} & \multicolumn{9}{r}{(two epochs)\quad[0.3--1.5\,keV]\quad$\mathcal{C}=1060\pm30$\quad$\mathcal{B}=11$\,\%} \\
\hline\noalign{\vskip 0.4ex}
\texttt{apecG} & 83 & 83\,(98) & $2.8_{-0.7}^{+0.7}$ & $<1.2$ & $-$ & $-$ & $375_{-17}^{+20}$ & 1\,\tablefootmark{$\dagger$} & 0 & $-$ & $-$ & $-13.20_{-0.02}^{+0.02}$\\
\texttt{bb} & 11 & 106\,(98) & $7.9_{-0.9}^{+0.9}$ & $7.3_{-0.4}^{+0.4}$ & $95.6_{-1.8}^{+1.9}$ & $12.4_{-1.0}^{+1.1}$ & $-$ & $-$ & 0 & $-$ & $-$ & $-11.84_{-0.04}^{+0.04}$\\
\texttt{bb$\ast$g} & 78 & 81\,(96) & $8.8_{-1.2}^{+1.4}$ & $7.7_{-0.5}^{+0.4}$ & $99.0_{-2.6}^{+2.7}$ & $13.8_{-1.5}^{+1.6}$ & $-$ & $-$ & 0 & $-$ & $1050_{-24}^{+23}$ & $-11.71_{-0.04}^{+0.04}$\\
\texttt{bb$\ast$gg} & 93 & 70\,(94) & $2.6_{-1.6}^{+3}$ & $0.8_{-0.2}^{+4}$ & $141_{-14}^{+18}$ & $0.65_{-0.20}^{+0.3}$ & $-$ & $-$ & 0 & $526_{-17}^{+13}$ & $1110_{-60}^{+50}$ & $-13.06_{-0.08}^{+0.07}$\\
\texttt{Obb$\ast$g} & 92 & 72\,(95) & $10.7_{-1.8}^{+3}$ & $8.4_{-0.7}^{+3}$ & $103_{-5}^{+5}$ & $20_{-3}^{+5}$ & $-$ & $-$ & $<1.2$ & $-$ & $1111_{-24}^{+21}$ & $-11.45_{-0.03}^{+0.15}$\\
\texttt{bb+bb} & 95 & 71\,(96) & $16.3_{-2.8}^{+1.9}$ & $15_{-3}^{+5}$ & $54.1_{-1.5}^{+1.5}$ & $1530_{-260}^{+340}$ & $180_{-40}^{+80}$ & $1.3_{-0.8}^{+1.9}$ & 0 & $-$ & $-$ & $-9.55_{-0.07}^{+0.07}$\\
\texttt{bb+pl} & 92 & 72\,(96) & $13.8_{-1.5}^{+1.5}$ & $11.6_{-0.8}^{+2.2}$ & $62.1_{-1.2}^{+1.3}$ & $360_{-30}^{+40}$ & $<4.6$ & $-$ & 0 & $-$ & $-$ & $-10.22_{-0.07}^{+0.07}$\\
\texttt{nsa} & 15 & 103\,(98) & $10.4_{-0.8}^{+0.4}$ & $8.2_{-0.3}^{+0.2}$ & $3.94_{-0.06}^{+0.11}$ & $0.039_{-0.003}^{+0.006}$ & $-$ & $-$ & 0 & $-$ & $-$ & $-10.90_{-0.05}^{+0.05}$\\
\texttt{nsa$\ast$gg} & 93 & 70\,(94) & $4.0_{-2.4}^{+4}$ & $1.9_{-1.2}^{+5}$ & $7.3_{-1.1}^{+1.7}$ & $0.9_{-0.4}^{+1.1}$ & $-$ & $-$ & 0 & $526_{-17}^{+13}$ & $1100_{-70}^{+60}$ & $-12.91_{-0.03}^{+0.03}$\\
\texttt{Onsa$\ast$g} & 93 & 71\,(95) & $10.0_{-0.8}^{+1.9}$ & $8.1_{-0.3}^{+1.6}$ & $4.11_{-0.21}^{+0.28}$ & $0.023_{-0.006}^{+0.008}$ & $-$ & $-$ & $<1.1$ & $-$ & $1112_{-25}^{+21}$ & $-10.42_{-0.06}^{+0.06}$\\
\hline\noalign{\vskip 0.4ex}
\multicolumn{4}{l}{\josff} & \multicolumn{9}{r}{(four epochs)\quad[0.3--1.5\,keV]\quad$\mathcal{C}=490\pm22$\quad$\mathcal{B}=28\%$} \\
\hline\noalign{\vskip 0.4ex}
\texttt{apecG} & 96 & 65\,(75) & $2.5_{-1.8}^{+1.0}$ & $1.2_{-1.0}^{+0.1}$ & $-$ & $-$ & $250_{-10}^{+11}$ & 1\,\tablefootmark{$\dagger$} & 0 & $-$ & $-$ & $-13.28_{-0.04}^{+0.04}$\\
\texttt{bb} & 100 & 47\,(75) & $3.6_{-1.0}^{+1.1}$ & $1.3_{-0.1}^{+0.7}$ & $110_{-4}^{+4}$ & $0.94_{-0.11}^{+0.12}$ & $-$ & $-$ & 0 & $-$ & $-$ & $-12.88_{-0.05}^{+0.05}$\\
\texttt{nsa} & 100 & 45\,(75) & $6.0_{-1.4}^{+1.4}$ & $3.7_{-1.7}^{+7}$ & $4.81_{-0.21}^{+0.21}$ & $0.31_{-0.06}^{+0.07}$ & $-$ & $-$ & 0 & $-$ & $-$ & $-11.94_{-0.08}^{+0.08}$\\
\hline\noalign{\vskip 0.4ex}
\end{tabular}
\tablefoot{
Results are shown for the five compelling XINS candidates discussed in Sect.~\ref{sec_disc}, excluding the confirmed source \ershiyi\ \citep{2022A&A...666A.148P}; the remaining sources are listed in Table~\ref{tab_spec_extra} and discussed in Appendix~\ref{sec_contaminants}. Quoted errors correspond to 1$\sigma$ confidence levels. All best-fit models shown assume no inter-epoch variability.
For each target, we list the number of available epochs; total counts ($\mathcal{C}$) refer to the indicated energy bands, while the mean background level per epoch, $\mathcal{B}$, is averaged over detectors.
Galactic and extragalactic APEC models (\texttt{apecG}, \texttt{apecX}) assume solar abundances at redshift $z = 0$ and a free abundance ($Z/Z_\odot$) at the fixed redshift given in the table, respectively. Models including \texttt{gabs} absorption features are denoted with labels \texttt{$\ast$g} (number of components), and models with overabundant oxygen are prefixed with \texttt{O}. The blackbody emission radius is computed using the distance $d_{\nh}$ listed in the table.
Neutron-star atmosphere models assume $M = 1.4$\,M$_\odot$, $R = 10$\,km, and a magnetic field of $B = 10^{13}$\,G. The \texttt{gabs} line width is fixed to 120\,eV (targets \si\ and \sanshiwu) and to 100\,eV and 200\,eV (target \ershi).
\tablefoottext{a}{Null-hypothesis probability (percent) that the data are consistent with the model.}
\tablefoottext{b}{C or $\chi^2$ statistic, with corresponding degrees of freedom in brackets (d.o.f.).}
\tablefoottext{c}{Hydrogen column density in units of $10^{21}$\,cm$^{-2}$.}
\tablefoottext{d}{Distance estimate from 3D extinction/absorption maps \citep{2024arXiv240303127D}, based on the best-fit $\nh$.}
\tablefoottext{e}{Logarithm of the unabsorbed flux in \fluxcgs\ (0.2--12\,keV).}
\tablefoottext{$\dagger$}{Parameter fixed during fitting.}
}
\end{table*}

Spectral fitting was performed in XSPEC v12.13.1 \citep{1996ASPC..101...17A}, using either the Cash C-statistic---suitable for low-count data---or the $\chi^2$ statistic, with a minimum of 20 counts per bin in the latter case. Unless otherwise specified, fit parameters were allowed to vary within physically reasonable bounds.

For the moderately bright candidate \sanshiwu, spectra were extracted from each individual \eROS\ scan and fitted jointly with the corresponding EPIC dataset. For the fainter sources \shi\ and \ershiyi, a single spectrum was extracted from the concatenated \eROS\ event file combining all available scans and fitted simultaneously with the available EPIC observations. The available \chan\ data for \ershi\ and \sanshijiu\ were also included in the spectral fits. A renormalisation factor between detectors was included in all simultaneous fits to account for cross-calibration uncertainties.

\begin{figure}[t]
\begin{center}
\includegraphics[width=\columnwidth]{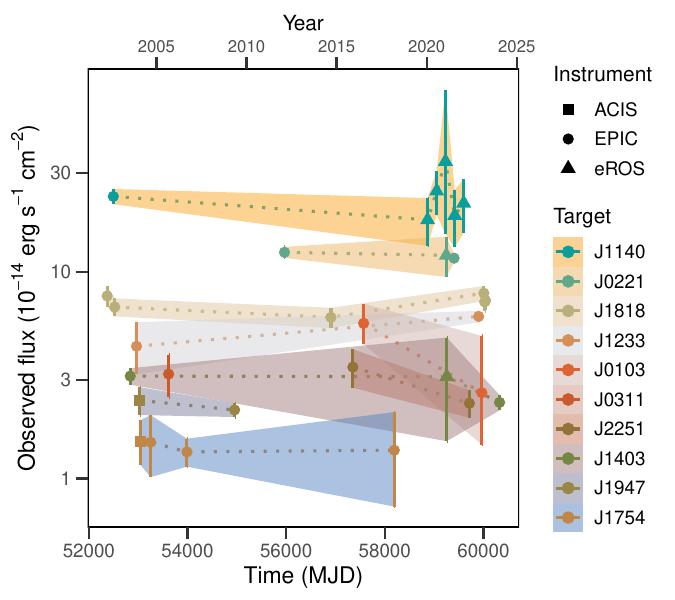}
\end{center}
\caption{Observed X-ray flux as a function of time for the ten XINS candidates. The colours denote the individual targets, while the symbols indicate the instrument. The shaded bands and error bars represent the $\pm2\sigma$ flux uncertainties.\label{fig_fxevol}}
\end{figure}

Interstellar absorption was modelled with the \texttt{tbabs} component, adopting the photoelectric cross-sections of \citet{2000ApJ...542..914W}. Assuming no spectral evolution across epochs, we first evaluated single-component models, including a blackbody (\texttt{bb}), a neutron-star hydrogen atmosphere (\texttt{nsa}; \citealt{1995ASIC..450...71P,1996A&A...315..141Z}), and an optically thin plasma (\texttt{apec}\footnote{\texttt{http://atomdb.org}}).

For sources with sufficient statistics and/or multiple observations, we then considered more complex models, such as a double blackbody (\texttt{bb+bb}), blackbody plus power law (\texttt{bb+pl}), and models including absorption features (\texttt{gabs}) or inter-epoch variability (free parameters, except for $\nh$). For sources exhibiting significant residuals around 530--650\,eV (typically more absorbed XINS candidates in the Galactic plane; see, for example, the XINS in the Carina Nebula \carINS; \citealt{2012A&A...544A..17P,2015A&A...583A.117P}), we allowed the oxygen abundance to exceed the solar value using the \texttt{vphabs} model. Improvements in fit quality were assessed using F-tests for nested models and the null-hypothesis probability (NHP).

The results are summarised in Tables~\ref{tab_spec} and \ref{tab_spec_extra} and discussed in Sections~\ref{sec_nature} and \ref{sec_contaminants}. For each target, we report the number of available epochs, total spectral counts, mean background fraction per epoch, and analysed energy band (see Table caption). Only the most relevant models are shown---those that are physically plausible and/or have the highest NHPs. For nested models, only statistically significant improvements over simpler alternatives are retained.

Alongside the fit statistic (C-statistic or $\chi^2$) and degrees of freedom, we report the best-fit $\nh$ and the distance estimate $d_{N_{\rm H}}$, derived from three-dimensional extinction and absorption maps \citep{2024arXiv240303127D}. Model parameters include: temperature $kT$ for \texttt{bb} and \texttt{apec}; effective temperature $T_{\rm eff}$ and distance $D_{\rm kpc}$ for \texttt{nsa}; blackbody emission radius $R_{\rm em}$ normalised to $d_{N_{\rm H}}$; photon index $\Gamma$ for power-law components; metal abundance $Z$ and assumed redshift $z$ for \texttt{apec}; oxygen abundance $Z_{\rm O}$ for \texttt{vphabs}; and line energy for \texttt{gabs}. Unabsorbed fluxes in the 0.2--12\,keV band are expressed as base-10 logarithms in units of \fluxcgs.

For all targets, we find no clear evidence for inter-epoch variability within the uncertainties. Figure~\ref{fig_fxevol} shows the observed flux evolution of the neutron star candidates, derived from simultaneous spectral fits in which the model parameters were allowed to vary between epochs, except for $\nh$. The statistically best-fitting spectral model for each source was adopted. The plotted fluxes include EPIC, eRASS, and \chan\ observations, with conservative $2\sigma$ uncertainties.

In addition, we retrieved upper limits and flux estimates from the High-Energy Lightcurve Generator (HILIGT\footnote{\texttt{http://xmmuls.esac.esa.int/upperlimitserver}}; \citealt{2022A&C....3800531S}) using archival \swift\ XRT and \ros\ observations \citep[see][and references therein]{2022MNRAS.511.4265R}. These measurements are not shown in Fig.~\ref{fig_fxevol}, but were used to assess variability at additional epochs.

\subsection{Timing analysis\label{sec_timing}}

We first verified the EPIC light curve statistics to assess general variability trends. The light curves were binned into $100$\,s intervals and corrected for bad pixels, dead time, exposure, and background counts using the SAS task \textsf{\small epiclccorr}. All exposures were found to be consistent with a constant flux.

We then searched for periodic signals that could be associated with neutron star rotation, typically ranging from hundreds of milliseconds to several hundred seconds. For this purpose, we considered only the longest \xmm\ observations of each target suitable for meaningful timing analysis (Table~\ref{tab_pflim}). To maximise sensitivity, we included all valid patterns and events from all three EPIC cameras in the 0.2--2\,keV energy band, without filtering for GTIs.

Photon arrival times were barycentred to the solar system barycentre using the SAS task \texttt{barycen}, adopting the DE405 ephemeris and the source coordinates listed in Table~\ref{tab_psfloc}. We performed the timing analysis in the frequency domain using the $Z^2_m$ technique \citep{1983A&A...128..245B}. When available, data from the three EPIC cameras were analysed jointly over the frequency range $\Delta\nu = (0.002\text{--}1.9)\times10^{-1}\,\mathrm{Hz}$. Thanks to its higher time resolution, the pn camera also allows a broader frequency search, albeit with reduced sensitivity compared to the full EPIC combination. In both cases (EPIC and pn), we adopted frequency steps of $8\,\mu$Hz, corresponding to an oversampling factor of at least $\sim3$.

No statistically significant signal ($>4\sigma$) was detected in the EPIC datasets. The corresponding $4\sigma$ upper limits on the pulsed fraction are listed in Table~\ref{tab_pflim}.

For the two XINS candidates observed with \fast, \liu\ and \ershi, we searched for both periodic and possible sporadic single-pulse radio emission using the pulsar search software package \textsc{PRESTO}\footnote{\url{https://github.com/scottransom/presto}} \citep{2002AJ....124.1788R}, following standard procedures. Radio frequency interference (RFI) was first identified and masked using the \texttt{rfifind} routine, with a data length of 2\,s used in the RFI search. The data were subsequently de-dispersed across a broad range of trial dispersion measures (DMs) using the \texttt{prepsubband} routine. A DM range of 0--1400\,pc\,cm$^{-3}$ was adopted, significantly exceeding the maximum expected values (e.g., $\sim$20\,pc\,cm$^{-3}$ for \liu\ and $\sim$500\,pc\,cm$^{-3}$ for \ershi) from the NE2001 \citep{2002astro.ph..7156C} and YMW16 \citep{2017ApJ...835...29Y} Galactic electron density models along the lines of sight. The DM step sizes were determined using the \texttt{DDplan.py} code, with values of 0.05, 0.10, 0.20, 0.30, and 0.50\,pc\,cm$^{-3}$ applied to the DM ranges of 0--113.7, 113.7--189.5, 189.5--341.1, 341.1--586.5, and 586.5--1404.5 \,pc\,cm$^{-3}$, respectively. 

Following de-dispersion, the time series were Fourier-transformed using the \texttt{realfft} routine, low-frequency noise was mitigated with the \texttt{rednoise} routine, and periodic signals were subsequently searched for via the \texttt{accelsearch} routine. The maximum number of Fourier bins allowing for signal drift ($z_{\rm max}$) was set to 20 and 200 to accommodate potential acceleration of the targets during the observation. 

Periodicities identified across all DM trials were sifted using a customised version of the \texttt{ACCEL\_sift.py} code, and promising candidates were folded with the \texttt{prepfold} routine to produce diagnostic plots for visual inspection. In parallel, the \texttt{single\_pulse\_search.py} code was used to search for single radio bursts in all de-dispersed time series, with a signal-to-noise (S/N) threshold of 8. Any potential single pulses were visualised using the \texttt{waterfaller.py} code and subjected to visual inspection.

No credible candidates for periodic or transient radio emission were identified across the full range of trial DMs in the two search schemes using the FAST $L$-band data.
Following the radiometer equation \citep[e.g.][]{2004hpa..book.....L}, we estimate an upper limit on the flux density for periodic emission as
\begin{equation}
S_{\rm periodic} = \beta
\frac{(S/N)\,T_{\rm sys}}
{G\sqrt{n_{\rm p} T_{\rm int} \Delta BW}}
\sqrt{\frac{W_{\rm obs}}{P - W_{\rm obs}}} ,
\end{equation}
where we adopt $\beta = 1$, a telescope gain of $G = 16~{\rm K\,Jy^{-1}}$, a system temperature of $T_{\rm sys} \approx 24~{\rm K}$, two summed polarisations ($n_{\rm p} = 2$), and an effective bandwidth of $\Delta BW = 400~{\rm MHz}$ \citep{2020RAA....20...64J}. A detection threshold of $S/N = 8$ is assumed, together with a duty cycle of $W_{\rm obs} = 0.1P$. Under these assumptions, we derive $8\sigma$ upper limits on the periodic flux density of 4.1~$\mu$Jy for \liu\ and 3.3~$\mu$Jy for \ershi.

For single-pulse emission, the corresponding peak flux density limit can be expressed as \citep[e.g.][]{2003ApJ...596.1142C}
\begin{equation}
S_{\rm single} = 2\beta
\frac{(S/N)_{\rm peak}\,T_{\rm sys}}
{G\sqrt{n_{\rm p} W_{\rm obs} \Delta BW}} ,
\end{equation}
where $(S/N)_{\rm peak} = 8$. Assuming a representative observed pulse width of $W_{\rm obs} = 1~{\rm ms}$, the resulting $8\sigma$ upper limit on the single-pulse flux density is approximately 26.8~mJy for both sources.

\section{Implications\label{sec_disc}}

\subsection{Source properties\label{sec_nature}}

In the following subsections, we discuss the five compelling XINS candidates (\sanshiwu, \si, \shi, \ershi, and \sanshijiu) identified in this work, together with the confirmed source (\ershiyi). The remaining sources are classified as uncertain (\shiyi, \ershisan) or likely contaminants (\liu, \sanshier), and are discussed separately in Appendix~\ref{sec_contaminants}.

\subsubsection{4XMM J114051.9$-$641848}

The brightest XINS candidate in our sample has an observed flux of $2.23(8)\times10^{-13}$\,\fluxcgs\ (0.2--2\,keV). It is a soft X-ray source detected serendipitously in a 2002 \xmm\ snapshot of the cataclysmic variable \object{V1033~Cen}, at an off-axis angle of 8\arcmin. Although it is the brightest X-ray emitter in the field, it has neither been studied further nor re-observed with \xmm. Based on the astrometrically corrected EPIC centroid, the absence of optical counterparts ($g>24.3$, $X/O>3.51$; $4\sigma$) makes it a viable candidate for an XINS or another compact Galactic X-ray source.

The source was re-detected in eRASS between 2020 and 2022 with comparable flux and spectral shape, although it was not flagged as an XINS candidate by \citet{2024A&A...687A.251K} due to source confusion. The stacked, astrometrically corrected \eROS\ position (90\% confidence error radius of 0.8\arcsec) lies $1.8\arcsec$ northwest of the EPIC coordinates. It coincides with a faint optical object in the DECaPS2 catalogue ($g=22.08$, $S/N>30$; Fig.~\ref{fig_fcxins}). Its colours ($g-r\sim1.6$, $r-i\sim0.8$, $i-z\sim0.5$) are consistent with an early to mid-M dwarf at a distance of $\sim$0.5--1\,kpc. Given the corresponding $X/O \sim 2.7$ ratio, which is too high for coronal emission, this object is likely a foreground, unrelated source. Improved X-ray localisation will be required to determine whether the two sources are physically associated.

The combined EPIC and \eROS\ spectra, comprising six epochs and eight datasets fitted simultaneously ($970 \pm 30$ net counts with a background contribution of $\sim$5\%), are well described by a soft, moderately absorbed blackbody with $kT = 84.5 \pm 1.8$\,eV and $\nh = (7.7 \pm 1.2) \times 10^{21}$\,cm$^{-2}$. The inferred $\nh$ suggests a large distance ($d_{\nh} \sim 2.8$--5.7\,kpc); at the best-fit distance of 3.6\,kpc, the implied luminosity is high for typical XINSs, $L_{\rm X} = (4.0 \pm 0.4) \times 10^{34}$\,\lumcgs\ (uncertainty reflecting only the error on the unabsorbed flux). Magnetised neutron star atmosphere models with $T_{\rm eff} = (2.4$--$3.4) \times 10^{5}$\,K and $B = 10^{12}$--$10^{13}$\,G provide fits of comparable quality. However, the model normalisations imply either unusually large blackbody emission radii or, for the atmosphere models, implausibly small distances.
The flux and spectral shape have remained stable over the past two decades. Alternative spectral models (e.g.~\texttt{apec}, bremsstrahlung) yield poorer fits or non-physical parameters, and the data do not require additional spectral components.

\subsubsection{4XMM J181844.3$-$120751}

The X-ray source was serendipitously detected multiple times by \xmm\ during observations of massive stellar systems in the OB association \object{Ser~OB2}, including \object{HD~168112} (2002, 2023) and \object{HD~167971} in \object{NGC~6604} (2014). Its lack of optical or near-infrared counterparts, soft spectrum, and absence of variability were first reported by \citet{2005A&A...437.1029D}, who designated it \object{[DRB2005]}. \citet{2012ApJ...756...27L} classified it as a compact object without obvious counterparts, while \citet{2022MNRAS.509.1217R} favoured an \texttt{apec} model with solar abundances over a blackbody, arguing against an XINS interpretation.

Our analysis incorporates more recent observations with the source positioned closer to the aimpoint and comprises six epochs, totalling $3.11(6)\times10^3$ net counts with an average background contribution of 8\% per epoch. The best-fit single-temperature blackbody model yields $kT = 100.0\pm1.2$\,eV and $N_\mathrm{H} = 9.6(7)\times10^{21}$\,cm$^{-2}$, with an emitting radius of $13.0_{-0.7}^{+0.8}$\,km at a distance of $2.0_{-0.1}^{+1.3}$\,kpc based on extinction maps \citep{2024arXiv240303127D}. The corresponding luminosity is $L_{\rm X} = 2.96(18)\times10^{33}$\,\lumcgs, with a stable observed flux of $7.05(14)\times10^{-14}$\,\fluxcgs\ (0.2--2\,keV). 
We find no evidence of long-term variability: spectral parameters remain consistent across epochs, and allowing for inter-epoch variability yields an F-test probability indicating that a constant model is statistically favoured.

A high-redshift ($z = 1.4$), low-metallicity ($Z < 0.02$\,Z$_\odot$) thermal plasma model with $kT = 312_{-4}^{+5}$\,eV provides a comparably good fit, whereas a Galactic \texttt{apec} model is clearly disfavoured (Table~\ref{tab_spec}). Neutron star atmosphere models with magnetic fields of $B = 10^{12}$--$10^{13}$\,G and effective temperatures $T_{\rm eff} \sim 4 \times 10^5$\,K also yield acceptable fits and outperform the non-magnetic case.

Adding a second thermal component ($kT \sim 205$\,eV) or a steep power-law ($\Gamma = 9.3_{-0.8}^{+0.5}$) significantly improves the fit relative to a single-temperature blackbody (F-test probability $< 6\times10^{-4}$), but the required high column densities imply distances beyond 10\,kpc and emitting regions exceeding the neutron star radius. In contrast, thermal models (\texttt{bb} or magnetised \texttt{nsa}, with $T_{\rm eff} \sim 6 \times 10^5$\,K) modified by a broad absorption feature at 510--520\,eV ($\sigma \sim 120$\,eV) yield similarly good fits, lower column densities of $\nh = (4$--$6)\times10^{21}$\,cm$^{-2}$, and more plausible distances of 1.3--1.5\,kpc.

The current $4\sigma$ upper limit on the logarithmic X-ray-to-optical flux ratio, based on Pan-STARRS PS1, is $X/O > 2.11$ ($g > 22.8$; Fig.~\ref{fig_fcxins}). This value exceeds those typically found in normal or star-forming galaxies \citep[e.g.][]{2003AJ....126..575H,2004AJ....128.2048B}, but remains consistent with extreme BL Lac objects or distant cataclysmic variables. However, the soft X-ray spectrum and lack of variability make these scenarios less likely.

Considering its spectral characteristics, lack of confirmed counterparts, and location near rich OB associations, the X-ray source \si\ is a strong XINS candidate. Targeted optical follow-up and optimised X-ray observations at the aimpoint, using the medium or thin filter, are required to search for pulsations and to confirm the presence of the tentative absorption feature.

\subsubsection{4XMM J140340.4$-$603007}

The X-ray source was first detected in 2003 at an off-axis angle of 6.8\arcmin\ from the bright, massive, and variable star \object{Beta~Cen}.  
That archival exposure used the thick EPIC filter, and the detection of this particularly soft emitter was affected by out-of-time events. Its characterisation improved with the 2024 follow-up \xmm\ observation, complemented by \eROS\ data obtained between 2020 and 2022.  
It was not included among the XINS candidates in the search by \citet{2024A&A...687A.251K} due to its faintness, with a flux below their selection limit of $10^{-13}$\,\fluxcgs\ in the 0.2--2\,keV energy band.

Spectral analysis shows a soft, stable energy distribution, well described by a blackbody with temperature $kT = 98.1 \pm 1.8$\,eV and absorption column density $\nh = 6.2^{+0.9}_{-0.8} \times 10^{21}$\,cm$^{-2}$. The observed flux is $2.58(8) \times 10^{-14}$\,\fluxcgs\ (0.2--2\,keV). Assuming a distance of $3.6^{+0.4}_{-0.7}$\,kpc from extinction maps, the inferred emitting radius is $7.6(6)$\,km and the X-ray luminosity is $L_{\rm X} = 1.49^{+0.13}_{-0.12} \times 10^{33}$\,\lumcgs. There is no significant evidence for additional spectral complexity or variability over the past twenty years. Magnetised \texttt{nsa} models with $B = 10^{12}$--$10^{13}$\,G and $T_{\rm eff} = 4 \times 10^{5}$\,K reproduce the spectrum equally well, yielding comparable fit statistics and NHP. All acceptable models imply $\nh \sim (6$--$9) \times 10^{21}$\,cm$^{-2}$, below the total Galactic column in this direction ($3.7 \times 10^{22}$\,cm$^{-2}$), disfavouring an extragalactic origin and suggesting a distance of $\sim$3--5\,kpc.

The X-ray-to-optical flux ratio from DeCAPS2 coverage is $X/O > 2.07$ ($4\sigma$; $g > 23.8$; Fig.~\ref{fig_fcxins}). While this moderately high value does not formally exclude extragalactic scenarios, the soft and stable thermal emission, low Galactic latitude, and long-term flux constancy favour an XINS interpretation. Deeper optical imaging is required to rule out faint companions and strengthen its classification as a new, more distant thermally emitting neutron star.

\begin{table*}
\small
\caption{Summary of the properties of \xmm-selected thermally emitting isolated neutron stars and candidates\label{tab_xins_summary}}
\centering
\begin{tabular}{lrcccccccc}
\hline\hline
Identifier & \multicolumn{1}{c}{Flux\,\tablefootmark{a}} & $d_{N_{\rm H}}$\,\tablefootmark{b} & $|h_z|$\,\tablefootmark{c} & $kT$\,\tablefootmark{d} & $\epsilon$\,\tablefootmark{e} & $p_{\rm f}$\,\tablefootmark{f} & $\log(L_{\rm X})$\,\tablefootmark{g} & Association/LOS\,\tablefootmark{h} & Status\,\tablefootmark{i} \\
2XMM | 4XMM & ($10^{-13}$ cgs) & (kpc) & (pc) & (eV) & (keV) & (\%) & (\lumcgs) & & \\ 
\hline
J114051.9$-$641848 & $2.23(8)$  & 2.8--5.7 & 70--80 & 85 & $-$ & $<19$ & 34.6 & Galactic plane & Candidate\\
J022141.5$-$735632 & $1.19(21)$ & $<1.9$ & $<1200$ & 60 & $-$ & $<11$ & 32.4 & Magellanic Bridge (LOS) & XINS\\
J104608.7$-$594306 & $1.17(3)$ & 2.3 & 24 & 135 & $(0.61,1.35)$ & $<14$ & 32.6 & Carina Nebula & XINS\\
J181844.3$-$120751 & $0.71(14)$ & 1.9--3.3 & 50--90 & 100 & $0.51$ & $<30$ & 33.5 & Ser OB2 & Candidate \\
J140340.4$-$603007 & $0.26(8)$ & 2.9--4.0 & 60--80 & 98 & $-$ & $<24$ & 33.2 & Galactic plane & Candidate\\
J194744.5$+$274220 & $0.22(7)$ & 6.9--7.7 & 140--160 & 96 & $(0.53,1.11)$ & $<21$ & 34.0 & SNR/PWN & Candidate\\
J175437.8$-$294148 & $0.14(8)$ & 1.2--2.0 & 40--70 & 110 & $-$ & $<44$ & 31.4 & Galactic bulge (LOS) & Candidate\\
\hline
\end{tabular}
\tablefoot{The sources are sorted by decreasing observed flux.
\tablefoottext{a}{Observed X-ray flux from spectral analysis, fitting all available data sets of each source simultaneously (Sect.~\ref{sec_spec}).}
\tablefoottext{b}{Distance range estimated from X-ray interstellar absorption, assuming the best-fit blackbody model and extinction maps (\citealt{2024arXiv240303127D}; Table~\ref{tab_spec}).}
\tablefoottext{c}{Vertical distance from the Galactic plane, $h_{z}=d_{N_{\rm H}}\sin|b|$.}
\tablefoottext{d}{Best-fit blackbody temperature (Table~\ref{tab_spec}).}
\tablefoottext{e}{Energy of tentative absorption features, when significant deviations from a simple continuum are detected (Sect.~\ref{sec_nature}).}
\tablefoottext{f}{$4\sigma$ upper limits on pulsations (Sect.~\ref{sec_timing}).}
\tablefoottext{g}{Corresponding X-ray luminosity, calculated from the blackbody model and the estimated distance range, in logarithmic units.}
\tablefoottext{h}{Large-scale stellar structures or prominent features projected close to, or along, the line-of-sight. Entries marked as LOS indicate projection against a given structure, not a physical association.}
\tablefoottext{i}{Current classification as a confirmed XINS or candidate, primarily based on X-ray spectral properties and X-ray-to-optical flux ratio constraints (Table~\ref{tab_optlim}).}}
\end{table*}

\subsubsection{4XMM J194744.5$+$274220}

The thermal X-ray source (observed flux $2.17(7)\times10^{-14}$\,\fluxcgs\ in the 0.2--2\,keV band) was first noted in a joint \xmm\ and \chan\ study of the evolved ($>8$\,kyr) plerionic supernova remnant \object{SNR~G063.7+01.1} \citep{1997AJ....114.2068W}, aimed at detecting its pulsar wind nebula (PWN) and associated neutron star \citep{2016ApJ...825..134M}. That study revealed clear nebular emission and identified the hard ($\Gamma = 1.1 \pm 0.9$, $N_\mathrm{H} = 1.6\times10^{22}$\,cm$^{-2}$), faint ($2.0\times10^{-14}$\,\fluxcgs\ in the 0.5--10\,keV band), possibly extended source \object{CXO~J194753.3+274351} as a potential neutron star powering the PWN. In contrast, the thermal source \object{CXO~J194744.5+274221}/\jonfs, our XINS candidate, lies 3.5\,arcmin from the remnant centre and 2.4\,arcmin from the PWN peak. 

The faint PWN itself is absent from the \xmm\ and \chan\ catalogues \citep{2020A&A...641A.136W,2024ApJS..274...22E} but is listed in the \fermi\ 14-year Source Catalog as \object{4FGL~J1947.7+2744}, with an energy flux of $1.03(15)\times10^{-11}$\,\fluxcgs\ in the 0.1--100\,GeV band. Our enhanced spectral source detection (Section~\ref{sec_sourcedetection}) reveals a blend of extended X-ray emission and a point-like source coincident with the PWN. The best-fit parameters are $N_\mathrm{H} = 2.3(7)\times10^{22}$\,cm$^{-2}$ and $\Gamma = 1.8 \pm 0.4$, consistent with \citet{2016ApJ...825..134M}. The point-like component of this blended detection, possibly the neutron star powering the nebula, is located at $\alpha = 19^{\mathrm{h}}47^{\mathrm{m}}53\fs0$, $\delta = +27\degr43\arcmin52\farcs7$ (J2000) with a $1\sigma$ uncertainty of $1.4\arcsec$, and has an observed flux of $6.8\times10^{-15}$\,\fluxcgs\ in the 0.2--12\,keV band.

Regarding the XINS candidate, the total number of counts is $1060\pm30$ in the 0.3--1.5\,keV band (background contribution of 11\%), combining the \xmm\ and \chan\ epochs. A simple blackbody fit with $kT = 95.6_{-1.8}^{+1.9}$\,eV yields an emitting radius of $R = 12.4_{-1.0}^{+1.1}$\,km and a column density of $\nh = 7.9(9)\times10^{21}$\,cm$^{-2}$, corresponding to a distance of $d = 7.3(4)$\,kpc from extinction maps and an X-ray luminosity of $L_{\rm X} = (9.1_{-0.7}^{+0.9})\times10^{33}$\,\lumcgs. The fit quality is poor ($\mathrm{NHP} = 11\%$), owing to residuals near 530\,eV and 1.1\,keV.

A Galactic \texttt{apec} model provides a statistically acceptable fit ($\mathrm{NHP} = 83\%$, $C = 83$ for 98 degrees of freedom) but is disfavoured by the lack of an optical counterpart for a coronal emitter ($g > 22.9$, $X/O > 1.72$). All other single-component models (\texttt{nsa}, low-metallicity or high-redshift \texttt{apec} variants) yield poor statistics ($\mathrm{NHP} \lesssim 20\%$) and show similar residuals. Adding a second thermal component ($kT \sim 140$--260\,eV) or a steep power-law ($\Gamma < 4.6$) improves the fit significantly (F-test probability $< 4\times10^{-9}$), but requires unrealistically large column densities, distances, and emitting regions.

Including one or two broad absorption features, with fixed widths of 100--200\,eV, at $526_{-17}^{+13}$\,eV and $1.05_{-0.24}^{+0.23}$\,keV significantly improves the fit ($\mathrm{NHP} \sim 70$--90\%) for both blackbody and \texttt{nsa} continua (Table~\ref{tab_spec}). The fitted temperatures are hotter than for single-component models ($kT \sim 140$\,eV; $T_{\rm eff} \sim 7\times10^5$\,K), while the inferred luminosity is reduced by more than three orders of magnitude; the corresponding column densities and distances are poorly constrained over wide ranges. Alternatively, a model with mildly overabundant oxygen ($Z_{\rm O} = 1.1$--1.2\,Z$_\odot$) and a single absorption feature at $1.10\pm0.20$\,keV provides an equally good description of the data while preserving the main continuum properties of the simpler fits.

No variability is detected between the \xmm\ and \chan\ epochs over a 5.3-year baseline, while only flux upper limits are available from \ros\ and \swift. A follow-up \fast\ observation reveals no significant periodic signal ($\sim$3.0\,$\mu$Jy at 1.4\,GHz). Deeper optical limits and dedicated X-ray observations are required to better characterise this source and to assess its possible association with the SNR/PWN system.

\subsubsection{4XMM J022141.5$-$735632}

The X-ray source, the first followed up by our programme, is a confirmed XINS located in the foreground of the Magellanic Bridge \citep{2022A&A...666A.148P}. It has also been independently identified in similar searches using 4XMM-DR10 and eRASS data \citep{2022MNRAS.509.1217R,2024A&A...687A.251K}. 

Consistent with previous analyses \citep{2022A&A...666A.148P,2022MNRAS.509.1217R}, the spectral fit combining two EPIC observations and the concatenated eRASS:4 epochs shows constant, soft, blackbody-like emission with $kT = 60.5 \pm 0.8$\,eV and $\nh = (7.0 \pm 1.0) \times 10^{20}$\,cm$^{-2}$. The extinction-derived distance is constrained to $\lesssim 1.9$\,kpc, placing an upper limit on the X-ray luminosity of $L_{\rm X} < 2.4 \times 10^{32}$\,\lumcgs. Based on LS coverage, the X-ray-to-optical flux ratio is $X/O > 3.01$ for a $4\sigma$ magnitude upper limit of $g > 25.4$.

A supersoft-source interpretation associated with the Magellanic Bridge or the SMC is disfavoured, as the implied luminosity would be too low for a classical supersoft source at such distances, while remaining consistent with a Galactic XINS. The flux stability and lack of optical counterparts further support this interpretation.

\subsubsection{4XMM J175437.8$-$294148}

The faintest X-ray source in our sample (observed flux $1.42(8)\times10^{-14}$\,\fluxcgs\ in the 0.2--2\,keV band), also selected as an XINS candidate by \citet{2022MNRAS.509.1217R}, was detected multiple times in \xmm\ observations of the Galactic bulge in 2004, 2006, and 2018. In all cases, the soft source was observed at moderately large off-axis angles of 8--13\,arcmin (Table~\ref{tab_obsdet}). It was also detected by \chan\ in 2004 (Table~\ref{tab_chanobs}).

Spectra from the four epochs, with total counts of $490 \pm 22$, are consistently fitted with a blackbody of temperature $kT = 110 \pm 4$\,eV, absorbed by $\nh = (3.6^{+1.1}_{-1.0}) \times 10^{21}$\,cm$^{-2}$. These parameters imply a distance of 1.2--2\,kpc and an emission radius of $R_{\rm em} = 0.94^{+0.12}_{-0.11}$\,km, corresponding to a luminosity of $2.7(3) \times 10^{31}$\,\lumcgs. Neutron-star atmosphere models provide similarly good fits, whereas \texttt{apec} models yield larger residuals. The lack of significant spectral or flux variability across epochs further disfavors a stellar coronal origin. No counterparts are detected at the stacked \xmm\ position, giving $X/O > 1.64$ ($g > 24.0$; $4\sigma$), though source confusion in this direction remains a concern without a dedicated X-ray observation.

\subsection{Population constraints\label{sec_population}}

Table~\ref{tab_xins_summary} presents the properties of the sources discussed in Section~\ref{sec_nature}, together with \carINS, an XINS initially identified in the 2XMMp catalogue \citep{2009A&A...498..233P}. The table summarises the observed X-ray flux, distance estimates based on interstellar absorption, height above the Galactic plane, blackbody temperature, energies of tentative absorption features, upper limits on pulsations, and the corresponding X-ray luminosity (see caption for details).

For comparison, Table~\ref{tab_xins_allsky} provides analogous quantities for previously known XINSs. This includes those identified in \ros\ observations (the \msev\ XINSs and Calvera; see \citealt{2024ApJ...969...53B,2024ApJ...976..228R} and references therein), as well as more recent eRASS discoveries at intermediate fluxes \citep{2023A&A...674A.155K,2026A&A...705A.148K}.

\begin{figure}[t]
\begin{center}
\includegraphics[width=\columnwidth]{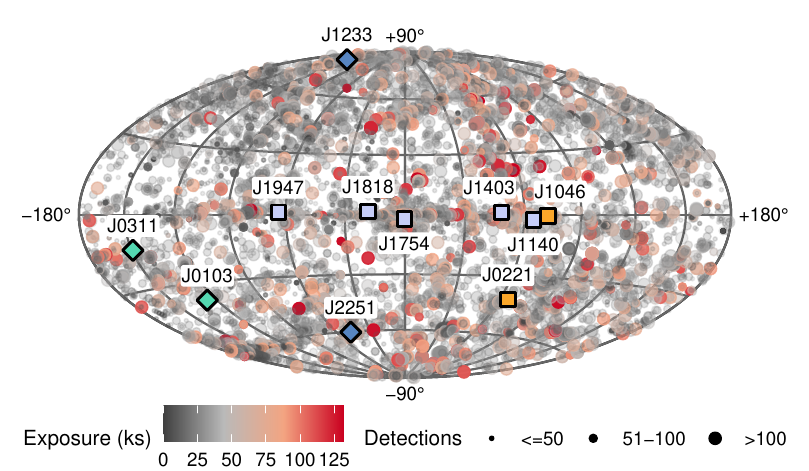}
\end{center}
\caption{Hammer--Aitoff projection in Galactic coordinates showing the footprint of 12,712 \xmm\ observations included in the 4XMM-DR12 catalogue (not scaled to the \xmm\ field of view). The size and colour of each pointing indicate the number of good (i.e.\ not flagged) detections and the filtered EPIC pn exposure. Confirmed (yellow) and compelling (lilac) XINS candidates from 4XMM-DR12 are shown as labelled squares. The XINS in the Carina Nebula \carINS\ is also shown for reference. Contaminants (dark blue) and poorly characterised sources (cyan) are shown as diamonds.}
\label{fig_aitoff}
\end{figure}

In addition, Table~\ref{tab_xins_summary} lists line-of-sight features, prominent nearby stellar structures, and possible associations, as well as the current classification of each source as either a confirmed or candidate thermally emitting XINS. This classification is primarily guided by limits on the X-ray-to-optical flux ratio (Table~\ref{tab_optlim}), as all sources discussed in Section~\ref{sec_nature} exhibit soft, stable X-ray spectra consistent with those of confirmed XINSs.

Ultimately, a direct measurement of the spin period (and spin-down rate) provides the most unambiguous means of classification within the XINS population. For the faintest and most distant sources, however, such measurements are likely to remain observationally challenging with current X-ray facilities, motivating the use of indirect diagnostics. 

A relevant example is \carINS, an XINS with properties similar to those of \msev, for which no spin period has been firmly established \citep{2012A&A...544A..17P,2015A&A...583A.117P}. The recent discovery of unusual bursts of coherent radio emission \citep{2025ApJ...985L...3R,2026arXiv260500720T} suggests a possible connection to radio-loud magnetars or repeating fast radio bursts (FRBs; \citealt{2021NatAs...5..414K,2021ApJ...907....7I}).

To date, two of the six compelling 4XMM sources (\ershiyi\ and \shi) have been followed up in X-rays. For the remaining candidates, source confusion affects the fields of \sanshiwu\ and \sanshijiu, while spectral complexity (similar to that seen in confirmed XINSs) is evident for \si\ and \ershi. On-axis observations with soft-response-appropriate optical blocking filters are therefore needed to robustly assess positional accuracy and characterise the spectra in detail.

Figure~\ref{fig_aitoff} shows the sample within the 4XMM-DR12 footprint in Galactic coordinates, together with the contaminants and poorly constrained candidates discussed in Appendix~\ref{sec_contaminants}. With the exception of the softest source, \ershiyi, the \xmm-selected XINSs and candidates are concentrated at low Galactic latitudes and remain close to the Galactic plane, with vertical distances of $\lesssim 200$\,pc. Their blackbody temperatures span a narrow range of 85--135\,eV, and the moderately high interstellar absorption is consistent with these low-$b$ locations. The median observed flux is $7\times10^{-14}$\,\fluxcgs, making these sources a factor of $\sim$3 and $\sim$40 fainter than the XINSs discovered with eRASS and \ros, respectively.

\begin{figure*}[t]
\begin{center}
\includegraphics[width=0.495\textwidth]{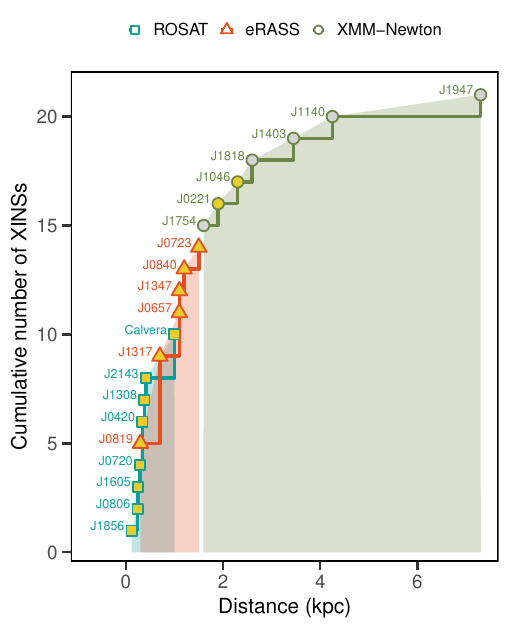}\hfill
\includegraphics[width=0.495\textwidth]{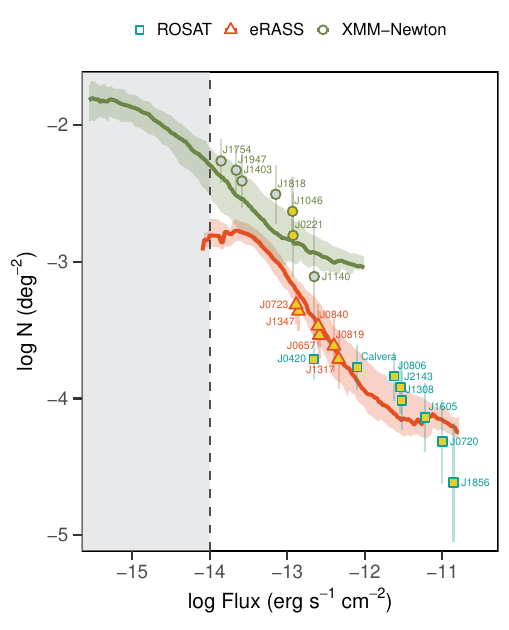}
\end{center}
\caption{Cumulative distributions of XINSs, including newly identified sources from this work and previously known objects from the \ros\ and \eROS\ surveys. The grey-filled symbols indicate candidate XINSs, while yellow-filled symbols denote confirmed objects (see Tables~\ref{tab_xins_summary} and \ref{tab_xins_allsky}). \textit{Left.} Cumulative distribution as a function of distance, with individual source labels indicated. \textit{Right.} Cumulative $\log N$--$\log S$ distributions for eRASS and \xmm. The curves show the predicted cumulative surface density $N(>S)$ from population synthesis simulations, normalised by the effective survey area of each survey. Solid lines indicate the median of 100 Monte Carlo realisations, with shaded regions showing the 16--84\% percentile range. The vertical dashed line indicates the 4XMM search flux limit ($10^{-14}\,\mathrm{erg\,s^{-1}\,cm^{-2}}$). Observed XINSs are shown as points with Poisson uncertainties on the cumulative counts. Multiple detections of the same source across different surveys are shown only once, assigned to the discovery survey.}
\label{fig_cumdist_lnls}
\end{figure*}

The left panel of Fig.~\ref{fig_cumdist_lnls} compares the cumulative distance distributions of confirmed and candidate XINSs across the three surveys. If the 4XMM candidates are confirmed, the known population extends to distances of $\sim$2--7\,kpc, well beyond those reached by the \ros\ and \eROS\ surveys.

To assess the observability of XINSs with \xmm, we employ a Galactic population synthesis model of thermally emitting isolated neutron stars previously used to forecast detections in the four-year eRASS \citep{2017AN....338..213P}. We apply this model to the 4XMM-DR12 footprint, enabling a direct comparison between the XINS population probed by \eROS\ and that serendipitously detected in cumulative \xmm\ observations. The implementation details are provided in Appendix~\ref{sec_popsynt}; here we summarise the key aspects relevant to constructing the observable sample.

The simulations yield neutron star realisations with predicted physical and observational properties, including spatial positions, distances, intrinsic luminosities, and absorbed X-ray fluxes. These quantities are converted into expected EPIC pn and MOS count rates using SSC information (nominal pointing coordinates, GTI exposures, observing modes, and filters\footnote{\url{http://xmmssc.irap.omp.eu/Catalogue/4XMM-DR12/4XMM_DR12.html}}), with pn and MOS treated separately. To simulate serendipitous detections, we restrict the analysis to full-frame imaging modes (and extended full-frame for pn). A vignetting correction is applied to sources within each field of view, as described in Appendix~\ref{sec_popsynt}.

A source is included in the mock catalogue if it yields at least ten EPIC counts (0.2--12\,keV), corresponding to a probability of $\lesssim5\times10^{-5}$ for a background fluctuation. This threshold is more stringent than the 4XMM-DR12 criterion of $\mathrm{DET\_ML}=6.5$ for any EPIC camera \citep{2020A&A...641A.136W}. As background is not explicitly modelled, this conservative cut ensures effective completeness at the adopted flux limit of $10^{-14}$\,\fluxcgs.
If a source is detected multiple times, only the detection with the highest EPIC counts is retained, while the number of detections is recorded for statistics. This procedure is repeated for 100 Monte Carlo realisations to capture population variance.

For each realisation, $\log N$--$\log S$ distributions are constructed by computing the cumulative number of detectable XINSs above flux $S$, normalised by the effective survey area. We derive these relations separately for the 4XMM-DR12 footprint (1283 deg$^2$, exposures $\ge 1$ ks, accounting for overlaps) and for eRASS, using four half-sky scans restricted to the western Galactic hemisphere ($\simeq 2\pi$ sr).
Each realisation is interpolated onto a common flux grid, and the final model prediction is defined by the median and 16th--84th percentiles across simulations after excluding extreme flux values (below the 1st percentile and above the 99.73rd percentile).

The right panel of Fig.~\ref{fig_cumdist_lnls} compares model predictions with observed XINS samples in \xmm\ and eRASS. The curves show the cumulative surface density $N(>S)$, while points include Poisson uncertainties. Additional data correspond to \ros-discovered \msev\ sources and Calvera, converted to all-sky equivalent surface densities.

Overall, the figure highlights how survey strategy and selection effects shape the accessible flux range. \xmm\ is dominated by faint detections over a small, heterogeneous footprint, while brighter sources are rare, partly because our model lacks nearby structures such as the Local Bubble and OB associations of the Gould Belt, resulting in a deficit of nearby high-latitude sources \citep[see e.g.][]{2005Ap&SS.299..117P,2008A&A...482..617P}. Moreover, targeted observations of known neutron stars and pulsars are usually taken in windowed or timing modes, which are excluded here. In contrast, eRASS provides wider sky coverage at higher flux limits, and the flattening around $\sim2.3\times10^{-14}$\,\fluxcgs\ is consistent with the shallower depth of the completed survey relative to its nominal design sensitivity.

\begin{table}
\small
\centering
\caption{Results of population synthesis simulations\label{tab_simresults}}
\begin{tabular}{@{}lc@{}ccc@{}}
\hline\hline
&& 4XMM-DR12\,\tablefootmark{a} & Sought\,\tablefootmark{b} & Remaining\,\tablefootmark{c} \\
\hline\noalign{\vskip 0.4ex}
$N_{\rm XINS}$ & & $20\pm5$ & $6_{-3}^{+2}$ & $13_{-4}^{+6}$ \\ 
$N_{\rm OBS}$ & & 1--5 & 1--3 & 1--4 \\
$N_{\rm eRASS}$ & & $2\pm2$ & $2\pm2$ & 0\\
$\mathcal{C}_{\rm min}$ & & $10.7_{-0.4}^{+1.4}$ & $52_{-25}^{+90}$ & $10.7_{-0.4}^{+1.5}$ \\
$\mathcal{C}/\mathcal{C}_{\rm eROS}$ & & $14_{-3}^{+4}$ & $11_{-5}^{+6}$ & $16_{-4}^{+6}$ \\
$f_{\rm X}$ & $10^{-14}$\,cgs & $0.41_{-0.14}^{+0.3}$ & $2.8_{-1.3}^{+2.3}$ & $0.23_{-0.07}^{+0.11}$ \\
$kT_{\rm min}$ & eV & $48_{-15}^{+13}$ & $75_{-21}^{+18}$ & $48_{-15}^{+13}$ \\
$kT$ & eV & $97_{-7}^{+6}$ & $103_{-17}^{+15}$ & $97_{-7}^{+6}$ \\
$kT_{\rm max}$ & eV & $154\pm19$ & $138_{-26}^{+30}$ & $138_{-12}^{+20}$ \\
$d$ & kpc & $2.7_{-0.7}^{+0.4}$ & $1.7_{-0.4}^{+1.1}$ & $3.2_{-0.6}^{+0.5}$ \\
$d_{\rm max}$ & kpc & $7.4_{-1.4}^{+1.5}$ & $4.1_{-1.8}^{+5}$ & $7.3_{-1.6}^{+1.4}$ \\
$\nh$ & $10^{21}$\,cm$^{-2}$ & $6.8_{-1.2}^{+1.1}$ & $5.9_{-1.8}^{+1.4}$ & $7.4_{-1.4}^{+1.6}$ \\
$t_{\rm coo}$ & kyr & $80_{-30}^{+50}$ & $50_{-40}^{+110}$ & $110\pm40$ \\
$v_{\rm 3d}$ & km\,s$^{-1}$ & $310_{-100}^{+120}$ & $290_{-130}^{+190}$ & $310_{-100}^{+140}$ \\
$t_{\rm EPIC}$ & ks & $29\pm6$ & $24\pm10$ & $31\pm9$ \\
\hline
\end{tabular}
\tablefoot{Uncertainties are given by the 16th and 84th percentiles.
\tablefoottext{a}{Median properties of simulated XINSs detected within the 4XMM-DR12 footprint and yielding more than 10 counts in the EPIC cameras (0.2--12\,keV).}
\tablefoottext{b}{Median properties of simulated XINSs selected in a flux-limited search with $f_{\rm X} > 10^{-14}$\,\fluxcgs\ (0.2--12\,keV).}
\tablefoottext{c}{Median properties of simulated XINSs that remain detectable in the catalogue but are not selected by the flux-limited search.}}
\end{table}

Table~\ref{tab_simresults} summarises the median properties of the simulated XINS population, distinguishing sources detected in the 4XMM-DR12 footprint, those selected by the flux-limited search, and those below the flux cut. The table lists the number of sources ($N_{\rm XINS}$), detections ($N_{\rm OBS}$) from repeated coverage, XINSs also detected in four completed eRASS scans ($N_{\rm eRASS}$)\footnote{Detection in eRASS corresponds to at least 30 counts (0.2--2\,keV) \citep{2017AN....338..213P}. Only detections in the western Galactic hemisphere are considered.}, minimum counts ($\mathcal{C}_{\rm min}$), count ratio relative to eRASS ($\mathcal{C}/\mathcal{C}_{\rm eROS}$), flux, temperature range, median and maximum distance, column density, cooling age ($t_{\rm coo}$), spatial velocity ($v_{\rm 3d}$), and EPIC exposure ($t_{\rm EPIC}$; averaged over active detectors).
Uncertainties correspond to 68\% confidence intervals and include only statistical errors; systematic effects from the neutron star birthrate, cooling, and absorption dominate the absolute normalisation and are not explored further.

The simulations yield a median of $20 \pm 5$ XINSs in the 4XMM-DR12 footprint, of which $6_{-3}^{+2}$ exceed the flux cut of $10^{-14}$\,\fluxcgs, consistent with the number of compelling candidates identified in our search. Up to four of these (68\% confidence) are expected to overlap with the western Galactic hemisphere of the completed eRASS. For comparison, the same model predicts $\sim100$ XINSs all-sky at the nominal survey depth (eight \eROS\ scans; \citealt{2017AN....338..213P}).

While the model broadly reproduces the properties of the local XINS population discovered with \ros, it does not recover the \msev\ XINSs in the simulated 4XMM-DR12 footprint. This reflects both the spatial distribution of synthetic neutron stars and the implemented simulation selection function, which excludes observational configurations such as calibration set-ups and partial-window modes used for \msev\ observations. Owing to repeated sky coverage, the median number of serendipitous detections per source above the flux cut is typically 1--3, indicating that many benefit from the stacked analysis adopted in this work.

A substantial fraction of the serendipitously detectable population ($13_{-4}^{+6}$ XINSs, $\sim70\%$) remains below the flux limit and is absent from the candidate sample. These sources have median minimum fluxes of $(2.3$--$8)\times10^{-16}$\,\fluxcgs, about 30 times below the faintest candidates selected in our search. Relative to the flux-limited sample, they are cooler, more distant, and more absorbed, with typical differences of factors $\sim1.6$ in temperature, $\sim1.8$ in distance, and $\sim2.1$ in column density.

How source confusion and the applied $P_{\rm no\text{-}id} > 50\%$ cut in optical/infrared cross-matching affect this hidden population remains to be investigated. Simulations may help optimise appropriate $P_{\rm no\text{-}id}$ thresholds for such faint sources.

\section{Summary and conclusions\label{sec_summary}}

We analysed a sample of ten isolated neutron star candidates drawn from the fourth-generation \xmm\ catalogues, combining archival and new X-ray observations with multiwavelength constraints to assess their nature. Using spectral modelling, stacked detections and astrometry, variability studies, and counterpart searches, we aimed to distinguish bona fide thermally emitting neutron stars (XINSs) from stellar and extragalactic contaminants. The observational results were then compared with predictions from a population synthesis model, thus allowing us to evaluate sample completeness, the detectability of faint sources, and the overall properties of the Galactic XINS population. Our main observational findings are as follows:

\begin{enumerate}
	\item Six multiply detected, generally well-localised sources (\sanshiwu, \ershiyi, \si, \shi, \ershi, and \sanshijiu), all at low Galactic latitudes except for the confirmed \ershiyi\ \citep{2022A&A...666A.148P}, are likely new XINSs or, more generally, compact Galactic thermally emitting objects at distances of $\sim$2--7\,kpc. These sources lack optical counterparts and exhibit soft ($kT \sim 60$--110\,eV), largely stable X-ray emission, with emission radii and luminosities consistent with established XINSs. Deeper optical limits or dedicated on-axis X-ray observations are required for confirmation in most cases. Source confusion affects \sanshiwu\ and \sanshijiu. Although these sources are also detected by \eROS\ and \chan, their positional accuracy is lower than that achieved with serendipitous \xmm\ data.
	
	\item Two candidates (\liu\ and \sanshier) are securely classified as extragalactic contaminants, supported by updated \xmm\ positions and deeper optical coverage. Two intermediate to high Galactic latitude sources (\shiyi\ and \ershisan) remain poorly characterised, and further follow-up is required to exclude an extragalactic origin.
	
	\item No source shows significant long-term variability in flux or spectral shape over timescales of 5 to 21 years. The observed flux of \shiyi\ shows a possible decline by a factor of $\sim$2--3 over seven years; however, this trend is only marginally significant and model-dependent. The source \ershisan\ is the only object in the sample not multiply detected.
	
	\item Significant deviations from the dominant thermal continuum, tentatively modelled as broad absorption features or enhanced oxygen abundances, are detected in the distant low-$b$ candidates \si\ and \ershi. Higher signal-to-noise spectra are required to constrain the properties and physical origin of these deviations.
	
	\item No significant pulsed emission was detected in any source, including in radio follow-up observations of two candidates with \fast\ ($\sim$3--4\,$\mu$Jy; 8$\sigma$). The resulting X-ray upper limits on the pulsed fraction (10--40\%) are unconstraining, owing to the faintness of the candidates and the shallow depth of both the serendipitous \xmm\ observations and the follow-up programme.
	
	\item Three sources in the western Galactic hemisphere were detected in eRASS. Two of these were not selected as XINS candidates in \citet{2024A&A...687A.251K}, either due to source confusion in the Galactic plane or their overall faintness.
\end{enumerate}

\noindent
In addition to the observational results, our population synthesis simulations provide complementary insights into the completeness and hidden fraction of the XINS population. 

The number and overall properties of the flux-limited candidates identified in this work broadly agree with the simulations, which suggests that the search is reasonably complete above the adopted flux threshold, although this depends on the assumed model and birthrate. The simulations indicate a median of 1--3 detections per source due to repeated sky coverage, implying that many sources are observed multiple times; combining serendipitous detections would result in great benefits. 

The simulations also suggest that a significant fraction of the serendipitously detectable population (approximately 70\%) remains hidden below the selection threshold. These sources, likely already present in the 4XMM-DR12 catalogue, are expected to be fainter, softer, more distant, and more strongly absorbed than the flux-limited sample. Identifying them is challenging due to source confusion, and requires more refined optical/infrared cross-matching, possibly incorporating counterpart properties as Bayesian priors. 

Finally, this work highlights the complementarity of \xmm\ and \textit{SRG}/\eROS. The serendipitous \xmm\ coverage extends the accessible parameter space of neutron star populations by enabling the detection of faint and distant sources, as well as objects in crowded Galactic fields. Future improvements to population synthesis models, with more accurate prescriptions for cooling, absorption, and birthrate, together with upcoming results from \eROS\ and \xmm, will tighten constraints on the Galactic evolution and population properties of XINSs.

\begin{acknowledgements}
	
We sincerely thank Pan Zhichen for their assistance with the FAST observations. We thank the anonymous referee for their constructive comments, which helped improve the clarity of the manuscript.
AMP acknowledges the Major Science and Technology Project of Guizhou Province (Grant No.~QKH-ZDZX(2024)-016). JK acknowledges support by the German DLR under contract number 50OR2408. Zhang Liyun acknowledges the Guizhou Provincial Natural Science Foundation project No.~ZD [2026] 058.
Based on observations obtained with \xmm, an ESA science mission with instruments and contributions directly funded by ESA Member States and NASA (fulfil programs 088419, 090126, 092282; archival data 0008820301, 0008820601, 0109460801, 0147390701, 0150020101, 0206590201, 0306680301, 0402280101, 0603380101, 0674110401, 0740990101, 0765030801, 0784450501, 0801683001, 0920040401, 0920040501, 0920040601).
This work is based on data from eROSITA, the soft X-ray instrument aboard SRG, a joint Russian-German science mission supported by the Russian Space Agency (Roskosmos), in the interests of the Russian Academy of Sciences represented by its Space Research Institute (IKI), and the Deutsches Zentrum f\"ur Luft- und Raumfahrt (DLR). The SRG spacecraft was built by Lavochkin Association (NPOL) and its subcontractors, and is operated by NPOL with support from the Max Planck Institute for Extraterrestrial Physics (MPE).
The development and construction of the eROSITA X-ray instrument was led by MPE, with contributions from the Dr. Karl Remeis Observatory Bamberg \& ECAP (FAU Erlangen-Nuernberg), the University of Hamburg Observatory, the Leibniz Institute for Astrophysics Potsdam (AIP), and the Institute for Astronomy and Astrophysics of the University of T\"ubingen, with the support of DLR and the Max Planck Society. The Argelander Institute for Astronomy of the University of Bonn and the Ludwig Maximilians Universit\"at Munich also participated in the science preparation for eROSITA.
The eROSITA data shown here were processed using the eSASS/NRTA software system developed by the German eROSITA consortium.
This work has used the data from the Five-hundred-meter Aperture Spherical radio Telescope (FAST). FAST is a Chinese national mega-science facility, operated by the National Astronomical Observatories of Chinese Academy of Sciences (NAOC).
This research has made use of data obtained from the Chandra Source Catalog, provided by the Chandra X-ray Center (CXC).
The XMM2ATHENA project has received funding from the European Union's Horizon 2020 research and innovation programme under grant agreement no.~101004168. 
This research has made use of data obtained from the 4XMM XMM-Newton serendipitous source catalogue compiled by the XMM-Newton Survey Science Centre consortium. 
This research has made use of data and/or software provided by the High Energy Astrophysics Science Archive Research Center (HEASARC), which is a service of the Astrophysics Science Division at NASA/GSFC.
This research has made use of the VizieR catalogue access tool, CDS, Strasbourg, France. The original description of the VizieR service was published in A\&AS 143, 23. 
The authors acknowledge the use of the following sets of astronomical software: \texttt{stilts} \citep{2006ASPC..351..666T}, \texttt{ds9} \citep{2003ASPC..295..489J}.

\end{acknowledgements}

\newpage
\bibliographystyle{aa}
\bibliography{dr9}

\begin{appendix}
	
\section{Extra material}

Table~\ref{tab_4xmm} summarises the properties of the ten XINS candidates, including their catalogued fluxes and hardness ratios, the total hydrogen column density along the line-of-sight \citep{2024arXiv240303127D}, $N_{\rm H}^{\rm Gal}$, and the number of detections by \xmm\ (both serendipitous and dedicated). Any additional coverage from \chan\ or \eROS\ is indicated in the table for each source.
	
Table~\ref{tab_obsdet} summarises the \xmm\ observations included in the analysis, detailing the observation date, instrumental set-up, target off-axis angle, net exposure time, percentage of GTIs, and total EPIC target counts in the 0.2--2\,keV energy band. Tables~\ref{tab_erass} and \ref{tab_chanobs} summarise the corresponding complementary \eROS\ and \chan\ observations.
	
\begin{figure*}[t]
\begin{center}
\includegraphics[width=0.33\textwidth]{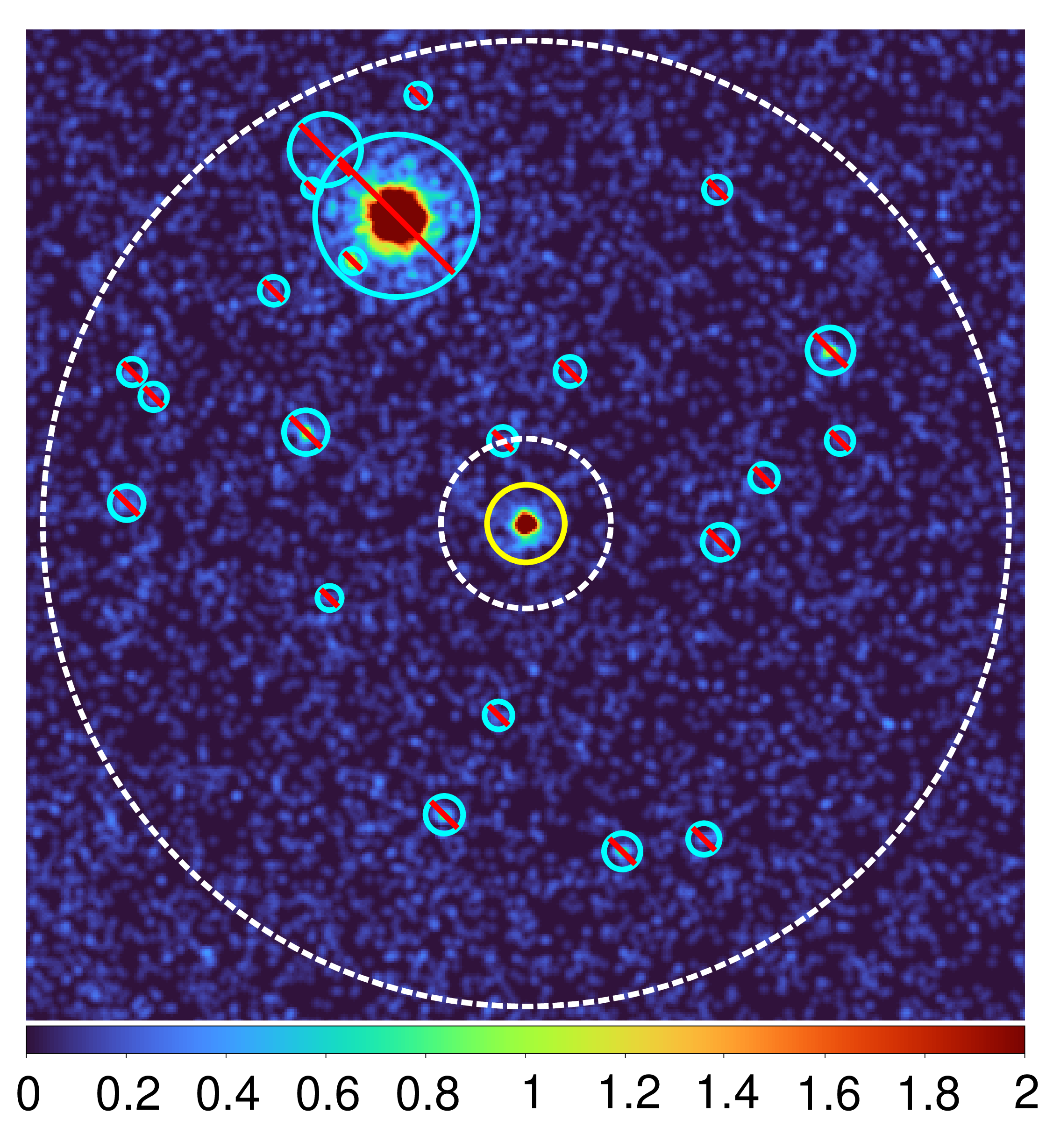}\hfill
\includegraphics[width=0.33\textwidth]{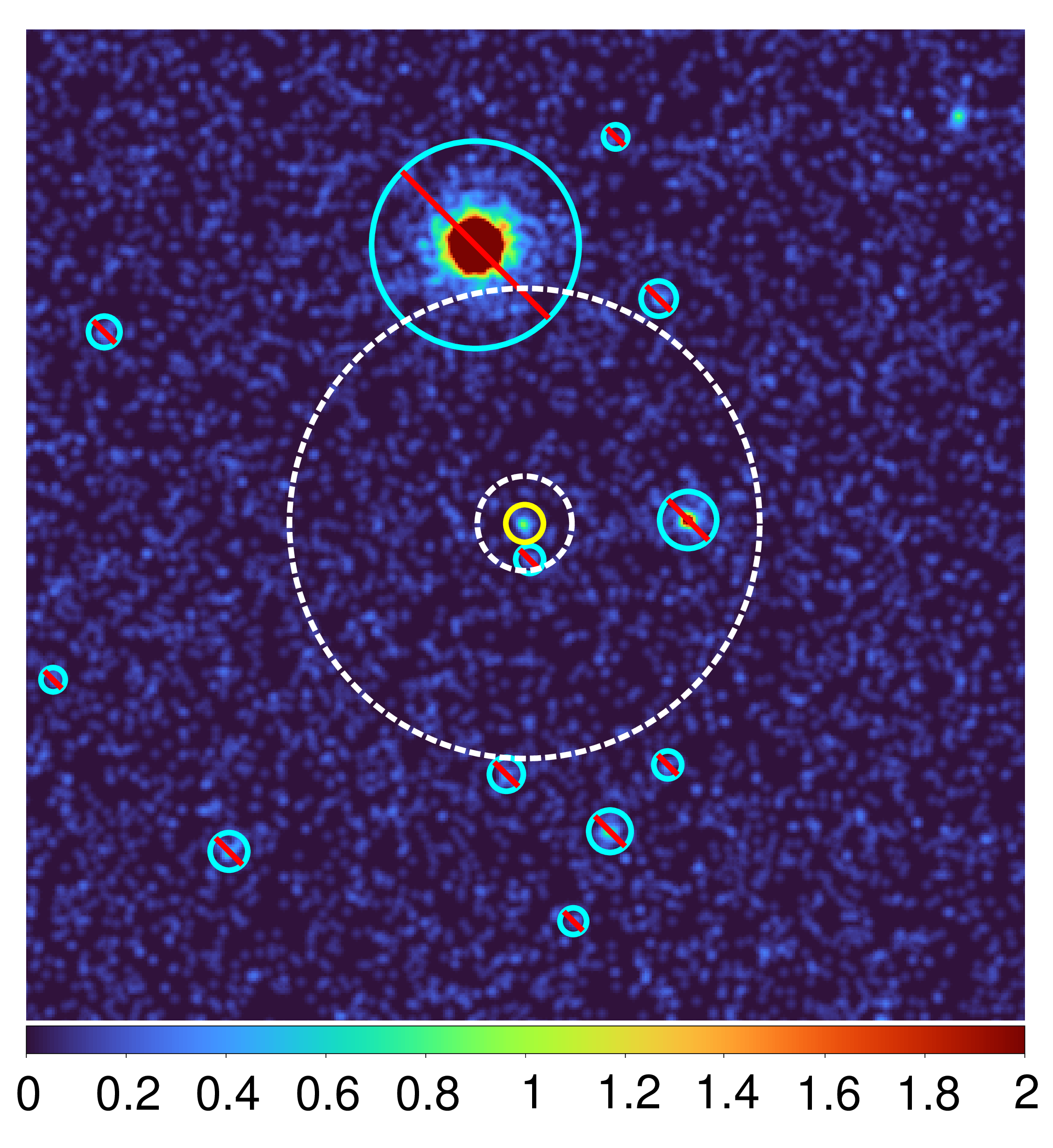}\hfill
\includegraphics[width=0.33\textwidth]{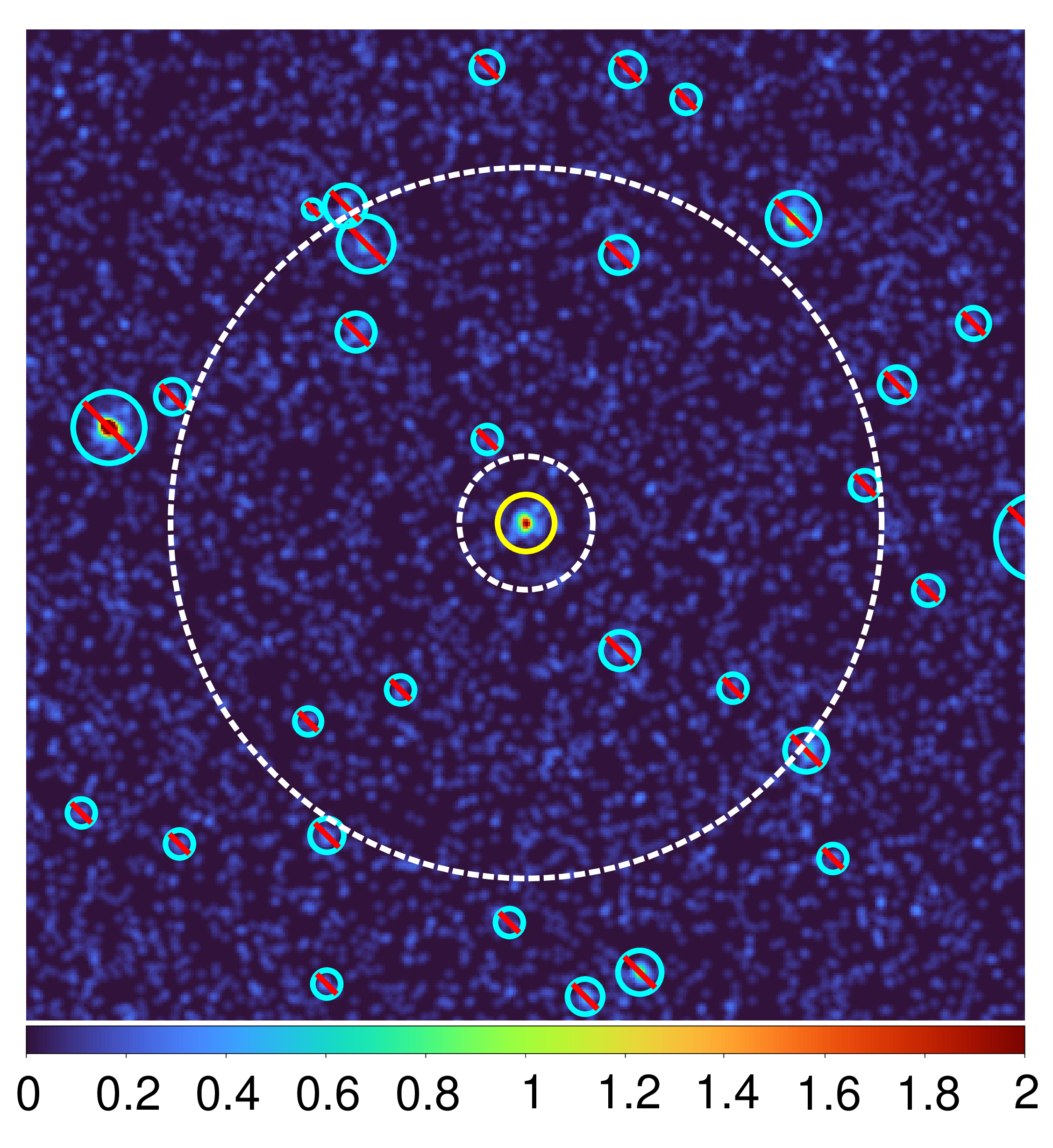}
\end{center}
\caption{\eROS\ extraction regions (eRASS:4/5, TM0, 0.2--5\,keV) for \sanshiwu\ (left), \shi\ (centre), and \ershiyi\ (right). The same intensity scale (in counts per pixel) is applied to all panels. Source regions (central yellow circles) have radii of 64\arcsec, 31\arcsec, and 47\arcsec, respectively. Background annuli (dashed white) have inner and outer radii of (140\arcsec, 796\arcsec), (78\arcsec, 388\arcsec), and (110\arcsec, 586\arcsec), respectively. Contaminating sources excluded from the background are marked with cyan circles crossed by red bars.\label{fig_eros}}
\end{figure*}

\begin{figure*}[t]
\sidecaption
\includegraphics[width=6cm]{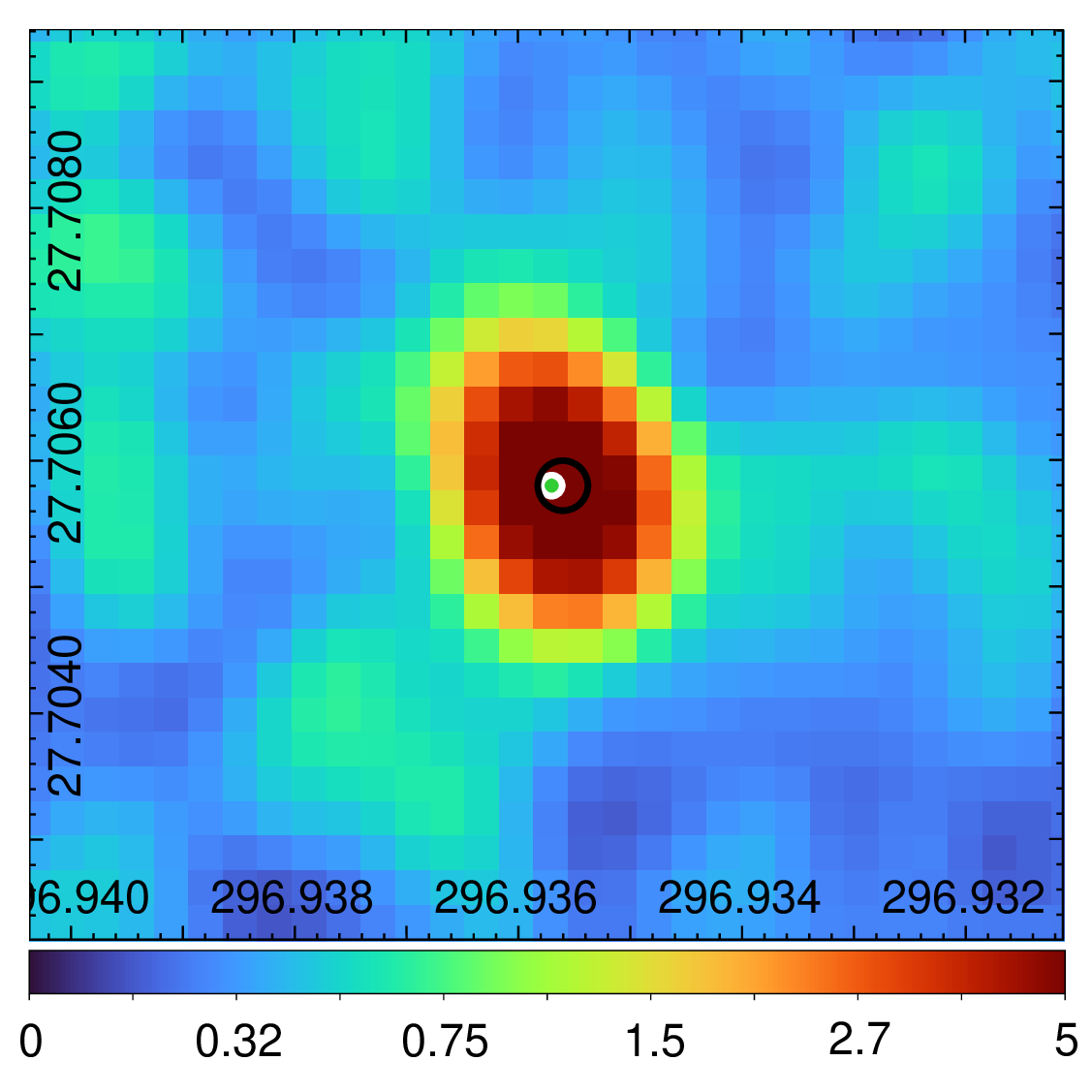}
\includegraphics[width=6cm]{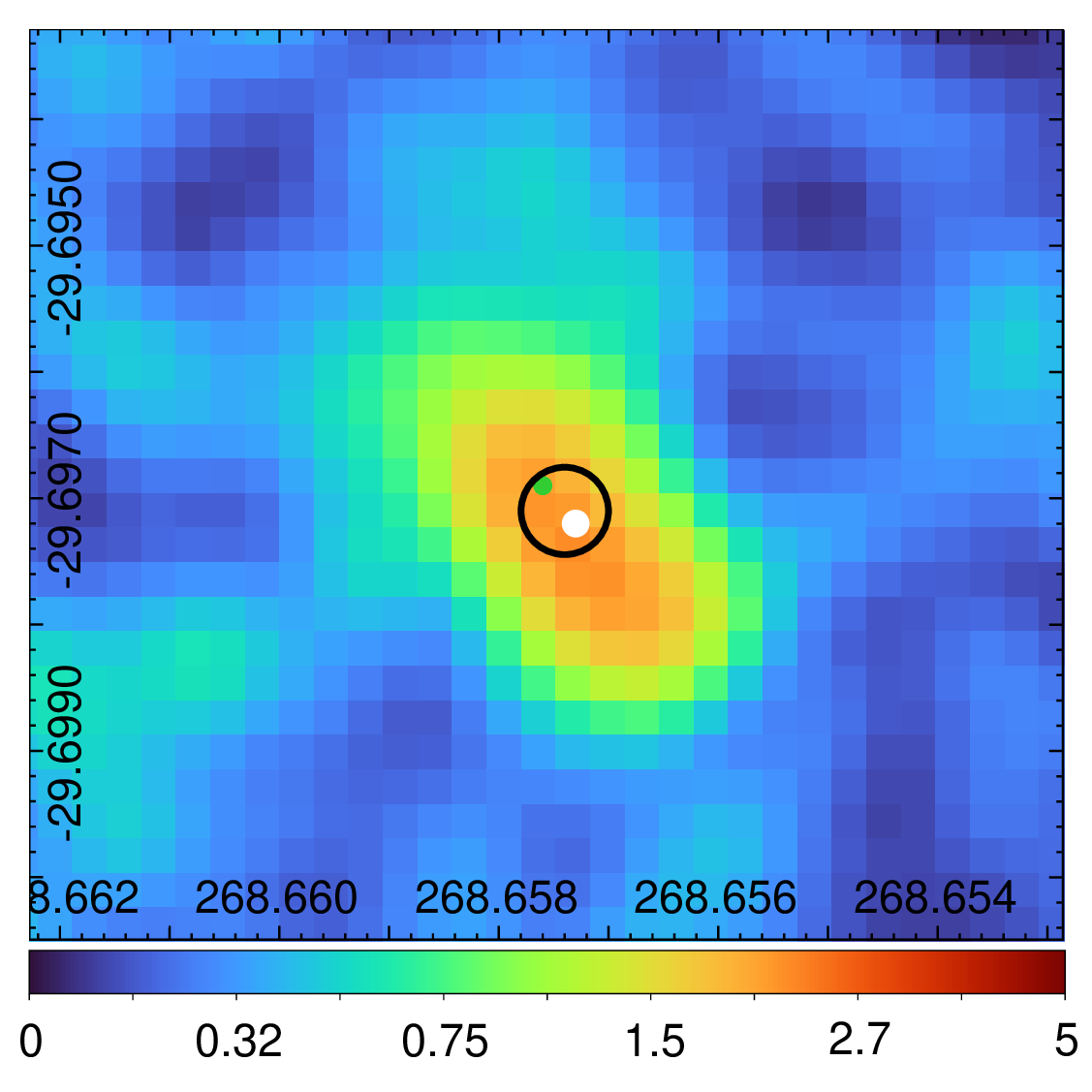}
\caption{Localisation of the targets \ershi\ (left; 4.7\arcmin\ off-axis) and \sanshijiu\ (right; 7.6\arcmin\ off-axis) using archival \chan\ ACIS imaging data in the 0.3--10\,keV band. The same intensity scale (in counts per pixel) is applied to both panels. The astrometrically corrected target positions derived from the \xmm\ stacked analysis are shown as filled green circles (99.994\% confidence level; cf.~Table~\ref{tab_psfloc}). The black circles indicate the positions from the \chan\ source catalogue (CSC Release~2.1.1; \citealt{2024ApJS..274...22E}), and the filled white circles mark the \texttt{wavdetect} detections (statistical uncertainties only). These three localisation levels are shown for direct comparison.\vspace{10pt}\label{fig_chanpsf}}
\end{figure*}

Source and background extraction regions for the \eROS-observed sources \sanshiwu, \shi, and \ershiyi\ are shown in Fig.~\ref{fig_eros}. The localisation of the targets observed by \chan, \ershi\ and \sanshijiu\ are shown in Fig.~\ref{fig_chanpsf}.

\begin{table*}[t]
\small
\caption{Properties of the sample of XINS candidates selected from 4XMM-DR9 and 4XMM-DR12
\label{tab_4xmm}}
\centering
\begin{tabular}{lcrrrrrcc}
\hline\hline
XINS candidate & $N_{\rm H}^{\rm Gal}$\,\tablefootmark{a} & Flux\,\tablefootmark{b} & HR$_1$\,\tablefootmark{c} & HR$_2$\,\tablefootmark{c} & HR$_3$\,\tablefootmark{c} & HR$_4$\,\tablefootmark{c} & $N_{\rm det}$\,\tablefootmark{d} & Additional data\,\tablefootmark{e}\\
DR9 & (cm$^{-2}$) & & & & & & & \\
\hline
\joeoe & $1.6\times10^{22}$ & 5.42(17) & 0.905(15) & $-0.353(25)$ & $-0.991(13)$ & $0.32(28)$ & 6 & $-$ \\
\jottt & $2.9\times10^{20}$ & 2.9(4) & 0.72(10) & $-0.31(11)$ & $-0.86(20)$ & $-1\pm3$ & 2 & $-$ \\
\jofzt & $3.7\times10^{22}$ & 1.98(12) & 0.85(3) & $-0.62(4)$ & $-1.00(7)$ & $0.6\pm1.8$ & 2 & \eROS \\
\jzozt & $6.8\times10^{20}$ & 1.93(20) & 0.19(9) & $-0.38(9)$ & $-0.99(6)$ & $0.3\pm1.2$ & 2 & $-$ \\
\jonfs & $1.9\times10^{22}$ & 1.66(6) & 0.835(23) & $-0.54(3)$ & $-0.986(24)$ & $-1.00(20)$ & 1 & \chan \\
\jztto & $4.7\times10^{20}$ & 1.59(11) & $-0.663(21)$ & $-0.91(4)$ & $-0.95(25)$ & 0.7(5) & 2 & \eROS \\
\jztoo & $1.3\times10^{21}$ & 1.48(19) & 0.27(10) & $-0.53(11)$ & $-0.83(22)$ & $-1.0(9)$ & 1 & $-$ \\
\jttfo & $4.1\times10^{20}$ & 1.10(17) & 0.15(11) & $-1.00(8)$ & $1\pm4$ & $0.9(4)$ & 2 & $-$ \\
\josff & $7.2\times10^{21}$ & 0.98(7) & 0.54(5) & $-0.76(5)$ & $-0.96(7)$ & $0.6(3)$ & 3 & \chan \\
\hline
DR12 & & & & & & & & \\
\hline
\joofz & $8.0\times10^{21}$ & 18.3(9) & 0.73(3) & $-0.73(3)$ & $-1.00(5)$ & $1.0(4)$ & 1 & \eROS \\
\hline
\end{tabular}
\tablefoot{
The sources from each catalogue release are sorted by decreasing flux in the 0.5--1\,keV energy band (\texttt{SC\_EP\_2\_FLUX}). Errors (significant figures in brackets) represent $1\sigma$ confidence levels. 
\tablefoottext{a}{Total hydrogen column density along the line-of-sight \citep{2024arXiv240303127D}.}
\tablefoottext{b}{Catalogued EPIC flux in units of $10^{-14}$\,\fluxcgs\ in the 0.5--1\,keV energy band.}
\tablefoottext{c}{Hardness ratios (HR) defined as the ratio of the difference to the total counts in the five \xmm\ energy bands: 0.2--0.5\,keV, 0.5--1\,keV, 1--2\,keV, 2--4.5\,keV, and 4.5--12\,keV.}
\tablefoottext{d}{Number of \xmm\ detections, including both serendipitous and dedicated observations.}
\tablefoottext{e}{Available coverage from additional observatories.}
}
\end{table*}

\begin{table*}
\small
\caption{Summary of target \xmm\ detections and astrometric corrections
\label{tab_obsdet}}
\centering
\begin{tabular}{@{}lcccccrcrrrr@{}}
\hline\hline
Target & OBSID\,\tablefootmark{a} & Start date & MJD & Set-up\,\tablefootmark{b} & Off-axis\,\tablefootmark{c} & \multicolumn{2}{c}{GTI\,\tablefootmark{d}} & Counts\,\tablefootmark{e} & \multicolumn{3}{c}{Boresight correction\,\tablefootmark{f}} \\ 
\cline{7-8}
\cline{10-12}
& & & (days) & & (arcmin) & \multicolumn{1}{c}{(s)} & \multicolumn{1}{c}{(\%)} & & $\Delta\alpha$ (arcsec) & $\Delta\delta$ (arcsec) & $N_{\rm ref}$ \\
\hline
J1140 & 0109460801 & 2002-08-10 & 52496.757 & FF\ 1 & 8 & 6\,267 & 49 & $775\pm30$ & $<0.6$ & $<0.6$ & 30\\
J1818 & 0008820301 & 2002-04-07 & 52372.058 & FF\ 3 & 3 & 8\,967 & 65 & $498\pm25$ & $<0.4$ & $-1.6(4)$ & 125\\
 & 0008820601 & 2002-09-09 & 52526.286 & FF\ 3 & 4 & 12\,567 & 90 & $602\pm28$ & $1.10(28)$ & $-0.5(3)$ & 181\\
 & 0740990101 & 2014-09-09 & 56909.235 & FF\ 2 & 12 & 23\,833 & 93 & $720\pm30$ & $-0.78(19)$ & $-0.89(18)$ & 435\\
 & 0920040401 & 2023-03-06 & 60009.317 & FF\ 3 & 1.8 & 12\,067 & 93 & $790\pm30$ & $<0.4$ & $0.8(4)$ & 177\\
 & 0920040501 & 2023-03-22 & 60025.481 & FF\ 3 & 1.8 & 11\,067 & 93 & $648\pm29$ & $1.85(29)$ & $0.4(3)$ & 230\\
 & 0920040601 & 2023-04-07 & 60041.955 & FF\ 3 & 1.9 & 11067 & 93 & $666\pm29$ & $2.6(4)$ & $0.5(4)$ & 179\\
J1233 & 0147390701 & 2003-11-24 & 52967.677 & FF\ 1 & 17 & 11\,667 & 94 & $121\pm14$ & $0.68(27)$ & $-1.29(27)$ & 188\\
 & 0901260401 & 2022-11-20 & 59903.594 & FF\ 1 & 0.2 & 30\,267 & 95 & $1250\pm40$ & $0.40(20)$ & $0.49(20)$ & 287\\
J1403 & 0150020101 & 2003-07-19 & 52839.638 & FF\ 3 & 7 & 47\,200 & 72 & $600\pm30$ & $<0.4$ & $0.5(4)$ & 80\\
 & 0922820201 & 2024-01-18 & 60327.372 & FF/LW\ 2 & 0.2 & 36\ 400 & 82 & $950\pm40$ & $0.8(4)$ & $<0.5$ & 62\\
J0103 & 0784450501 & 2016-07-02 & 57571.377 & FF\ 1 & 10 & 15\,467 & 77 & $211\pm18$ & $-0.6(5)$ & $<0.5$ & 88\\
 & 0901261001 & 2023-01-24 & 59968.497 & FF\ 1 & 0.2 & 8\,700 & 29 & $85\pm13$ & $-2.7\pm1.2$ & $<1.2$ & 36\\
J1947 & 0603380101 & 2009-05-04 & 54955.891 & FF\ 2 & 4 & 66\,400 & 83 & $1060\pm40$ & $0.91(29)$ & $<0.3$ & 170\\  
J0221 & 0674110401 & 2012-02-09 & 55967.076 & FF\ 2 & 9 & 28\,867 & 90 & $1570\pm40$ & $-0.4(3)$ & $-0.56(27)$ & 83\\
 & 0884190401 & 2021-07-09 & 59405.013 & FF\ 1 & 0.3 & 29\,533 & 71 & $2650\pm60$ & $<0.3$ & $-1.4(4)$ & 88\\
J0311 & 0306680301 & 2005-09-04 & 53617.519 & FF\ 2 & 14 & 12\,500 & 20 & $130\pm16$ & $-0.7(3)$ & $<0.3$ & 94\\
J2251 & 0765030801 & 2015-11-27 & 57353.085 & FF\ 2 & 9 & 11\,167 & 94 & $123\pm14$ & $<0.5$ & $<0.5$ & 98\\
 & 0901260701 & 2022-05-15 & 59715.173 & FF\ 2/1 & 0.2 & 19\,267 & 61 & $184\pm19$ & $0.71(29)$ & $<0.3$ & 302\\
J1754 & 0206590201 & 2004-09-05 & 53253.293 & FF\ 2 & 8 & 19\,067 & 91 & $160\pm19$ & $-0.83(25)$ & $-0.64(26)$ & 158\\
 & 0402280101 & 2006-09-10 & 53989.086 & FF\ 2 & 8 & 42\,233 & 96 & $361\pm28$ & $-0.90(20)$ & $1.36(19)$ & 406\\
 & 0801683001 & 2018-03-16 & 58193.293 & FF\ 2 & 13 & 17\,167 & 61 & $58\pm12$ & $<0.3$ & $1.14(29)$ & 138\\
\hline
\end{tabular}
\tablefoot{
The sources are sorted by decreasing flux in the 0.5--1\,keV energy band (\texttt{SC\_EP\_2\_FLUX}). Errors (significant figures in brackets) indicate $1\sigma$ confidence levels. 
\tablefoottext{a}{Observation identifier.}
\tablefoottext{b}{Instrumental set-up of the EPIC cameras. The science mode is either full frame (FF) or large window (LW). The optical blocking filters of the pn and MOS cameras are denoted as: 1 = thin, 2 = medium, and 3 = thick.}
\tablefoottext{c}{Target off-axis angle in arcminutes, averaged across all active EPIC cameras.}
\tablefoottext{d}{Good-time intervals in seconds and as a percentage of the total observation duration, averaged across all active EPIC cameras.}
\tablefoottext{e}{Total EPIC counts in the 0.2--12\,keV energy band from PSF fitting.}
\tablefoottext{f}{Boresight corrections in right ascension ($\Delta\alpha$) and declination ($\Delta\delta$) in arcseconds, based on a number of $N_{\rm ref}$ potential matches with the GSC~2.4.2 catalogue.}
}
\end{table*}

\begin{table*}[t]
\small
\caption{\textit{SRG}/\eROS\ observations used in the analysis
\label{tab_erass}}
\centering
\begin{tabular}{llcllccccc}
\hline\hline
Target & \eROS\ source & Sky tile & Start date & End date & Scans & Exposure & Counts & \multicolumn{1}{c}{$\mathcal{L}$} & Separation\\
4XMM & 1eRASS & & & & & (s) & & & (\arcsec)\\
\hline
J114051.9$-$641848 & J114052.0$-$641849 & 174153 & 2020-01-12 & 2022-01-27 & 5 & 2\,423 & $354\pm20$ & 1000 & 1.8\\
J140340.4$-$603037 & J140340.3$-$603004 & 209150 & 2020-02-12 & 2022-02-15 & 5 & 1\,750 & $35\pm7$ & 46 & 2.0\\
J022141.5$-$735632 & J022140.9$-$735631 & 037165 & 2020-04-25 & 2021-11-10 & 4 & 1\,816 & $106\pm11$ & 275 & 2.6\\
\hline
\end{tabular}
\tablefoot{Shown are the \eROS\ IAU identifier and sky tile, the number of sky scans covering the source position within the given date range, the vignetting-corrected exposure time (in seconds), the net counts (0.2--5\,keV), the detection likelihood ($\mathcal{L}$), and the angular separation (in arcseconds) between the astrometrically corrected \eROS\ and \xmm\ coordinates.}
\end{table*}

\begin{table*}
\small
\caption{\chan\ observations used in the analysis\label{tab_chanobs}}
\begin{tabular}{llcccccrrcc}
\hline\hline
Target & \chan\ source & OBSID & Start date & MJD & Set-up & $\Delta\theta$ & \multicolumn{1}{c}{Exposure} & \multicolumn{1}{c}{Counts} & $\mathcal{L}$ & Separation \\
4XMM & 2CXO & & & (days) & & (\arcmin) & \multicolumn{1}{c}{(s)} & & & (\arcsec) \\ 
\hline\noalign{\vskip 0.4ex}
J194744.5$+$274220 & J194744.5$+$274220 & 4601 & 2004-01-29 & 53033.3 & ACIS-S & 5 & 32\,140 & $175\pm14$ & 12 & 0.4\\
J175437.8$-$294148 & J175437.7$-$294149 & 4547 & 2004-02-14 & 53049.5 & ACIS-I & 8 & 83\,167 & $66\pm10$ & 9 & 1.0\\
\hline
\end{tabular}
\tablefoot{Shown are the \chan\ source IAU identifier and observation ID, the observation start date and corresponding MJD, the instrumental set-up, the target off-axis angle, the exposure time, the net counts in the 0.3--10\,keV band, the detection likelihood ($\mathcal{L}$), and the angular separation (in arcseconds) between the astrometrically corrected \chan\ and \xmm\ positions.}
\end{table*}

\section{Source properties\label{sec_contaminants}}

\begin{figure*}[t]
\begin{center}
\includegraphics[width=0.245\textwidth]{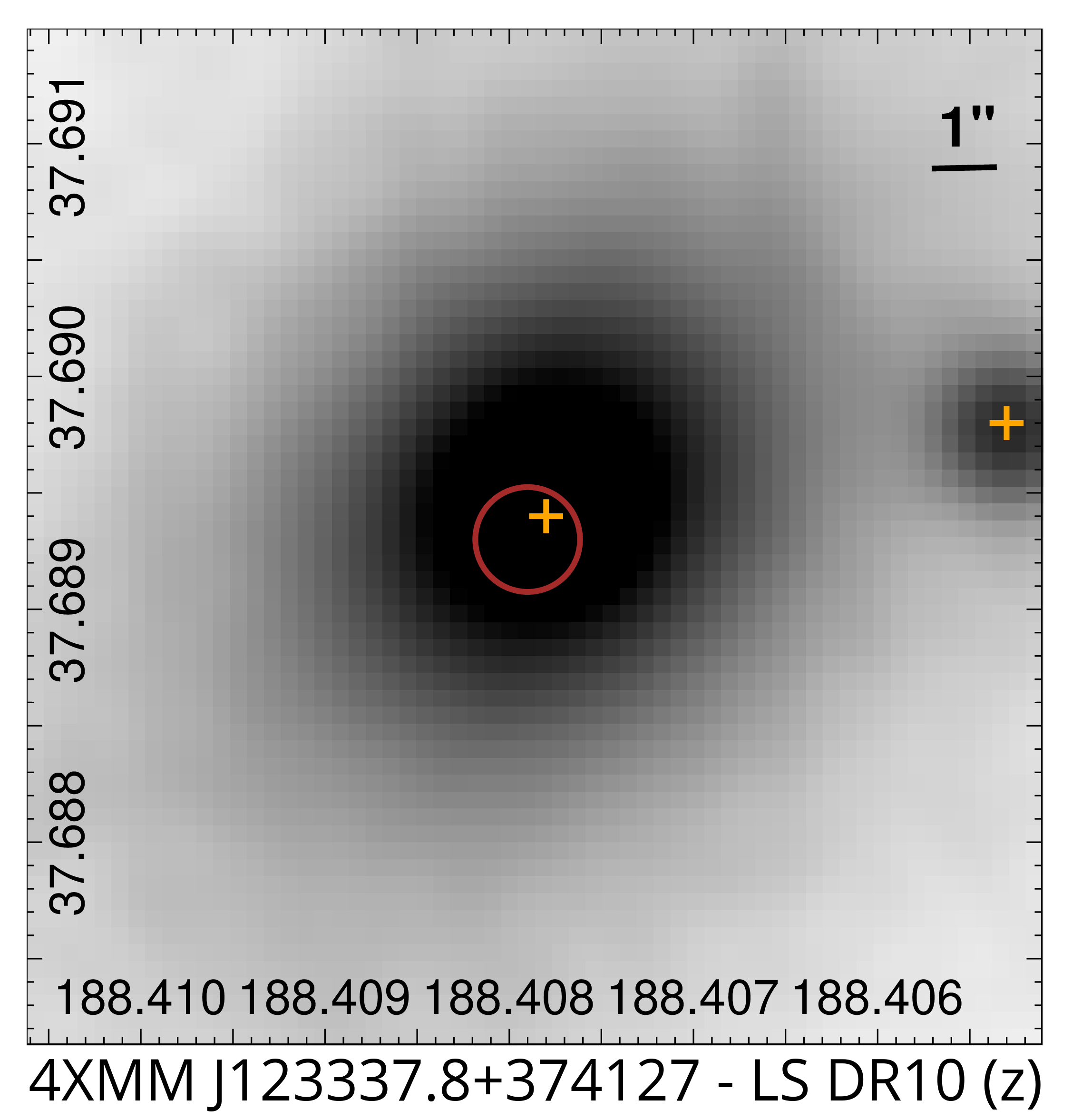}\hfill
\includegraphics[width=0.245\textwidth]{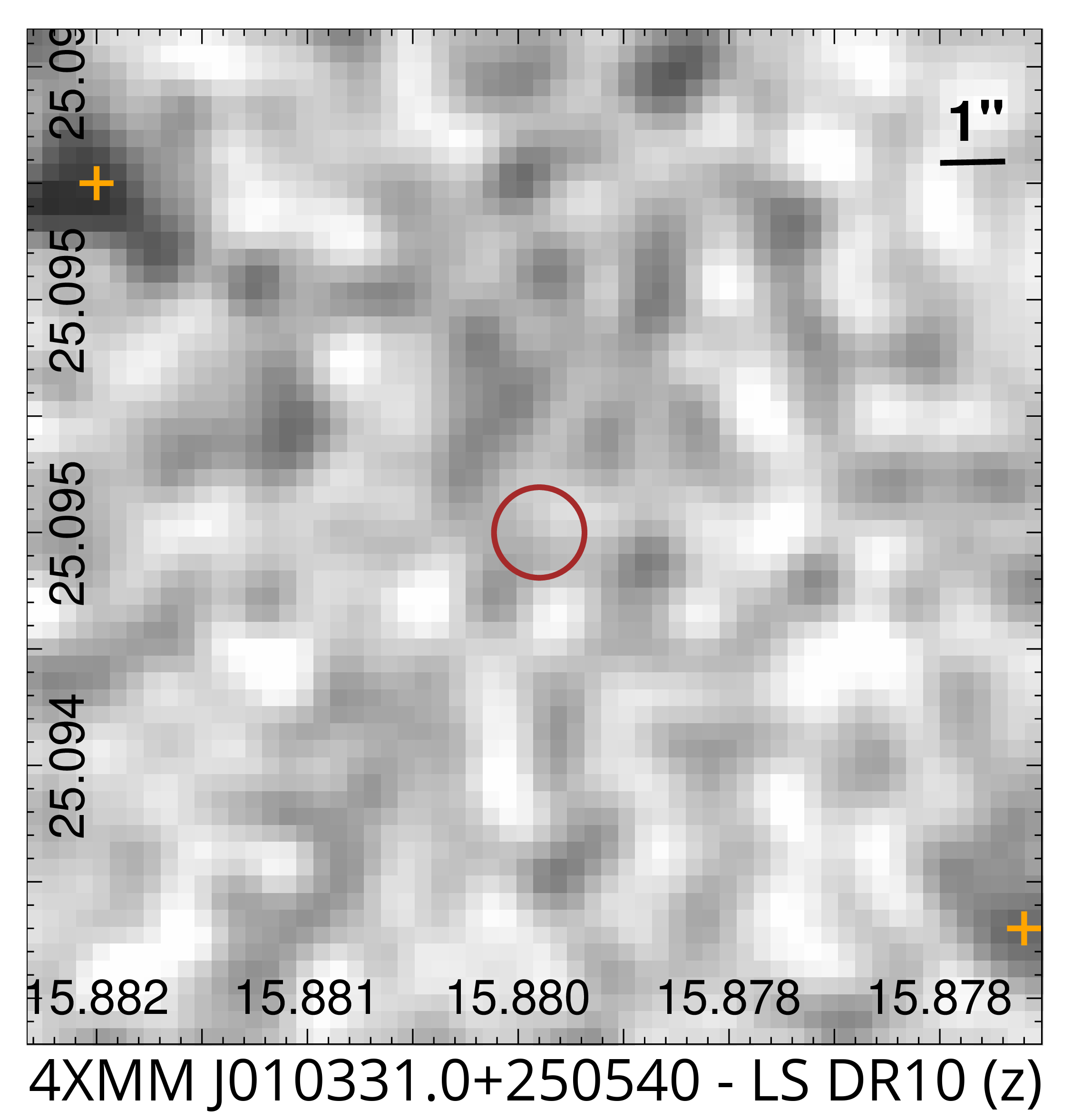}\hfill
\includegraphics[width=0.245\textwidth]{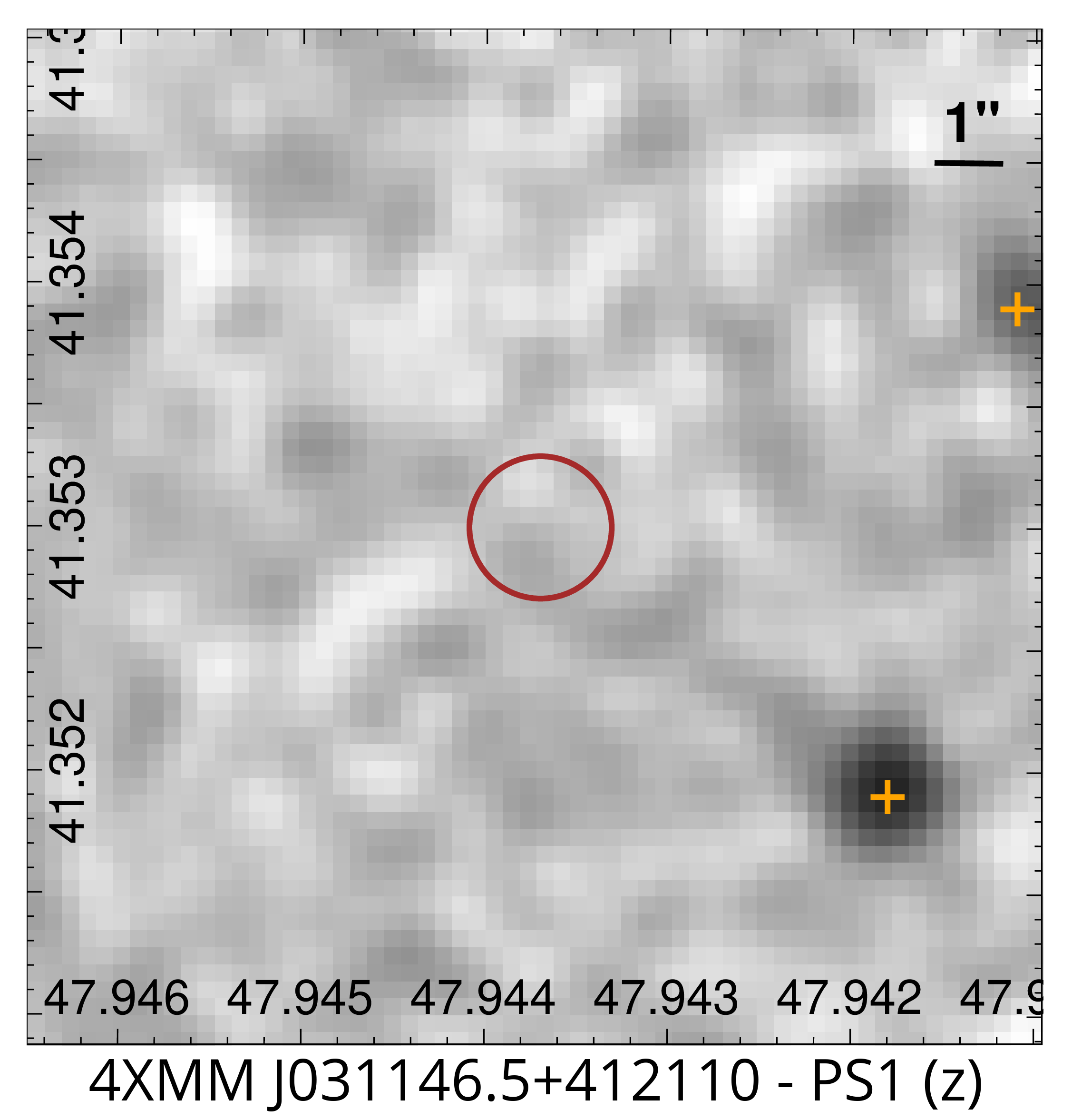}\hfill
\includegraphics[width=0.245\textwidth]{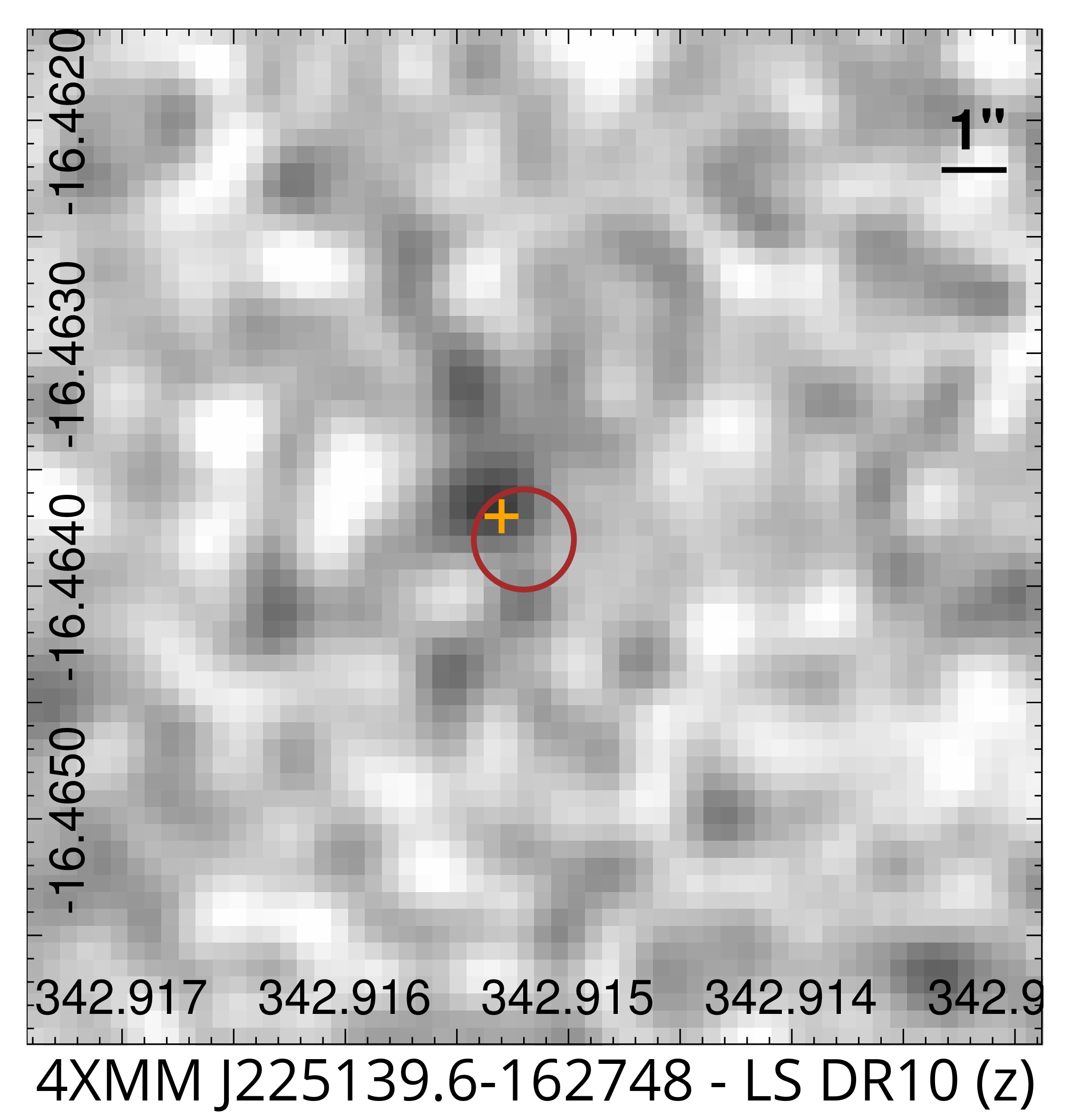}
\end{center}
\caption{Optical finding charts in the $z$ filter for the less certain 4XMM XINS candidates and likely contaminants: \liu\ and \sanshier\ (extragalactic sources), \shiyi\ (marginally flux-variable, no confirmed counterpart), and \ershisan\ (poorly constrained, possibly extragalactic). The solid circles indicate the 99.994\% confidence level uncertainties in the (stacked) \xmm\ source positions. The crosses mark the positions of catalogued optical objects in the $z$ band from the deepest available photometric survey.\label{fig_fcextra}}
\end{figure*}

In the following subsections, we discuss the poorly characterised 4XMM XINS candidates \shiyi\ and \ershisan, together with the contaminant sources \liu\ and \sanshier. Figure~\ref{fig_fcextra} shows the $z$-band finding charts for all four sources. The results of the spectral analysis are summarised in Table~\ref{tab_spec_extra}.

\begin{table*}[!t]
\small
\caption{Results of spectral analysis for contaminants or less compelling XINS candidates
\label{tab_spec_extra}}
\centering
\begin{tabular}{@{}lcrrr@{\hspace{10pt}}rr@{\hspace{10pt}}r@{\hspace{10pt}}rc@{\hspace{10pt}}c@{\hspace{3pt}}cr@{}}
\hline\hline\noalign{\vskip 0.4ex}
Model & NHP\,\tablefootmark{a} & $C$ || $\chi^2$\,\tablefootmark{b} & \multicolumn{1}{c}{$\nh$\,\tablefootmark{c}} & \multicolumn{1}{c}{$d_{N_{\rm H}}$\,\tablefootmark{d}} & \multicolumn{4}{c}{Continuum model parameters} & $z$ | $Z_{\odot}^{\rm O}$ & \multicolumn{2}{c}{\texttt{gabs} (eV)} & $\log(F_{\rm X})$\,\tablefootmark{e}\\
\cline{6-9}\cline{11-12}
\noalign{\vskip 0.4ex}
& (\%) & & & \multicolumn{1}{c}{(kpc)} & $kT_{\rm eV}$ || $T^{\rm eff}_{\rm 10^5\,K}$ & $R_{\rm km}^{\rm em}$ || $D_{\rm kpc}$ & $kT_{\rm eV}$ || $\Gamma$ & $R_{\rm km}^{\rm em}$ || $Z_{\rm Z_\odot}$ & & $\epsilon_1$ & $\epsilon_2$ & \\
\hline\noalign{\vskip 0.4ex}
\multicolumn{4}{l}{\jottt} & \multicolumn{9}{r}{(two epochs)\quad[0.2--2\,keV]\quad$\mathcal{C}=1590\pm40$\quad$\mathcal{B}=14$\,\%} \\
\hline\noalign{\vskip 0.4ex}
\texttt{apecX} & 20 & 75\,(65) & $0.3_{-0.2}^{+0.2}$ & $>3.3$ & $-$ & $-$ & $1122_{-22}^{+20}$ & $0.88_{-0.13}^{+0.16}$ & 0.1\,\tablefootmark{$\dagger$} & $-$ & $-$ & $-13.16_{-0.01}^{+0.01}$\\
\texttt{bb} & 0 & 277\,(66) & $0.29$\,\tablefootmark{$\dagger$} & $>3.3$ & $244_{-5}^{+5}$ & $0.20_{-0.01}^{+0.01}$ & $-$ & $-$ & 0 & $-$ & $-$ & $-13.25_{-0.01}^{+0.01}$\\
\hline\noalign{\vskip 0.4ex}
\multicolumn{4}{l}{\jzozt} & \multicolumn{9}{r}{(two epochs)\quad[0.2--2\,keV]\quad$\mathcal{C}=260\pm20$\quad$\mathcal{B}=21$\,\%} \\
\hline\noalign{\vskip 0.4ex}
\texttt{bb} & 81 & 48\,(46) & $0.68$\,\tablefootmark{$\dagger$} & $>1.9$ & $165_{-10}^{+11}$ & $0.17_{-0.02}^{+0.03}$ & $-$ & $-$ & 0 & $-$ & $-$ & $-13.35_{-0.04}^{+0.04}$\\
\texttt{diskbb} & 90 & 43\,(46) & $0.68$\,\tablefootmark{$\dagger$} & $>1.9$ & $-$ & $-$ & $238_{-19}^{+22}$ & $1.03(16)\times10^4$ & $0.02$\,\tablefootmark{$\dagger$} & $-$ & $-$ & $-13.27_{-0.04}^{+0.04}$\\
\texttt{nsa} & 90 & 43\,(46) & $0.68$\,\tablefootmark{$\dagger$} & $>1.9$ & $9.5_{-1.0}^{+1.1}$ & $5.7_{-1.4}^{+1.6}$ & $-$ & $-$ & 0 & $-$ & $-$ & $-13.26_{-0.05}^{+0.05}$\\
\texttt{pl} & 92 & 41\,(46)  & $0.68$\,\tablefootmark{$\dagger$} & $>1.9$ & $-$ & $-$ & $2.87_{-0.19}^{+0.19}$ & $-$ & 0 & $-$ & $-$ & $-13.01_{-0.05}^{+0.06}$\\
\hline\noalign{\vskip 0.4ex}
\multicolumn{4}{l}{\jztoo} & \multicolumn{9}{r}{(one epoch)\quad[0.2--2\,keV]\quad$\mathcal{C}=122\pm11$\quad$\mathcal{B}=19\%$} \\
\hline\noalign{\vskip 0.4ex}
\texttt{apecX} & 52 & 18\,(22) & 1.3\,\tablefootmark{$\dagger$} & $>1.9$ & $-$ & $-$ & $1440_{-180}^{+240}$ & 0.2\,\tablefootmark{$\dagger$} & 1.5\,\tablefootmark{$\dagger$} & $-$ & $-$ & $-13.04_{-0.05}^{+0.05}$\\
\texttt{bb} & 84 & 14\,(19) & $1.3_{-0.9}^{+1.2}$ & $1.2_{-0.9}^{+0.7}$ & $160_{-13}^{+15}$ & $0.37_{-0.07}^{+0.09}$ & $-$ & $-$ & 0 & $-$ & $-$ & $-13.16_{-0.06}^{+0.06}$\\
\texttt{nsa} & 88 & 12\,(19) & $2.6_{-1.3}^{+1.5}$ & $>1.9$ &$6.8_{-1.0}^{+1.2}$ & $1860_{-650}^{+1000}$ & $-$ & $-$ & 0 & $-$ & $-$ & $-12.82_{-0.09}^{+0.09}$\\
\hline\noalign{\vskip 0.4ex}
\multicolumn{4}{l}{\jttfo} & \multicolumn{9}{r}{(two epochs)\quad[0.2--2\,keV]\quad$\mathcal{C}=390\pm20$\quad$\mathcal{B}=16\%$} \\
\hline\noalign{\vskip 0.4ex}
\texttt{apecX} & 35 & 71\,(62) & 0.41\,\tablefootmark{$\dagger$} & $>2.0$ & $-$ & $-$ & $1005_{-60}^{+60}$ &  $0.2$\,\tablefootmark{$\dagger$} & 1.5\,\tablefootmark{$\dagger$} & $-$ & $-$ & $-13.39_{-0.03}^{+0.03}$\\
\texttt{bb} & 73 & 50\,(54) & $<0.1$ & $>2.0$ & $122_{-6}^{+6}$ & $0.56_{-0.06}^{+0.07}$ & $-$ & $-$ & 0 & $-$ & $-$ & $-13.58_{-0.03}^{+0.03}$\\
\texttt{pl} & 88 & 48\,(54) & $0.8_{-0.3}^{+0.4}$ & $>2.0$ & $-$ & $-$ & $4.42_{-0.18}^{+0.19}$ & $-$ & 0 & $-$ & $-$ & $-12.85_{-0.05}^{+0.06}$\\
\hline\noalign{\vskip 0.4ex}
\end{tabular}
\tablefoot{
Quoted errors correspond to 1$\sigma$ confidence levels. All best-fit models shown assume no inter-epoch variability. For each target, we list the number of available epochs; total counts ($\mathcal{C}$) refer to the indicated energy bands, while the mean background level per epoch, $\mathcal{B}$, is averaged over detectors.
Galactic and extragalactic APEC models (\texttt{apecG}, \texttt{apecX}) assume solar abundances at redshift $z = 0$ and a free abundance ($Z/Z_\odot$) at the fixed redshift given in the table, respectively. For \texttt{diskbb}, the face-on inner radius (in km) is computed using the distance $d_{\nh}$ listed in the table for Galactic sources, or assuming a fixed redshift of $z = 0.02$ for extragalactic cases. The blackbody emission radius is computed using the distance $d_{\nh}$ listed in the table. Neutron star atmosphere models assume $M = 1.4$\,M$_\odot$, $R = 10$\,km, and a non-magnetic configuration ($B = 0$\,G) as implemented in the model grid.
\tablefoottext{a}{Null-hypothesis probability (percent) that the data are consistent with the model.}
\tablefoottext{b}{C or $\chi^2$ statistic, with corresponding degrees of freedom in brackets (d.o.f.).}
\tablefoottext{c}{Hydrogen column density in units of $10^{21}$\,cm$^{-2}$.}
\tablefoottext{d}{Distance estimate from 3D extinction/absorption maps \citep{2024arXiv240303127D}, based on the best-fit $\nh$.}
\tablefoottext{e}{Logarithm of the unabsorbed flux in \fluxcgs\ (0.2--12\,keV).}
\tablefoottext{$\dagger$}{Parameter fixed during fitting.}
}
\end{table*}

\subsection{4XMM J123337.8$+$374127}

The X-ray source was detected solely by the EPIC pn camera in 2003, at a large off-axis angle of 16.8\arcmin. A follow-up observation in 2022 from our programme, with the source positioned at the aimpoint, reveals a significant positional shift of 5.8\arcsec and clear evidence of spatial extent. The revised position unambiguously identifies the source as the X-ray counterpart of the radio galaxy \object{NVSS~J123337+374122} (Fig.~\ref{fig_fcextra}), confirming it as a contaminant in our sample of XINS candidates. 

Among the thermal models tested, the source spectrum---with an observed flux of $5.96(19)\times10^{-14}$\,\fluxcgs\ (0.2--2\,keV)---is best fit by an \texttt{apec} model with $\nh \sim 3 \times 10^{20}$\,cm$^{-2}$, $kT \sim 1.1$\,keV, and abundance $Z \sim 0.9$\,Z$_\odot$, assuming a fixed redshift of $z = 0.1$ \citep{2020ApJS..249....3A}. There is no evidence of long-term X-ray variability. A search for coherent pulsations in the follow-up \fast\ observation yields no candidate periodicities (down to $\sim4$\,$\mu$Jy), and no single-pulse modulations are detected. The X-ray-to-optical flux ratio of $X/O\sim-0.47$ is based on the optical counterpart magnitudes from \gaia, PS1, and LS surveys at different epochs.

\subsection{4XMM J010331.0$+$250540}

The high Galactic latitude X-ray source was first detected in a 2016 observation of the galaxy cluster \object{PSZ2 G126.61$-$37.63}. The 2023 follow-up observation was affected by strong background flaring, resulting in a loss of over 70\% of the exposure.

A steep power-law model with photon index $\Gamma = 2.87(19)$, a neutron star atmosphere model with temperature $(1.0$--$1.3)\times10^{6}$\,K at a distance of 6--11\,kpc, or an optically thick, geometrically thin accretion disk model (\texttt{diskbb}; \citealt{1984PASJ...36..741M,1986ApJ...308..635M}) with inner disk temperature $\sim240$\,eV all provide comparably good fits to the X-ray spectrum ($\mathrm{NHP}\sim90\%$; Table~\ref{tab_spec_extra}). Adding further spectral components yields only marginal improvements and remains poorly constrained. 
Despite the limited photon statistics---only $\sim$260 spectral counts across the two epochs---we find no evidence for changes in spectral shape. The observed flux appears to decline by a factor of $\sim$2--3 over seven years; however, this trend is only marginally significant and model dependent.
No optical counterpart is found down to a $4\sigma$ limit of $g = 24.7$ (Fig.~\ref{fig_fcextra}), implying a high X-ray-to-optical flux ratio of $X/O > 2.03$ ($4\sigma$).

Given the low photon statistics and the absence of a secure multiwavelength counterpart, the physical nature of this source remains unclear. While its soft X-ray spectrum and optical faintness make it potentially interesting, deeper X-ray and optical observations are required to clarify its origin.

\subsection{4XMM J031146.5$+$412110}

The intermediate Galactic latitude source at $b = -14$\degr\ was detected in a 2005 \xmm\ observation of the Seyfert~1 galaxy \object{ICRF~J031301.9+412001}, at a large off-axis angle of 14.2\arcmin. Its observed X-ray flux is $3.2(4) \times 10^{-14}$\,\fluxcgs\ in the 0.2--2\,keV band. The source has not been re-observed and remains one of the least well characterised in our sample, with only $122 \pm 11$ photon counts recorded in a single epoch. The archival detection was severely affected by high background, resulting in 80\% data loss and a net exposure of 12.5\,ks. 
Its localisation is relatively poor compared to other XINS candidates, at $\sim$1.1\arcsec, though it lies in a sparsely populated region of the sky where source confusion is unlikely (Fig.~\ref{fig_fcextra}).
No optical counterparts are known down to $g < 22.7$ ($4\sigma$), implying an X-ray-to-optical flux ratio of $X/O > 1.10$, within the range observed for extragalactic sources.

The spectrum is well described either by a simple blackbody model with $kT = 160^{+15}_{-13}$\,eV and a small radius of 0.3--0.5\,km at $1.2_{-0.9}^{+1.2}$\,kpc, or by a non-magnetised \texttt{nsa} model with $T_{\rm eff} = (6.8^{+1.2}_{-1.0}) \times 10^{5}$\,K at a model distance of $1.9^{+1.0}_{-0.7}$\,kpc; both fits yield NHPs in the range 84--88\%. Magnetised models with $B = 10^{12}-10^{13}$\,G result in higher temperatures of $\sim10^{6}$\,K and larger distances of $4.4^{+2.0}_{-1.4}$\,kpc, with comparable fit quality. The inferred column density, $\nh = (1.0-2.6) \times 10^{21}$\,cm$^{-2}$, is consistent with the Galactic value along the line-of-sight to within a factor of two, supporting a distance of $\gtrsim1-2$\,kpc. The implied luminosity for these models is $(1.3-2.6) \times 10^{32}$\,\lumcgs. 
Thermal plasma models (\texttt{apec}) with free redshift or abundance are not well constrained; a model with fixed $\nh = 1.3 \times 10^{21}$\,cm$^{-2}$, $z = 1.5$, and $Z = 0.2\,Z_\odot$ yields $kT = 1.44^{+0.24}_{-0.18}$\,keV and an NHP of 52\%.

\subsection{4XMM J225139.6$-$162748}

The X-ray source (observed flux: $2.59_{-0.16}^{+0.17}\times10^{-14}$\,\fluxcgs; 0.2--2\,keV) was first detected at a large off-axis angle in a 2015 observation of the galaxy cluster \object{ACO~2496}, with a follow-up in 2022 as part of our \xmm\ programme. The emission is soft and shows no significant flux or spectral variability over seven years. The updated X-ray position, offset by 1.4\arcsec\ relative to the 4XMM-DR9 coordinates, unambiguously identifies a faint optical counterpart ($X/O \sim 1.92$; Fig.~\ref{fig_fcextra}) catalogued in LS DR10 as a faint optical source ($g \approx 23.7$, $r \approx 22.4$, $i \approx 22.7$, $z \approx 22.3$). It shows a red $g-r = 1.35$ and an unusually negative $r-i = -0.33$, possibly due to photometric uncertainties or spectral features such as emission lines. The 0.63\arcsec\ X-ray/optical separation supports a robust association.

The steep X-ray spectrum (power-law fit with $\Gamma = 4.42_{-0.18}^{+0.19}$; $\nh$ and distance unconstrained), along with thermal plasma fits favouring low metallicity and moderate redshift, strongly suggest an extragalactic origin---likely a soft-excess quasar or distant AGN. With only $\sim$390 net counts, we cannot reliably constrain more complex models (e.g.~a soft excess plus a hard tail), leaving open the possibility that we are detecting only the soft component of a broader AGN spectrum.

\section{Upper limits on X-ray pulsations}

Upper limits on pulsations derived from the EPIC-pn and EPIC datasets for all ten sources are reported in Table~\ref{tab_pflim}.

\begin{table*}[t]
\caption{Pulsed fraction limits from X-ray timing analysis
\label{tab_pflim}}
\small
\centering
\begin{tabular}{lccccccccccc}
\hline\hline
\multicolumn{1}{l}{Searched dataset} & Frequency range (Hz) & J1140 & J1818 & J1233 & J1403 & J0103 & J1947 & J0221 & J0311 & J2251 & J1754 \\
\hline
pn & 0.002--6.8 & 26 & 40 & 28 & 30 & 49 & 28 & 14 & 35 & 28 & 63\\
EPIC & 0.002--0.19 & 19 & 30 & 21 & 24 & 39 & 21 & 11 & 31 & 24 & 44\\
\hline
\end{tabular}
\tablefoot{All values refer to the 0.2--2\,keV energy band. Pulsed fraction limits in percentage are quoted at the 4$\sigma$ confidence level.}
\end{table*}

\section{ROSAT and SRG/eROSITA XINSs}

Table~\ref{tab_xins_allsky} summarises the properties of XINSs discovered from both wide-area \ros\ and eRASS observations. For a description of the XINSs identified from \xmm\ observations, see Table~\ref{tab_xins_summary} in Section~\ref{sec_population}.

\begin{table*}
\small
\caption{Summary of the properties of XINSs from \ros\ and \textit{SRG}/\eROS\label{tab_xins_allsky}}
\centering
\begin{tabular}{@{}l@{}cccccccccc@{}}
\hline\hline
Identifier & Flux & Distance & $|h_{z}|$\,\tablefootmark{a} & $kT$\,\tablefootmark{b} & Energy\,\tablefootmark{c} & $p_{\rm f}$\,\tablefootmark{d} & Period\tablefootmark{e} & $\log(L_{\rm X})$\,\tablefootmark{f} & Class & References \\
RX | eRASSU & ($\times10^{-13}$ cgs) & (kpc) & (pc) & (eV) & (keV) & (\%) & (s) & (\lumcgs) & & \\ 
\hline
J1856.5$-$3754 & 140 & 0.12 & 40 & 62 & $-$ & 1 & 7.06 & 31.7--31.9 & \msev & [1]\\
J0720.4$-$3125 & 100 & 0.29 & 40 & 87 & $(0.36,0.55,0.77)$ & 11 & 8.39 & 32.0--32.5 & \msev & [1]\\
J1605.3$+$3249 & 60 & 0.1--0.4 & 70--300 & 93 & $(0.44,0.56,0.85)$ & $<1.3$ & ? & 30.8--32.7 & \msev & [1]\\
J1308.6$+$2127 & 30 & 0.38 & 400 & 102 & $(0.21,0.54,0.78)$ & 18 & 10.31 & 32.4--32.6 & \msev & [1]\\
J2143.0$+$0654 & 29 & 0.41 & 220 & 104 & $(0.39,0.55,0.74)$ & 4 & 9.43 & 31.7--32.2 & \msev & [1]\\
J0806.4$-$4123 & 24 & 0.24 & 20 & 92 & 0.3 & 6 & 11.37 & 31.2--31.4 & \msev & [1]\\
Calvera & 8 & 1.0 & 2100 & 200 & $-$ & 18 & 0.06 & 32.3--33.7 & CCO & [1]\\
J131716.9$-$402647 & 4.6 & $<0.7$ & 300 & 88 & $(0.26,0.59)$ & 23 & 12.76 & $<32.6$ & \msev-like & [2,3]\\
J081952.1$-$131930 & 4.0 & $<0.3$ & $<70$ & 48 & $-$ & $<12$ & ? & $<31.5$& XINS & [4]\\ 
J065715.3$+$260428 & 2.6 & 0.7--1.5 & 150--350 & 93 & 0.3 & 15 & 0.26 & 30.6--32.5 & RPP & [5] \\
J084046.2$-$115222 & 2.5 & $<1.2$ & $<400$ & 67 & 0.28 & $<13$ & ? & $<32.6$ & XINS & [4] \\
J0420.0$-$5022 & 2.2 & 0.34 & 240 & 45 & $-$ & 17 & 3.45 & 30.6--30.9 & \msev & [1]\\
J134725.4$-$363415 & 1.4 & $<1.1$ & $<500$ & 80 & 0.36 & $<16$ & ? & $< 31.9$ & XINS & [4] \\
J072302.3$-$194225 & 1.3 & $<1.5$ & $<60$ & 71 & $-$ & $<14$ & ? & $< 32.5$ & XINS & [4] \\
\hline
\end{tabular}
\tablefoot{The sources are sorted by decreasing observed flux. References: [1] \citet{2020MNRAS.496.5052P}, [2] \citet{2024A&A...683A.164K}, [3] Kurpas et al.~(in prep.), [4] \citet{2026A&A...705A.148K}, [5] \citet{2025A&A...694A.160K}.
\tablefoottext{a}{Vertical distance from the Galactic plane, $h_{z}=d\sin|b|$.}
\tablefoottext{b}{Blackbody temperature of the dominant thermal component.}
\tablefoottext{c}{Energy of continuum deviations modelled as absorption features, when reported.}
\tablefoottext{d}{Measured pulsed fraction or $4\sigma$ upper limits.}
\tablefoottext{e}{Spin period of the neutron star.}
\tablefoottext{f}{X-ray luminosity from \citet{2020MNRAS.496.5052P}, or calculated from the blackbody model and estimated distance range, in logarithmic units.}}
\end{table*}

\section{Population synthesis model\label{sec_popsynt}}

The population synthesis model used to simulate the Galactic XINS population is based on the evolutionary framework of \citet{2006ApJ...643..332F}, originally developed for radio pulsars. Its implementation for cooling neutron stars is described in \citet[][for technical details and references]{2009PhDT.......222P}, and was previously used to forecast detections in the \eROS\ All-Sky Survey \citep{2017AN....338..213P}. Here, we adapt the same framework to assess the detectability of XINSs within the cumulative \xmm\ observational footprint and to compare the resulting population with sources detected in eRASS.

Neutron stars are generated according to a Galactic birth distribution and evolved in time within a realistic Galactic potential. The potential includes disk, bulge, and halo components \citep{1987gady.book.....B,1992ApJS...83..111W,1996A&A...313L..21R,2005ApJ...631..838W,2005MNRAS.358.1325C}. Spiral arms are included as a perturbation to the axisymmetric disk potential \citep{1991A&A...243..373P}, with birth sites following a four-arm logarithmic spiral structure \citep{2006ApJ...643..332F}.

The spatial birth distribution follows the radial profile of the Galactic radio pulsar population \citep{2004A&A...422..545Y}, with a vertical scale height of 50\,pc. Natal kick velocities are drawn from an isotropic exponential distribution with a mean of 380\,km\,s$^{-1}$ \citep{2006ApJ...643..332F}, which governs the subsequent dynamical evolution. A constant birthrate of approximately two neutron stars per century is adopted, consistent with estimates for thermally emitting isolated neutron stars \citep{2007MNRAS.381...52G}. Only neutron stars younger than $10^8$\,yr, relevant for thermal X-ray emission, are retained.

Thermal evolution follows a standard cooling curve for magnetised hadronic matter \citep{2000smcy.confE..15H}. All neutron stars are assumed to have canonical parameters, with a mass of 1.4\,M$_\odot$ and a radius of 13.8\,km, corresponding to a surface gravity of $10^{14}$\,cm\,s$^{-2}$ and a gravitational redshift of $z_g = 0.2$. Emission is modelled as isotropic blackbody radiation from a fixed dipolar magnetic field of $B = 10^{14}$\,G. Magnetic field decay and magneto-thermal coupling are neglected.

Interstellar absorption towards each source is computed using a model of atomic and molecular hydrogen \citep{1990ARA&A..28..215D}, together with standard photoelectric absorption cross-sections \citep{1983ApJ...270..119M}. The absorbed spectrum is folded through the instrument response matrices to obtain detector- and energy-band-dependent count rates:
\begin{equation}
	S_{d,k} = \int_{\epsilon_k} \frac{f_\epsilon}{\epsilon} A^{(d)}_\epsilon \exp\!\big[-\sigma_\epsilon N_{\mathrm{H}}(l,b)\big]\, d\epsilon,
\end{equation}
where $f_\epsilon=(R_\infty/d)^2 F_\epsilon^\infty$, $R_\infty$ is the redshifted emission radius, and $F_\epsilon^\infty$ is the isotropic blackbody flux at infinity. For \xmm, $A^{(d)}_\epsilon$ denotes the energy-dependent effective area and redistribution matrix for detector $d \in \{\mathrm{pn}, \mathrm{MOS1}, \mathrm{MOS2}\}$, evaluated separately for each optical blocking filter (thin, medium, thick). The index $k$ corresponds to the five standard 4XMM energy bands (0.2--0.5, 0.5--1, 1--2, 2--4.5, 4.5--12\,keV).

For each source, we considered all 4XMM-DR12 observations in which it lies within the EPIC field of view (off-axis angle $\theta \leq 15$ arcmin), restricting to full-frame and extended full-frame observing modes. The relevant detector configuration, filter setting, and GTI exposure are taken from the 4XMM-DR12 summary table provided by the SSC\footnote{Key details of the observations included in the \xmm\ footprint are distributed with each catalogue release; cf.~\url{http://xmmssc.irap.omp.eu/Catalogue/4XMM-DR12/4XMM_DR12.html}.}. The corresponding count rates are corrected for off-axis vignetting using the multiplicative factor $V^{(\mathrm{obs},d)}(\theta)$, derived from Moffat-function approximations of the corresponding camera vignetting Current Calibration File (CCF) at 0.5\,keV\footnote{Specifically, the calibration files XRT1\_XAREAEF\_0010.CCF, XRT2\_XAREAEF\_0011.CCF, and XRT3\_XAREAEF\_0014.CCF are used for the MOS1, MOS2, and pn detectors, respectively.}.

The total EPIC counts of the synthetic neutron star $i$ across active detectors for a given observation follow directly from Eq.~\ref{eq_counts}. If a source is covered by multiple observations, the detection corresponding to the maximum total EPIC counts is retained:
\begin{equation}
	\mathcal{C}_i = \max_{\mathrm{obs}} \left(
	\sum_{d}
	\sum_{k}
	S_{d,k}\, V^{(\mathrm{obs},d)}(\theta)\,
	\Delta t_{\mathrm{exp}}^{(\mathrm{obs},d)}
	\right),
	\label{eq_counts}
\end{equation}
where $\Delta t_{\mathrm{exp}}^{(\mathrm{obs},d)}$ denotes the GTI exposure time for each detector.

A source is classified as detectable if $\mathcal{C}_i \ge \mathcal{C}_{\mathrm{lim}}$, where $\mathcal{C}_{\mathrm{lim}} = 10$ EPIC counts (0.2--12\,keV), applied uniformly across the 4XMM-DR12 footprint. The number of detected XINSs is then
\begin{equation}
	N_{\rm XINS} = \sum_i \delta_i,
\end{equation}
where $\delta_i = 1$ if $\mathcal{C}_i \ge \mathcal{C}_{\mathrm{lim}}$ and $\delta_i = 0$ otherwise.

To account for stochastic variations in birth properties, kick velocities, dynamical evolution, and interstellar absorption, we performed 100 Monte Carlo realisations. In each realisation, a synthetic Galactic neutron-star population is generated and propagated through the \xmm\ and \eROS\ selection functions. These ensembles are used to derive median values and statistical uncertainties reported in Table~\ref{tab_simresults}, and to construct the cumulative $\log N$--$\log S$ distributions shown in Fig.~\ref{fig_cumdist_lnls}.

\end{appendix}


\end{document}